\documentclass[prd,preprint,showpacs,nofootinbib,preprintnumbers]{revtex4}
 \usepackage{graphicx, epsfig,color, amsmath, bm} 
\usepackage{multirow} 

\def\to{\rightarrow}

\def\bi{\begin{itemize}}
\def\ei{\end{itemize}}

\def\tu{\tilde u} 
\def\tc{\tilde c}

\def\tf{\tilde f}
\def\td{\tilde d} 
\def\tQ{\tilde Q}

\def\tst{\tilde t} 
 
\def\tmu{\tilde \mu}
\def\tg{\tilde g} 
\def\tnu{\tilde\nu} 

\def\tq{\tilde q} 
\def\tw{\tilde W} 
\def\tz{\tilde Z}

 \def\alt{\stackrel{<}{\sim}}
 
\def\be{\begin{equation}}
\def\ee{\end{equation}} 
\def\sn{s} 
\def\cs{c}

\def\dblone{\hbox{$\bm{1}\hskip -1.2pt\vrule depth 0pt height 1.6ex width 0.7pt
                  \vrule depth 0pt height 0.3pt width 0.12em$}}

\newcommand\jhep[3]{{\it J. High Energy Phys.\ }{\bf #1} (#2) #3}

\newcommand\npb[3]{{\it Nucl.\ Phys.\ }{\bf B #1} (#2) #3}

\newcommand{\hepph}[1]{hep-ph/#1}

\newcommand{\msb}{\overline{\mathrm{MS}}}
\newcommand{\drb}{\overline{\mathrm{DR}}}
\newcommand{\GeV}{\mathrm{GeV}}
\newcommand{\by}{\mathbf{Y}}
\newcommand{\bu}{\mathbf{U}}
\newcommand{\bv}{\mathbf{V}}
\newcommand{\bw}{\mathbf{W}}
\newcommand{\bx}{\mathbf{X}}

\newcommand{\bdf}{\mathbf{f}}
\newcommand{\bdg}{\mathbf{g}}
\newcommand{\ba}{\mathbf{a}}

\newcommand{\boldl}{\bm{\lambda}}

\newcommand{\bgt}{\tilde{\mathbf{g}}}
\newcommand{\bbet}{\bm{\beta}}
\newcommand{\bdm}{\mathbf{m}}
\newcommand{\bdmsf}{\mathbf{\mathsf{m}}}
\newcommand{\bdLm}{\bm{\Lambda}}
\newcommand{\bmfx}{\left(\mathbf{m}_{X}-i\mathbf{m}'_{X}\right)}
\newcommand{\bmfxd}{\left(\mathbf{m}_{X}+i\mathbf{m}'_{X}\right)}
\newcommand{\bftuq}{(\tilde{\mathbf{f}}^Q_u)}
\newcommand{\bftqq}{(\tilde{\mathbf{f}}^Q_q)}

\newcommand{\bftur}{(\tilde{\mathbf{f}}^{u_R}_u)}
\newcommand{\bftqr}{(\tilde{\mathbf{f}}^{q_R}_q)}

\newcommand{\bftdq}{(\tilde{\mathbf{f}}^Q_d)}
\newcommand{\bftdr}{(\tilde{\mathbf{f}}^{d_R}_d)}
\newcommand{\bftel}{(\tilde{\mathbf{f}}^L_e)}
\newcommand{\bfter}{(\tilde{\mathbf{f}}^{e_R}_e)}
\newcommand{\bgtpq}{(\tilde{\mathbf{g}}'^Q)}

\newcommand{\bgtpl}{(\tilde{\mathbf{g}}'^L)}
\newcommand{\bgtpur}{(\tilde{\mathbf{g}}'^{u_R})}
\newcommand{\bgtpurd}{(\tilde{\mathbf{g}}'^{u_{R}\dagger})}

\newcommand{\bgtpdr}{(\tilde{\mathbf{g}}'^{d_R})}
\newcommand{\bgtpdrd}{(\tilde{\mathbf{g}}'^{d_{R}\dagger})}
\newcommand{\bgtper}{(\tilde{\mathbf{g}}'^{e_R})}
\newcommand{\bgtq}{(\tilde{\mathbf{g}}^Q)}

\newcommand{\bgtl}{(\tilde{\mathbf{g}}^L)}
\newcommand{\bgtsq}{(\tilde{\mathbf{g}}_s^{Q})}

\newcommand{\bgtsur}{(\tilde{\mathbf{g}}_s^{u_R})}

\newcommand{\bgtsdr}{(\tilde{\mathbf{g}}_s^{d_R})}

\newcommand{\gtphu}{\tilde{g}'^{h_u}}

\newcommand{\gtphd}{\tilde{g}'^{h_d}}

\newcommand{\gthu}{\tilde{g}^{h_u}}
\newcommand{\gthd}{\tilde{g}^{h_d}}

\newcommand{\fuhu}{(\mathbf{f}_{u})}
\newcommand{\fuhut}{(\mathbf{f}_{u})^{T}}
\newcommand{\fuhus}{(\mathbf{f}_{u})^{*}}
\newcommand{\fdhd}{(\mathbf{f}_{d})}
\newcommand{\fdhdt}{(\mathbf{f}_{d})^{T}}
\newcommand{\fdhds}{(\mathbf{f}_{d})^{*}}
\newcommand{\fehd}{(\mathbf{f}_{e})}
\newcommand{\fehdt}{(\mathbf{f}_{e})^{T}}
\newcommand{\fehds}{(\mathbf{f}_{e})^{*}}

\newcommand{\fuul}{(\mathbf{f}_{u})}
\newcommand{\fuult}{(\mathbf{f}_{u})^{T}}
\newcommand{\fuuls}{(\mathbf{f}_{u})^{*}}
\newcommand{\fuur}{(\mathbf{f}_{u})}
\newcommand{\fuurt}{(\mathbf{f}_{u})^{T}}
\newcommand{\fuurs}{(\mathbf{f}_{u})^{*}}
\newcommand{\fudl}{(\mathbf{f}_{u})}

\newcommand{\fddl}{(\mathbf{f}_{d})}
\newcommand{\fddlt}{(\mathbf{f}_{d})^{T}}
\newcommand{\fddls}{(\mathbf{f}_{d})^{*}}
\newcommand{\fddr}{(\mathbf{f}_{d})}
\newcommand{\fddrs}{(\mathbf{f}_{d})^{*}}
\newcommand{\fddrt}{(\mathbf{f}_{d})^{T}}
\newcommand{\fdul}{(\mathbf{f}_{d})}

\newcommand{\feel}{(\mathbf{f}_{e})}
\newcommand{\feelt}{(\mathbf{f}_{e})^{T}}
\newcommand{\feels}{(\mathbf{f}_{e})^{*}}
\newcommand{\feer}{(\mathbf{f}_{e})}
\newcommand{\feers}{(\mathbf{f}_{e})^{*}}
\newcommand{\feert}{(\mathbf{f}_{e})^{T}}

\newcommand{\glp}{g'}
\newcommand{\gtwl}{g_{2}}
\newcommand{\gthl}{g_{3}}
\newcommand{\au}{(\mathbf{a}_{u})}
\newcommand{\aut}{(\mathbf{a}_{u})^{T}}
\newcommand{\aus}{(\mathbf{a}_{u})^{*}}
\newcommand{\ad}{(\mathbf{a}_{d})}
\newcommand{\adt}{(\mathbf{a}_{d})^{T}}
\newcommand{\ads}{(\mathbf{a}_{d})^{*}}
\renewcommand{\ae}{(\mathbf{a}_{e})}
\newcommand{\aet}{(\mathbf{a}_{e})^{T}}
\newcommand{\aes}{(\mathbf{a}_{e})^{*}}
\newcommand{\mt}{\tilde{\mu}}
\newcommand{\mtsq}{\left|\tilde{\mu}\right|^{2}}

\newcommand{\mhusq}{m^{2}_{H_{u}}}
\newcommand{\mhdsq}{m^{2}_{H_{d}}}
\newcommand{\musq}{\left(\bdm^{2}_{U}\right)}
\newcommand{\mqsq}{\left(\bdm^{2}_{Q}\right)}
\newcommand{\mdsq}{\left(\bdm^{2}_{D}\right)}
\newcommand{\mlsq}{\left(\bdm^{2}_{L}\right)}
\newcommand{\mesq}{\left(\bdm^{2}_{E}\right)}
\newcommand{\h}{\theta_h}
\newcommand{\Hh}{\theta_H}

\newcommand{\sh}{\theta_{\tilde{h}}}

\newcommand{\sqi}{\theta_{{\tilde{Q}_i}}}
\newcommand{\sqj}{\theta_{{\tilde{Q}_j}}}
\newcommand{\sqk}{\theta_{{\tilde{Q}_k}}}
\newcommand{\sql}{\theta_{{\tilde{Q}_l}}}
\newcommand{\sui}{\theta_{{\tilde{u}_i}}}
\newcommand{\suj}{\theta_{{\tilde{u}_j}}}
\newcommand{\suk}{\theta_{{\tilde{u}_k}}}
\newcommand{\sul}{\theta_{{\tilde{u}_l}}}
\newcommand{\sdi}{\theta_{{\tilde{d}_i}}}
\newcommand{\sdj}{\theta_{{\tilde{d}_j}}}
\newcommand{\sdk}{\theta_{{\tilde{d}_k}}}
\newcommand{\sdl}{\theta_{{\tilde{d}_l}}}
\newcommand{\sLi}{\theta_{{\tilde{L}_i}}}
\newcommand{\sLj}{\theta_{{\tilde{L}_j}}}
\newcommand{\sll}{\theta_{{\tilde{L}_l}}}
\newcommand{\slk}{\theta_{{\tilde{L}_k}}}
\newcommand{\sei}{\theta_{{\tilde{e}_i}}}
\newcommand{\sej}{\theta_{{\tilde{e}_j}}}
\newcommand{\sek}{\theta_{{\tilde{e}_k}}}
\newcommand{\sel}{\theta_{{\tilde{e}_l}}}
\newcommand{\sbi}{\theta_{\tilde{B}}}
\newcommand{\swi}{\theta_{\tilde{W}}}
\newcommand{\sgl}{\theta_{\tilde{g}}}
\newcommand{\mgtphusq}{\left|\gtphu\right|^{2}}
\newcommand{\mgtphdsq}{\left|\gtphd\right|^{2}}
\newcommand{\mgthusq}{\left|\gthu\right|^{2}}
\newcommand{\mgthdsq}{\left|\gthd\right|^{2}}
\newcommand{\mtsfuhu}{\left(\tilde{\mu}^{*}\mathbf{f}^{h_{u}}_{u}\right)}
\newcommand{\mtsfdhd}{\left(\tilde{\mu}^{*}\mathbf{f}^{h_{d}}_{d}\right)}
\newcommand{\mtsfehd}{\left(\tilde{\mu}^{*}\mathbf{f}^{h_{d}}_{e}\right)}

\newcommand{\mtsfuhut}{\left(\tilde{\mu}^{*}\mathbf{f}^{h_{u}}_{u}\right)^{T}}
\newcommand{\mtsfdhdt}{\left(\tilde{\mu}^{*}\mathbf{f}^{h_{d}}_{d}\right)^{T}}
\newcommand{\mtsfehdt}{\left(\tilde{\mu}^{*}\mathbf{f}^{h_{d}}_{e}\right)^{T}}
\newcommand{\mtfuhus}{\left(\tilde{\mu}^{*}\mathbf{f}^{h_{u}}_{u}\right)^{*}}
\newcommand{\mtfdhds}{\left(\tilde{\mu}^{*}\mathbf{f}^{h_{d}}_{d}\right)^{*}}
\newcommand{\mtfehds}{\left(\tilde{\mu}^{*}\mathbf{f}^{h_{d}}_{e}\right)^{*}}
\newcommand{\nbmtsfuhut}{(\tilde{\mu}^{*}\mathbf{f}^{h_{u}}_{u})^{T}}
\newcommand{\nbmtsfdhdt}{(\tilde{\mu}^{*}\mathbf{f}^{h_{d}}_{d})^{T}}
\newcommand{\nbmtsfehdt}{(\tilde{\mu}^{*}\mathbf{f}^{h_{d}}_{e})^{T}}
\newcommand{\nbmtfuhus}{(\tilde{\mu}^{*}\mathbf{f}^{h_{u}}_{u})^{*}}
\newcommand{\nbmtfdhds}{(\tilde{\mu}^{*}\mathbf{f}^{h_{d}}_{d})^{*}}
\newcommand{\nbmtfehds}{(\tilde{\mu}^{*}\mathbf{f}^{h_{d}}_{e})^{*}}

\begin{document}

\preprint{UH-511-1135-08}

\title{Threshold and Flavour Effects in the Renormalization Group
  Equations of the MSSM II: Dimensionful couplings}

\author{Andrew D. Box}
 \email{abox@phys.hawaii.edu}
\author{Xerxes Tata}%
 \email{tata@phys.hawaii.edu}
\affiliation{%
\mbox{Dept. of Physics and Astronomy}, University of Hawaii, Honolulu, HI 96822, U.S.A.\\
}%
\vspace{-7cm}
\begin{abstract}
\noindent We re-examine the one-loop renormalization group equations
(RGEs) for the dimensionful parameters of the minimal supersymmetric
Standard Model with broken supersymmetry, allowing for arbitrary flavour
structure of the soft SUSY breaking (SSB) parameters. We include
threshold effects by evaluating the $\beta$-functions in
a sequence of (non-supersymmetric) effective theories with heavy
particles decoupled at the scale of their mass. 
We present the
most general form for high scale SSB parameters that obtains if we
assume that the supersymmetry breaking mechanism does not introduce new
inter-generational couplings.
This form, possibly
amended to allow additional sources of flavour-violation, serves as a
boundary condition for solving the RGEs for the dimensionful MSSM
parameters. We then present illustrative examples of numerical solutions
to the RGEs. We find that in a SUSY GUT with the scale of SUSY scalars
split from that of gauginos and higgsinos, the gaugino mass
unification condition may be violated by ${\cal O}$(10\%). As another
illustration, we show that in mSUGRA, the rate for the flavour-violating
$\tst_1\to c\tz_1$ decay obtained using the complete RGE solution is
smaller than that obtained using the commonly-used ``single-step''
integration of the RGEs by a factor 10-25, and so may qualitatively
change expectations for topologies from top-squark pair production at
colliders.  Together with the RGEs for dimensionless couplings presented
in a companion paper, the RGEs in Appendix \ref{sec:rgeapp} of this
paper form a complete set of one-loop MSSM RGEs that include threshold
and flavour-effects necessary for two-loop accuracy.

\end{abstract}

\pacs{11.30.Pb, 11.10.Hi, 14.80.Ly}


\maketitle

\section{Introduction}\label{intro}

Renormalization group equations (RGEs) have played a vital role for the
extraction of phenomenological predictions of theories where the physics
is specified at a scale much higher than the energy scale directly
relevant for phenomenology \cite{gutbounds}. A very familiar example of
this is the prediction of the weak mixing angle $\theta_W$ in grand
unified theories (GUTs), widely regarded as one of the important
successes of the idea of grand unification in the context of
supersymmetry (SUSY) \cite{amaldi}. RGE methods have become ubiquitous
for many studies of physics beyond the Standard Model (SM), where
relationships between various model parameters, {\em renormalized at
some very high scale}, are specified by the new physics.

Of interest to us here are supersymmetric models of particle physics
\cite{dimgeor,wss,drees,binetruy,primer} where supersymmetry fixes the
dimensionless couplings of superpartners in terms of the corresponding
gauge and Yukawa couplings of SM particles. Model-dependence arises via
the potentially very large number of undetermined dimensionful soft SUSY
breaking (SSB) parameters that reflects our ignorance about how
supersymmetry is broken \cite{SSB}.  In practice, one resorts to models
with underlying assumptions that fix all the SSB masses and couplings,
renormalized at a high scale, in terms of a handful of parameters
\cite{mSUGRA,GMSB,AMSB,gaugino}. These high scale parameters have, of
course, to be evolved to the low scale relevant for phenomenology in
order to sum the large logarithms that may otherwise invalidate any
(fixed-order) perturbative calculation \cite{earlyrges}.

This paper is a follow-up of an earlier paper, hereafter referred to as
Paper~I \cite{rge1}, where we used the seminal papers of Machacek and
Vaughn \cite{mv} to obtain the one-loop RGEs for the dimensionless
couplings of the Minimal Supersymmetric Standard Model (MSSM), carefully
including threshold effects \cite{ramond2,sakis} that are essential to
include for true two-loop precision. Here, we extend the existing
literature, and present the one-loop RGEs for the dimensionful (in
general, complex) parameters including both flavour-mixing effects and
threshold corrections. As discussed in Ref.~\cite{rge1}, these RGEs can
then be augmented by the two-loop terms of the MSSM RGEs
\cite{martv,bbo,yamada,jackjones}, without the need to include threshold
corrections to these for the required precision. Just as in Paper~I,
where we found that SUSY breaking threshold effects cause the
gaugino-fermion-sfermion [higgsino-fermion-sfermion] couplings to evolve
differently from the corresponding gauge [Yukawa] couplings, resulting
in an enlarged system of RGEs for the dimensionless couplings, we will
see that as we decouple heavy super-partners and Higgs bosons, a similar
situation obtains for the RGEs of dimensionful parameters. For instance,
the super-potential higgsino mass $\mu$ evolves differently from $\mt$,
that enters via the corresponding Higgs boson mass squared parameter,
$|\mt|^2$. Also, at scales below the masses of the heavier particles in
the Higgs boson sector, where just the SM Higgs boson, $h$, and the
associated would-be-Goldstone bosons that complete the scalar doublet
(see Paper~I) remain in the theory, we will see that only certain
combinations of model parameters can be evolved in the corresponding low
energy theory. These complications, to our knowledge, have not been
included in previous studies.

We use the RGEs for the dimensionful parameters of a general field
theory with real spin-zero fields and two-component spinor fields given
in Ref.~\cite{luo} to derive our results. In Sec.~\ref{sec:formal} we
first re-cast the results of Ref.~\cite{luo} in a form suitable for use
by phenomenologists more used to the formalism with four-component
spinors coupled to complex scalar fields and, of course, also gauge
fields.  As explained in Paper~I, this re-casting does much of the work
needed for the derivation of the MSSM RGEs. We discuss particle
decoupling in Sec.~\ref{sec:dec} and describe our derivation of the RGEs
in Sec.~\ref{sec:mssm}. In Sec.~\ref{sec:flav}, we numerically analyse
the solutions of (some of) the RGEs and discuss squark flavour
mixing. We apply our results to obtain the rate for the
flavour-violating decay of the lighter $t$-squark, $\tst_1 \to c\tz_1$
in Sec.~\ref{sec:stop}. We conclude in Sec.~\ref{summary} with general
remarks and a summary of our results. The set of RGEs that we obtain are
listed in Appendix~\ref{sec:rgeapp}.

\section{Formalism} \label{sec:formal}

Luo, Wang and Xiao \cite{luo}, that we use as the starting point, write
the Lagrangian density for a general field theory as,
\begin{equation}\begin{split}\label{eq:twolag}
\mathcal{L}_{(2)}= &\ i\psi^\dagger_p\sigma^\mu D_\mu\psi_p + \frac{1}{2}D_\mu\phi_a D^\mu\phi_a -\frac{1}{4}F_{\mu\nu A}F^{\mu\nu}_A\\
&-\frac{1}{2}\left[\left(\bdm_{f}\right)_{pq}\psi^{T}_{p}\zeta\psi_{q}+\mathrm{h.c.}\right]-\frac{1}{2!}\bdmsf^{2}_{ab}\phi_{a}\phi_{b}\\
&-\left(\frac{1}{2}\by^a_{pq}\psi^T_p\zeta\psi_q\phi_a+\mathrm{h.c.}\right)- \frac{1}{3!}\mathbf{h}_{abc}\phi_{a}\phi_{b}\phi_{c}-\frac{1}{4!}\bm{\lambda}_{abcd}\phi_{a}\phi_{b}\phi_{c}\phi_{d}\;,
\end{split}\end{equation}
where the matrix $\zeta=i\sigma_{2}$ is introduced to make the spinor
bilinear Lorentz invariant. The gauge interactions for the real fields
$\phi_a$ and two-component spinors $\psi_p$ are contained in the
corresponding covariant derivatives, the matrices ${\bdm_{f}}$ and ${\bf
Y}^a$ are symmetric (but not Hermitian), while the entries of the scalar
mass matrix $\bdmsf^2$ as well as of the trilinear and quartic scalar
couplings ${\bf h}$ and $\bm{\lambda}$ are symmetric in all their
indices as well as real.

The Lagrangian density can also be written in terms of complex scalar
fields $\Phi_a$ and four-component Dirac and Majorana spinor fields as
\cite{rge1},
\begin{equation}\begin{split}\label{eq:fourlag}
\mathcal{L}_{(4)} = &\ \frac{i}{2}\bar{\Psi}_{j} \gamma^\mu D_\mu \Psi_j + \left(D_{\mu}\Phi_{a}\right)^{\dagger}\left(D^{\mu}\Phi_{a}\right) - \frac{1}{4}F_{\mu\nu A}F^{\mu\nu}_A\\
&-\frac{1}{2}\left[\left(\mathbf{m}_{X}\right)_{jk}\bar{\Psi}_{Mj}\Psi_{Mk}+i\left(\mathbf{m}'_{X}\right)_{jk}\bar{\Psi}_{Mj}\gamma_{5}\Psi_{Mk}\right]+\left[\frac{1}{2!}\bm{\mathcal{B}}_{ab}\Phi_a\Phi_b+\mathrm{h.c.}\right]-\mathbf{m}^{2}_{ab}\Phi^\dagger_a\Phi_b\\
&-\left[\left(\bu^{1}_{a}\right)_{jk}\bar{\Psi}_{Dj}P_L\Psi_{Dk}\Phi_a+\left(\bu^{2}_{a}\right)_{jk}\bar{\Psi}_{Dj}P_L\Psi_{Dk}\Phi^\dagger_a\right.\\
&\quad\ +\left(\bv_{a}\right)_{jk}\bar{\Psi}_{Dj}P_L\Psi_{Mk}\Phi_a+\left(\bw_{a}\right)_{jk}\bar{\Psi}_{Mj}P_L\Psi_{Dk}\Phi^\dagger_a\\
&\left.\quad\ +\frac{1}{2}\left(\bx^{1}_{a}\right)_{jk}\bar{\Psi}_{Mj}P_L\Psi_{Mk}\Phi_a+\frac{1}{2}\left(\bx^{2}_{a}\right)_{jk}\bar{\Psi}_{Mj}P_L\Psi_{Mk}\Phi^\dagger_a+\mathrm{h.c.}\right]\\
&+\left[\frac{1}{2!}\Phi^\dagger_a\mathbf{H}_{abc}\Phi_b\Phi_c+\mathrm{h.c.}\right]-\frac{1}{2!}\frac{1}{2!}\bm{\Lambda}_{abcd}\Phi^\dagger_a\Phi^\dagger_b\Phi_c\Phi_d-\left[\frac{1}{3!}\bm{\Lambda}'_{abcd}\Phi^{\dagger}_{a}\Phi_{b}\Phi_{c}\Phi_{d}+\mathrm{h.c.}\right]\;.
\end{split}\end{equation}
The Yukawa coupling matrices $\bu^{1,2}_{a}$ that
couple Dirac spinor fields to complex scalars, and the matrices
$\bv_{a}$ and $\bw_{a}$ that couple these scalar fields to one Majorana and
one Dirac field, have no particular symmetry (or Hermiticity) properties
under interchange of the fermion field indices $j$ and $k$. These
indices label the fermion field type (quark, lepton, gaugino, higgsino)
and also carry information of flavour and other quantum numbers
(\textit{e.g.} weak isospin and colour). On the other hand, the matrices
$\bx_a^{1,2}$ that couple scalars to two Majorana fields are
symmetric under $j \leftrightarrow k$ because of the symmetry properties
\cite{wss} of the Majorana spinor bilinears that appear in
(\ref{eq:fourlag}).  The scalar mass squared matrix $\mathbf{m}^2$ ---
notice that we have used a different font in (\ref{eq:fourlag}) to
differentiate this mass term for complex scalar fields from the
corresponding one for real scalars coupled to two-component spinors in
\eqref{eq:twolag} --- is Hermitian in the scalar field indices, $a,b$,
while the matrix $\bm{\mathcal{B}}$ is symmetric.  The $CP$ even and odd
Majorana fermion mass matrices \cite{wss} $\mathbf{m}_X$ and
$\mathbf{m}'_X$ are both Hermitian and symmetric.  The trilinear and
quartic scalar couplings $\bm{\Lambda}$, $\bm{\Lambda}'$ and
$\mathbf{H}$ are symmetric under interchanges $a \leftrightarrow b$
and/or $c \leftrightarrow d$ for $\mathbf{\Lambda}$, under interchanges
of $b,c,d$ for $\mathbf{\Lambda}'$, and finally under $b\leftrightarrow
c$ for $\mathbf{H}$.
 As noted in Paper~I, while (\ref{eq:fourlag}) is not the most
general form for the Lagrangian density, it suffices for the
derivation of the RGEs of the MSSM with $R$-parity conservation. 

The evolution of the fermion mass parameters in (\ref{eq:twolag}) is
determined by the $\beta$-function \cite{luo},
\begin{equation}\begin{split}\label{eq:fermassrge}
\left(4\pi\right)^{2}\left.\bbet_{\mathbf{m}_{f}}\right|_{\rm
1-loop}=&\frac{1}{2}\left[\by^{T}_{2}(F)\mathbf{m}_{f}+\mathbf{m}_{f}\by_{2}(F)\right]+2\by^{b}\mathbf{m}^{\dagger}_{f}\by^{b}+\frac{1}{2}\by^{b}\mathrm{Tr}\left\{\by^{b\dagger}\mathbf{m}_{f}+\mathbf{m}^{\dagger}_{f}\by^{b}\right\}\\
&-3g^{2}\left\{\mathbf{C}_{2}(F),\mathbf{m}_{f}\right\}\;.
\end{split}\end{equation}
 Here, ${\mathbf{C}}_{2}(F)={\bf t}^A{\bf t}^A$ is the quadratic Casimir
  operator for the fermions, and $\by_{2}(F)=\by^{b\dagger}\by^b$.
  Notice that we have altered $\by_{2}^{\dagger}(F)$ (recall
  $\by_{2}(F)$ is a Hermitian matrix) that appears in the first term of
  the corresponding equation in Ref.~\cite{luo} to $\by_{2}^{T}(F)$.
  This ensures that the $\beta$-function for $\mathbf{m}_{f}$ has the
  same symmetry property as $\mathbf{m}_{f}$ in (\ref{eq:twolag}).  We
  can write the ``Yukawa couplings'' on the right hand side of
  (\ref{eq:fermassrge}) in terms of the corresponding couplings in the
  four-component notation by introducing
$$\vec{\psi}\equiv\left(\begin{array}{c}\psi_L\\\psi_R\\\psi_M\end{array}\right),$$
and replacing $\by^a$ with a $(3\times 3)$ block matrix 
containing $\bu^{1}_{a}$, $\bu^{2}_{a}$, $\bv_{a}$, $\bw_{a}$,
  $\bx^{1}_{a}$ and $\bx^{2}_{a}$ as worked out in Paper~I.

The fermion mass matrix can similarly be written (in this same basis)
as,
\begin{equation}
\bdm_{f}=\left(\begin{array}{ccc}
\mathbf{0}&(\bdm^{T}_{U}-i\bdm'^{T}_{U})&(\bdm^{T}_{W}-i\bdm'^{T}_{W})\\
(\bdm_{U}-i\bdm'_{U})&\mathbf{0}&(\bdm_{V}-i\bdm'_{V})\\
(\bdm_{W}-i\bdm'_{W})&(\bdm^{T}_{V}-i\bdm'^{T}_{V})&(\bdm_{X}-i\bdm'_{X})
\end{array}\right),
\end{equation}
where in the context of the $R$-parity-conserving MSSM, all the
entries except $(\bdm_{X}-i\bdm'_{X})$ vanish because gauge invariance
precludes the corresponding fermion bilinears. From now on, we retain
just $\bdm_{X}$ and
$\bdm'_{X}$ in our analysis. We can now substitute $\bdm_{f}$ as well as
the form of the ``Yukawa'' matrices from Paper~I into
(\ref{eq:fermassrge})
to obtain the RGEs for the matrices ${\bf m}_X$ and ${\bf m}_X^{\prime}$. We
find, 
\begin{equation}\label{eq:RGEmx}\begin{split}
\left(4\pi\right)^2\frac{d\mathbf{m}_X}{dt}=&\frac{1}{4}\left[\left(\bw_b\bw^\dagger_b+\bv^T_b\bv^*_b+\bx^{1}_{b}\bx^{1\dagger}_{b}+\bx^{2}_{b}\bx^{2\dagger}_{b}\right)\bmfx\right.\\
&\quad\ +\bmfxd\left(\bw_b\bw^\dagger_b+\bv^T_b\bv^*_b+\bx^{1}_{b}\bx^{1\dagger}_{b}+\bx^{2}_{b}\bx^{2\dagger}_{b}\right)\\
&\quad\ +\bmfx\left(\bw^*_b\bw^T_b+\bv^\dagger_b\bv_b+\bx^{1\dagger}_{b}\bx^{1}_{b}+\bx^{2\dagger}_{b}\bx^{2}_{b}\right)\\
&\left.\quad\ +\left(\bw^*_b\bw^T_b+\bv^\dagger_b\bv_b+\bx^{1\dagger}_{b}\bx^{1}_{b}+\bx^{2\dagger}_{b}\bx^{2}_{b}\right)\bmfxd\right]\\
&+\left[\bx^{1}_{b}\bmfxd\bx^{2}_{b}+\bx^{2}_{b}\bmfxd\bx^{1}_{b}\right.\\
&\left.\quad\ +\bx^{2\dagger}_{b}\bmfx\bx^{1\dagger}_{b}+\bx^{1\dagger}_{b}\bmfx\bx^{2\dagger}_{b}\right]\\
&+\frac{1}{4}\left[\bx^{1}_{b}\mathrm{Tr}\left\{\bx^{1\dagger}_{b}\bmfx+\bmfxd\bx^{2}_{b}\right\}\right.\\
&\qquad\ +\bx^{2}_{b}\mathrm{Tr}\left\{\bx^{2\dagger}_{b}\bmfx+\bmfxd\bx^{1}_{b}\right\}\\
&\qquad\ +\bx^{1\dagger}_{b}\mathrm{Tr}\left\{\bmfxd\bx^{1}_{b}+\bx^{2\dagger}_{b}\bmfx\right\}\\
&\left.\qquad\ +\bx^{2\dagger}_{b}\mathrm{Tr}\left\{\bmfxd\bx^{2}_{b}+\bx^{1\dagger}_{b}\bmfx\right\}\right]\\
&-6g^2\mathbf{C}^{M}_{2}(F)\mathbf{m}_X\;, 
\end{split}\end{equation}
and 
\begin{equation}\label{eq:RGEmxp}\begin{split}
\left(4\pi\right)^2\frac{d\mathbf{m}'_X}{dt}=&\frac{i}{4}\left[\left(\bw_b\bw^\dagger_b+\bv^T_b\bv^*_b+\bx^{1}_{b}\bx^{1\dagger}_{b}+\bx^{2}_{b}\bx^{2\dagger}_{b}\right)\bmfx\right.\\
&\quad\ -\bmfxd\left(\bw_b\bw^\dagger_b+\bv^T_b\bv^*_b+\bx^{1}_{b}\bx^{1\dagger}_{b}+\bx^{2}_{b}\bx^{2\dagger}_{b}\right)\\
&\quad\ +\bmfx\left(\bw^*_b\bw^T_b+\bv^\dagger_b\bv_b+\bx^{1\dagger}_{b}\bx^{1}_{b}+\bx^{2\dagger}_{b}\bx^{2}_{b}\right)\\
&\left.\quad\ -\left(\bw^*_b\bw^T_b+\bv^\dagger_b\bv_b+\bx^{1\dagger}_{b}\bx^{1}_{b}+\bx^{2\dagger}_{b}\bx^{2}_{b}\right)\bmfxd\right]\\
&+i\left[\bx^{1}_{b}\bmfxd\bx^{2}_{b}+\bx^{2}_{b}\bmfxd\bx^{1}_{b}\right.\\
&\left.\qquad-\bx^{2\dagger}_{b}\bmfx\bx^{1\dagger}_{b}-\bx^{1\dagger}_{b}\bmfx\bx^{2\dagger}_{b}\right]\\
&+\frac{i}{4}\left[\bx^{1}_{b}\mathrm{Tr}\left\{\bx^{1\dagger}_{b}\bmfx+\bmfxd\bx^{2}_{b}\right\}\right.\\
&\qquad\ +\bx^{2}_{b}\mathrm{Tr}\left\{\bx^{2\dagger}_{b}\bmfx+\bmfxd\bx^{1}_{b}\right\}\\
&\qquad\ -\bx^{1\dagger}_{b}\mathrm{Tr}\left\{\bmfxd\bx^{1}_{b}+\bx^{2\dagger}_{b}\bmfx\right\}\\
&\left.\qquad\ -\bx^{2\dagger}_{b}\mathrm{Tr}\left\{\bmfxd\bx^{2}_{b}+\bx^{1\dagger}_{b}\bmfx\right\}\right]\\
&-6g^2\mathbf{C}_{2}^M(F)\mathbf{m}'_X\;,
\end{split}\end{equation}
where $t=\ln{Q}$, ${\bf C}_{2}^M(F)$ is the quadratic Casimir for the Majorana fermions
as defined in Ref.~\cite{rge1}, and there is a sum over all gauge group factors
in the final term of both (\ref{eq:RGEmx}) and (\ref{eq:RGEmxp}).
Note also that in the MSSM with $R$-parity
conservation, the trace terms in (\ref{eq:RGEmx}) and (\ref{eq:RGEmxp})
vanish. This is because, as seen in Eq.~(\ref{eq:fourlag}), $\bx$
only connects higgsinos to gauginos, while $\mathbf{m}^{(\prime)}_{X}$ never
connects higgsinos to gauginos. Therefore, a trace of the
product of these two matrices is always zero.

Turning to the RGE for
the trilinear scalar couplings, we first write the one-loop RGE for these
couplings in (\ref{eq:twolag}) as \cite{luo}, 
\begin{equation}\label{eq:2comph}
\left(4\pi\right)^{2}\left.\bbet_{\mathbf{h}_{abc}}\right|_{1-loop}=\bdLm^{2}_{abc}-4\bm{\mathcal{H}}_{abc}+\bdLm^{Y}_{abc}-3g^{2}\bdLm^{S}_{abc}\;,
\end{equation}
with
\begin{align}
\bdLm^{2}_{abc}=&\frac{1}{2}\sum_{\rm perms}\bm{\lambda}_{abef}\mathbf{h}_{efc}\;,\\
\bm{\mathcal{H}}_{abc}=&\frac{1}{2}\sum_{\rm perms}\mathrm{Tr}\{\bdm_{f}\by^{\dagger}_{a}\by_{b}\by^{\dagger}_{c}+\by_{a}\bdm^{\dagger}_{f}\by_{b}\by^{\dagger}_{c}\}\;,\\
\begin{split}
\label{eq:hy2}\bdLm^{Y}_{abc}=&\frac{1}{2}\left[\mathrm{Tr}\left\{\by^{\dagger}_{a'}\by_{a}+\by^{\dagger}_{a}\by_{a'}\right\}\mathbf{h}_{a'bc}+\mathrm{Tr}\left\{\by^{\dagger}_{b'}\by_{b}+\by^{\dagger}_{b}\by_{b'}\right\}\mathbf{h}_{ab'c}\right.\\
&\left.\quad\
+\mathrm{Tr}\left\{\by^{\dagger}_{c'}\by_{c}+\by^{\dagger}_{c}\by_{c'}\right\}\mathbf{h}_{abc'}\right]\;,\end{split}\\
\bdLm^{S}_{abc}=&\left[C_{2}(a)+C_{2}(b)+C_{2}(c)\right]\mathbf{h}_{abc}\;,
\end{align}
where $a'$, $b'$, $c'$, and the indices $e$ and $f$, are summed over all
the scalars in the theory. Also, there is an implied sum over all gauge
group factors in the final term of (\ref{eq:2comph}), and
$C_{2}(S)=\bf{t}^{A}\bf{t}^{A}$ is the quadratic Casimir operator for
the scalars. All this is exactly as in
Ref.~\cite{luo} except that we have written the expression in
(\ref{eq:hy2}) in a more transparent form. We note that the
$\beta$-function in (\ref{eq:2comph}) is manifestly symmetric under any
permutation of the three indices.

In order to convert this into our complex, four-component
notation, we first expand out the trilinear scalar term in (\ref{eq:fourlag})
in terms of real spin-zero fields ({\it i.e.} the real and imaginary
parts of our complex fields), and
compare this with (\ref{eq:twolag}) in order to obtain
$\mathbf{H}_{abc}$ in terms of $\mathbf{h}_{abc}$ as,
\begin{equation}\label{eq:hconv}
\mathbf{H}_{abc}=\frac{1}{\sqrt{2}}\left[-\mathbf{h}_{a_{R}b_{R}c_{R}}+\mathbf{h}_{a_{R}b_{I}c_{I}}+i\mathbf{h}_{a_{R}b_{R}c_{I}}+i\mathbf{h}_{a_{R}b_{I}c_{R}}\right]\;.
\end{equation}
We can now write down the RGE for $\mathbf{H}_{abc}$ using the 
same relations for the two-component Yukawa matrices  as before, 
and substituting the trilinear and quartic scalar couplings 
$\mathbf{h}_{abc}$ and $\bm{\lambda}_{abcd}$ in terms of the
corresponding couplings $\mathbf{H}_{abc}$ and $\bdLm_{abcd}$ or
$\bdLm'_{abcd}$ on the right hand side of (\ref{eq:2comph}), taking
care to sum over real and imaginary components of
the scalar fields.
This RGE then becomes, 
\begin{equation}\begin{split}
\left(4\pi\right)^2\frac{d\mathbf{H}_{abc}}{dt}=&
\left[2\left(\bdLm_{afbe}+\bdLm'_{fabe}\right)\mathbf{H}_{ecf}+
\bdLm'_{abef}\mathbf{H}^{*}_{cef}+\left(b\leftrightarrow c\right)\right]\\
&+\bdLm_{efbc}\mathbf{H}_{aef}+2\bdLm'_{ebcf}\left(\mathbf{H}^{*}_{eaf}
+\mathbf{H}_{fae}\right)\\
&+2\mathrm{Tr}\left\{\left(\bdm_{X}-i\bdm'_{X}\right)\left[\left(\bv^
\dagger_a\left\{\bu^1_b\bw^\dagger_c+\bv_b\bx^{2\dagger}_c\right\}+
\left(\bx^{1\dagger}_a+\bx^{2\dagger}_a\right)\bx^1_b\bx^{2\dagger}_c
\right.\right.\right.\\
&\qquad\qquad\qquad\qquad\qquad+\bw^*_b\bw^T_a\bx^{2\dagger}_c+
\bx^{2\dagger}_b\left\{\bw_a\bw^\dagger_c+\left(\bx^1_a+\bx^2_a\right)
\bx^{2\dagger}_c\right\}\\
&\qquad\qquad\qquad\qquad\qquad\left.+\bw^*_b\bu^{1T}_c\bv^*_a+
\bx^{2\dagger}_b\left\{\bv^T_c\bv^*_a+\bx^1_c\left(\bx^{1\dagger}_a+
\bx^{2\dagger}_a\right)\right\}\right)\\
&\qquad\qquad\qquad\qquad\quad\ \left.\left.+\left(b\leftrightarrow c
\right)\right]\right\}\\
&+2\mathrm{Tr}\left\{\left(\bw^T_a\bmfxd\left\{\bv^T_b\bu^{2*}_c+
\bx^1_b\bw^*_c\right\}\right.\right.\\
&\qquad\qquad+\left(\bx^1_a+\bx^2_a\right)\bmfxd\bx^1_b\bx^{2\dagger}_c
+\bv_b\bmfxd\bw_a\bu^{2\dagger}_c\\
&\qquad\qquad+\bx^1_b\bmfxd\left\{\bw_a\bw^\dagger_c+\left(\bx^1_a+
\bx^2_a\right)\bx^{2\dagger}_c\right\}\\
&\qquad\qquad+\bv_b\bmfxd\bx^1_c\bv^\dagger_a\\
&\qquad\qquad\left.+\bx^1_b\bmfxd\left\{\bv^T_c\bv^*_a+\bx^1_c
\left(\bx^{1\dagger}_a+\bx^{2\dagger}_a\right)\right\}\right)\\
&\qquad\quad\ \left.+\left(b\leftrightarrow c\right)\right\}\\
&+\mathrm{Tr}\left\{\bu^{2\dagger}_{a'}\left(\bu^1_a+\bu^2_a\right)+
\left(\bu^{1\dagger}_a+\bu^{2\dagger}_a\right)\bu^1_{a'}+\bv^\dagger_a
\bv_{a'}+\bw^\dagger_{a'}\bw_a\right.\\
&\qquad\quad\left.+\frac{1}{2}\left\{\bx^{2\dagger}_{a'}\left(\bx^1_a+
\bx^2_a\right)+\left(\bx^{1\dagger}_a+\bx^{2\dagger}_a\right)\bx^1_{a'}
\right\}\right\}\mathbf{H}_{a'bc}\\
&+\mathrm{Tr}\left\{\bu^{1\dagger}_{b'}\bu^1_b+\bu^{2\dagger}_b
\bu^2_{b'}+\bv^\dagger_{b'}\bv_b+\bw^\dagger_b\bw_{b'}+\frac{1}{2}\left
\{\bx^{1\dagger}_{b'}\bx^1_b+\bx^{2\dagger}_b\bx^2_{b'}\right\}\right\}
\mathbf{H}_{a{b'}c}\\
&+\mathrm{Tr}\left\{\bu^{2\dagger}_{b'}\bu^1_b+\bu^{2\dagger}_b
\bu^1_{b'}+\frac{1}{2}\left\{\bx^{2\dagger}_{b'}\bx^1_b+
\bx^{2\dagger}_b\bx^1_{b'}\right\}\right\}\left(\mathbf{H}_{b'ac}+
\mathbf{H}^*_{cab'}\right)\\
&+\mathrm{Tr}\left\{\bu^{1\dagger}_{c'}\bu^1_c+\bu^{2\dagger}_c
\bu^2_{c'}+\bv^\dagger_{c'}\bv_c+\bw^\dagger_c\bw_{c'}+\frac{1}{2}\left
\{\bx^{1\dagger}_{c'}\bx^1_c+\bx^{2\dagger}_c\bx^2_{c'}\right\}\right\}
\mathbf{H}_{ac'b}\\
&+\mathrm{Tr}\left\{\bu^{2\dagger}_{c'}\bu^1_c+\bu^{2\dagger}_c
\bu^1_{c'}+\frac{1}{2}\left\{\bx^{2\dagger}_{c'}\bx^1_c+
\bx^{2\dagger}_c\bx^1_{c'}\right\}\right\}\left(\mathbf{H}_{c'ab}+
\mathbf{H}^*_{bac'}\right)\\
&-3g^2\left[C_2(a)+C_2(b)+C_2(c)\right]\mathbf{H}_{abc}\;, \label{eq:4compH}
\end{split}\end{equation}
where, as with \eqref{eq:RGEmx} and \eqref{eq:RGEmxp}, a sum over all 
gauge group factors is implied in the last term.
The origin of the various terms in this long equation when compared with
the two-component form of (\ref{eq:2comph}) should be clear. The terms
involving the quartic couplings clearly originate in the first term of
(\ref{eq:2comph}), the next set of terms involving the mass matrices for
the Majorana spinors in the second term, the terms with the summation
over $\{a',b',c'\}$ appear in both forms, and the last term involving
the gauge couplings is clearly the corresponding term in
(\ref{eq:2comph}).  In the context of the MSSM, where we can without
loss of generality choose for the index $a$ in $\mathbf{H}_{abc}$ to
always refer to a sfermion, the RGE can be somewhat simplified. This is
because, when $a$ is a sfermion index, $\bu^{1,2}_{a}$ and
$\bx^{1,2}_{a}$ are always zero, and the corresponding terms drop out
from the equation. We remark that the first trace term not involving 
Majorana fermion masses in (\ref{eq:4compH}) is clearly symmetric under 
$b \leftrightarrow c$ while the last two trace
terms ({\it i.e.} those with $c'$ as a dummy index) are  simply the
$b\leftrightarrow c$ equivalent of the previous two trace terms involving the dummy
index $b'$. Thus the RGE for $\mathbf{H}_{abc}$  
preserves the $b\leftrightarrow c$ symmetry of this coupling  
noted below Eq.~(\ref{eq:fourlag}).

The RGE for the real scalar mass squared parameters in (\ref{eq:twolag}) 
 reads \cite{luo}, 
\begin{equation} 
\left(4\pi\right)^{2}\left.\bbet_{\mathbf{\mathsf{m}}^{2}_{ab}}\right|_{\rm
  1-loop}
=\bm{\lambda}_{abef}\mathbf{\mathsf{m}}^{2}_{ef}+\mathbf{h}_{aef}
\mathbf{h}_{bef}-2\bm{\mathcal{H}}_{ab}-3g^{2}
\bdLm^{S}_{ab}+\bdLm^{Y}_{ab} \;,
\label{eq:2compmb}
\end{equation}
with 
\begin{align}
\label{eq:2comphab}\begin{split}
\bm{\mathcal{H}}_{ab}=&\mathrm{Tr}\left\{\left(\by_{a}\by^{\dagger}_{b}+\by_{b}\by^{\dagger}_{a}\right)\bdm_{f}\bdm^{\dagger}_{f}+\left(\by^{\dagger}_{a}\by_{b}+\by^{\dagger}_{b}\by_{a}\right)\bdm^{\dagger}_{f}\bdm_{f}\right.\\
&\qquad\left.+\by_{a}\bdm^{\dagger}_{f}\by_{b}\bdm^{\dagger}_{f}+\bdm_{f}\by^{\dagger}_{a}\bdm_{f}\by^{\dagger}_{b}\right\}\;,
\end{split}\\
\bdLm^{S}_{ab}=&\left[C_{2}(a)+C_{2}(b)\right] \bdmsf^{2}_{ab}\;,\\
\bdLm^{Y}_{ab}=&\frac{1}{2}\left[\mathrm{Tr}\left\{\by^{\dagger}_{a'}\by_{a}+\by^{\dagger}_{a}\by_{a'}\right\}\bdmsf^{2}_{a'b}+\mathrm{Tr}\left\{\by^{\dagger}_{b'}\by_{b}+\by^{\dagger}_{b}\by_{b'}\right\}\bdmsf^{2}_{ab'}\right]\;,
\label{eq:2complyabc}
\end{align}
where in (\ref{eq:2complyabc}) we have once again made the notation more
explicit. Writing the complex fields $\Phi_a$ as $\Phi_a=
\frac{\phi_{aR}+i\phi_{aI}}{\sqrt{2}}$ and comparing the coefficients of
the bilinear terms in the {\it real} scalar fields $\phi_{\bullet}$ then gives,
%
\begin{subequations}\label{eq:mconv}\begin{eqnarray}
\bdm^{2}_{ab}-\bm{\mathcal{B}}_{ab}=&\bdmsf^{2}_{a_{R}b_{R}}-i\bdmsf^{2}_{a_{R}b_{I}}\;,\\
\bdm^{2}_{ab}+\bm{\mathcal{B}}_{ab}=&\bdmsf^{2}_{a_{I}b_{I}}+i\bdmsf^{2}_{a_{I}b_{R}}\;.
\end{eqnarray}\end{subequations}
In the $R$-parity conserving MSSM, we never have non-zero entries
in the Lagrangian corresponding to both $\bdm^{2}_{ab}$ and
$\bm{\mathcal{B}}_{ab}$ for the same $a,b$. It suffices, therefore, to 
write the RGE for just one of these combinations, which we take to be
the first one. We then have, 
\begin{equation}\label{eq:4compmb}\begin{split}
\left(4\pi\right)^2\frac{d\left[\bdm^{2}_{ab}-\bm{\mathcal{B}}_{ab}\right]}{dt}=&\left\{2\left(\bdLm_{afbe}+\bdLm'_{fabe}\right)\bdm^{2}_{ef}-\bdLm'_{abef}\bm{\mathcal{B}}_{ef}^*-\left(\bdLm_{efab}+\bdLm'^{*}_{baef}\right)\bm{\mathcal{B}}_{ef}\right\}\\
&+\left\{\mathbf{H}_{aef}\mathbf{H}^*_{bef}+2\left(\mathbf{H}^*_{eaf}+\mathbf{H}_{fae}\right)\mathbf{H}_{ebf}\right\}\\
&-2\left[2\;\mathrm{Tr}\left\{\left[\bv^T_b\bv^*_a+\bw_a\bw^\dagger_b+\left(\bx^1_a+\bx^2_a\right)\bx^{2\dagger}_b+\bx^1_b\left(\bx^{1\dagger}_a+\bx^{2\dagger}_a\right)\right]\right.\right.\\
&\qquad\qquad\qquad\qquad\qquad\qquad\qquad\qquad\quad\left.\times\bmfx\bmfxd\right\}\\
&\qquad\ +\mathrm{Tr}\left\{\left(\bx^1_a+\bx^2_a\right)\bmfxd\bx^1_b\bmfxd\right\}\\
&\qquad\ \left.+\mathrm{Tr}\left\{\left(\bx^{1\dagger}_a+\bx^{2\dagger}_a\right)\bmfx\bx^{2\dagger}_b\bmfx\right\}\right]\\
&-3g^2\left[C_2(a)+C_2(b)\right]\left(\bdm^{2}_{ab}-\bm{\mathcal{B}}_{ab}\right)\\
&+\left[\mathrm{Tr}\left\{\bu^{2\dagger}_{a'}\left(\bu^1_a+\bu^2_a\right)+\left(\bu^{1\dagger}_a+\bu^{2\dagger}_a\right)\bu^1_{a'}+\bv^\dagger_a\bv_{a'}\right.\right.\\
&\qquad\quad\ \left.+\bw^\dagger_{a'}\bw_a+\frac{1}{2}\bx^{2\dagger}_{a'}\left(\bx^1_a+\bx^2_a\right)+\frac{1}{2}\left(\bx^{1\dagger}_a+\bx^{2\dagger}_a\right)\bx^1_{a'}\right\}\bdm^{2}_{a'b}\\
&\quad\ \ -\mathrm{Tr}\left\{\bu^{1\dagger}_{a'}\left(\bu^1_a+\bu^2_a\right)+\left(\bu^{1\dagger}_a+\bu^{2\dagger}_a\right)\bu^2_{a'}+\bv^\dagger_{a'}\bv_a\right.\\
&\qquad\qquad\ \left.+\bw^\dagger_a\bw_{a'}+\frac{1}{2}\bx^{1\dagger}_{a'}\left(\bx^1_a+\bx^2_a\right)+\frac{1}{2}\left(\bx^{1\dagger}_a+\bx^{2\dagger}_a\right)\bx^2_{a'}\right\}\bm{\mathcal{B}}_{a'b}\\
&\quad\ \ +\mathrm{Tr}\left\{\bu^{1\dagger}_{b'}\bu^1_b+\bu^{2\dagger}_b\bu^2_{b'}+\bv^\dagger_{b'}\bv_b\right.\\
&\qquad\qquad\ \left.+\bw^\dagger_b\bw_{b'}+\frac{1}{2}\bx^{1\dagger}_{b'}\bx^1_b+\frac{1}{2}\bx^{2\dagger}_b\bx^2_{b'}\right\}\left(\bdm^{2}_{ab'}-\bm{\mathcal{B}}_{ab'}\right)\\
&\quad\ \ \left.+\mathrm{Tr}\left\{\bu^{2\dagger}_{b'}\bu^1_b+\bu^{2\dagger}_b\bu^1_{b'}+\frac{1}{2}\bx^{2\dagger}_{b'}\bx^1_b+\frac{1}{2}\bx^{2\dagger}_b\bx^1_{b'}\right\}\left(\bdm^{2}_{ab'}-\bm{\mathcal{B}}_{ab'}\right)^*\right]\;,\\
\end{split}\end{equation}
where, as before, a sum over gauge group factors in the 
term proportional to $C_{2}(S)$ is implied.
Just as in the RGE of Eq.~(\ref{eq:4compH}), we have the ordered the
terms in (\ref{eq:4compmb}) to make evident the correspondence with the
terms in (\ref{eq:2compmb}). The length of this equation is somewhat
deceptive because many terms vanish in the case of the $R$-parity
conserving MSSM.  In a similar manner to the RGE for $H_{abc}$ earlier,
and as also seen in Paper~I, when $a$ and $b$ are sfermion indices,
$\bu^{1,2}_{a,b}$ and $\bx^{1,2}_{a,b}$ are zero.  Likewise, when $a$
and $b$ label Higgs fields, $\bv_{a,b}$ and $\bw_{a,b}$ are zero.  The
derivation of the MSSM RGEs is, therefore, considerably less cumbersome
than it appears at first sight.

\section{Particle Decoupling}\label{sec:dec}
In the previous section, we have obtained the RGEs for the Majorana
fermion and scalar mass parameters, as well as the bilinear and
trilinear scalar coupling parameters for a gauge field theory with
spin-0 and spin-1/2 fields. While these RGEs are not applicable to a
completely general gauge field theory, as we have explained in
Sec.~\ref{sec:formal}, they certainly apply to the MSSM with a conserved
$R$-parity quantum number. Since we did not assume supersymmetry in our
derivation, it is straightforward to include SUSY-breaking threshold
effects in essentially the same way as in Paper~I. Specifically, we
implement SUSY and Higgs particle thresholds as step functions in the
evaluation of the $\beta$-functions; {\it i.e.}  we include
the particle $\mathcal{P}$ in the effective theory only if the scale, $Q$,
is larger than the mass of $\mathcal{P}$.

To implement particle decoupling using this procedure clearly requires a
knowledge of the particle spectrum which, in many models, is obtained
using the RGEs. Fortunately, because the results depend only
logarithmically on the scale at which we decouple the particles, an
approximate knowledge of the particle spectrum suffices in order to
implement particle decoupling. Here, for reasons detailed in Sec.~III of
Paper~I, we will decouple both higgsinos at $Q=|\mu|$, the gauginos at
the scale $|M_i|$, and the heavy Higgs bosons at the scale $m_H$. The
decoupling of sfermions, if mixing effects are negligible, is also
straightforward. Since one of the main reasons for this analysis is to
study flavour-physics in the squark sector \cite{flavour}, we clearly
must include mixing among the squarks. We will, therefore, defer the
discussion of squark threshold corrections to a later point in the
paper.

It is evident that for any discussion of thresholds we need to know the
mass parameters in the MSSM. 
In addition to the SSB scalar and gaugino bilinears  given in (16) of Paper~I,
we have 
mass terms for the higgsinos as well as Higgs scalars from the
superpotential. These higgsino terms are,
\begin{equation}\label{eq:hinomass}\begin{split}
\mathcal{L}\ni&-\frac{1}{2}\left\{\frac{1}{2}\left(\mu+\mu^{*}\right)\left[\bar{\Psi}_{h^{0}_{u}}\Psi_{h^{0}_{d}}+\bar{\Psi}_{h^{0}_{d}}\Psi_{h^{0}_{u}}+\bar{\Psi}_{h^{+}_{u}}\Psi_{h^{-}_{d}}+\bar{\Psi}_{h^{-}_{d}}\Psi_{h^{+}_{u}}\right]\right\}\\
&+\frac{i}{2}\left\{\frac{1}{2i}\left(\mu-\mu^{*}\right)\left[\bar{\Psi}_{h^{0}_{u}}\gamma_{5}\Psi_{h^{0}_{d}}+\bar{\Psi}_{h^{0}_{d}}\gamma_{5}\Psi_{h^{0}_{u}}+\bar{\Psi}_{h^{+}_{u}}\gamma_{5}\Psi_{h^{-}_{d}}+\bar{\Psi}_{h^{-}_{d}}\gamma_{5}\Psi_{h^{+}_{u}}\right]\right\}\;,
\end{split}\end{equation}
resulting in two Majorana higgsino states of mass $|\mu|$, in the
approximation that any gaugino-higgsino mixing can be neglected.

The superpotential, $\hat{f}$, also leads to Higgs scalar bilinears along with
trilinear and quartic scalar terms in the potential. The so-called
$D$-term contributions to the Lagrangian density also lead to quartic
scalar couplings that, in the supersymmetric limit, are fixed by the gauge
couplings. In the notation of Ref.~\cite{wss}, these scalar potential
terms are given by, 
\begin{equation}\label{eq:quartlag}
\mathcal{L}\ni-\frac{1}{2}\sum_{A}\left|\sum_{i}\mathcal{S}^{\dagger}_{i}g_{\alpha}t_{\alpha A}\mathcal{S}_{i}\right|^{2}-\sum_{i}\left|\frac{\partial\hat{f}}{\partial\hat{\mathcal{S}}_{i}}\right|^{2}_{\hat{\mathcal{S}}=\mathcal{S}}\;.
\end{equation}
Superpotential interactions can result in scalar bilinear, trilinear and
quartic terms. As an illustration, if we take the second term
in (\ref{eq:quartlag}) and choose to differentiate with respect to the
up-type Higgs superfield, \textit{i.e.} $\hat{\mathcal{S}}=\hat{H}_{u}$, we
see that,
\begin{equation}\label{eq:quarteg}\begin{split}
\mathcal{L}\ni&-\left|\tilde{\mu}\right|^{2}h^{0\dagger}_{d}h^{0}_{d}
-\tilde{u}^{\dagger}_{Rk}\tilde{u}^{\dagger}_{Ll}\left(\bdf_{u}\right)^{T}_{kn}\left(\bdf_{u}\right)^{*}_{lm}\tilde{u}_{Rm}\tilde{u}_{Ln}\\
&-\left(\tilde{u}^{\dagger}_{Lk}\mtfuhus_{kl}\tilde{u}_{Rl}h^{0}_{d}+\mathrm{h.c.}\right)\;.
\end{split}\end{equation}
Note that we have inserted a tilde over the $\mu$ in writing these
terms. This is to allow for the fact that the higgsino mass $\mu$ and
the corresponding superpotential parameter $\tmu$ in the scalar sector
will, in general, evolve differently once SUSY-breaking threshold
effects are included. This is the analogue of the corresponding
situation for dimensionless parameters. As we saw in Paper~I,
gaugino-quark-squark couplings $\tilde{\bdg}^q$ evolve differently from
the corresponding gauge couplings (and in fact develop flavour-violating
components), while higgsino-quark-squark coupling matrices
$\tilde{\bdf}^{q}_{u,d}$ evolve differently from the corresponding
Yukawa coupling matrices $\bdf_{u,d}$ once the renormalization scale $Q$
is below the mass of the heaviest sparticle.  We should emphasize that
we can never get an RGE for the parameter $\tmu$ discussed above. It
enters via the SSB Higgs squared mass parameters in the combination
$m_{h_{u,d}}^2+|\tmu|^2$ or, as in (\ref{eq:quarteg}) above, via
combinations like $\mtfuhus_{lk}$, the coefficients of ``non-analytic''
trilinear scalar interactions \cite{hr}, denoted generically by
$\mathbf{c}_{ij}$ in Ref.~\cite{wss}. It is only for these combinations
of coefficients that enter the Lagrangian that we can (and do) obtain an
RGE, not for the separate pieces. Above all thresholds, the RGEs for
these combinations agree with the RGEs obtained from their component
pieces.  Notice that we have added a superscript $\Phi$ to the ``Yukawa
coupling constant'' that enters the trilinear scalar interaction in
(\ref{eq:quarteg}).
Here, where this term in the scalar potential originates
in the derivative of the superpotential with respect to
${\hat{h}}_u$, we set $\Phi=h_{u}$. 

The reader may legitimately wonder why we include a superscript $h_u$ on
the coupling $\bdf_u^{h_u}$ in the trilinear scalar coupling term but write
the quartic scalar coupling constant as the square of the usual quark
Yukawa coupling even below the scale of SUSY breaking. The reason is
that we are going to ignore the difference in the renormalization of the
quartic scalar couplings, which is also why we did not derive the
corresponding RGEs in Paper~I. These couplings are less important from a
phenomenological perspective (even though some of them enter the squark
and slepton mass matrices). It is possible that quartic squark couplings
that arise from the first term in (\ref{eq:quartlag}) may also develop a
non-trivial flavour-structure. These couplings will not affect the
two-body (flavour-violating) decays of squarks except at the loop-level,
leading us to believe that threshold corrections to these couplings can
be sensibly neglected. In the following, we have set {\it all} quartic
scalar couplings, irrespective of whether they originate in the
superpotential or in the $D$-term, to their supersymmetric values, {\it i.e.}
 equal to the ``square'' of
the usual Yukawa coupling or of the corresponding gauge coupling, but
have retained threshold effects in the trilinear couplings that cause
phenomenologically important mixing between left- and
right-squarks.\footnote{The quartic terms in the potential yield the
so-called $D$-term contributions to squark masses, and also the term
that makes the squark mass the same as the quark mass in the
supersymmetric limit. In most cases these are both subdominant
contributions to squark masses, giving futher credence to our
approximation.}  Since we do not, therefore, need independent RGEs for
these quartic couplings, we have not listed here equations to convert
the $\bdLm$ and $\bdLm'$ in (\ref{eq:fourlag}) to the corresponding
$\bm{\lambda}$ in (\ref{eq:twolag}), equivalent to (\ref{eq:hconv}) and
(\ref{eq:mconv}). Of course, the inverse of these equations is needed in
order to write the right hand sides of (\ref{eq:4compH}) and
(\ref{eq:4compmb}), but this is considerably simpler.

The decoupling of fields whose mixing can be ignored is straightforward:
below their mass scale, we simply remove all contributions from these
fields when evaluating any RGE. This covers: gauginos; higgsinos,
because they are approximately degenerate \cite{rge1}; and in many
models, also sleptons.

Next, we turn to the decoupling of the spin-zero
particles in the Higgs sector where we cannot disregard mixing effects,
and then discuss how we treat decoupling in the squark sector where
evaluation of flavour effects is one of our important goals. Before
proceeding, we mention one more complication that arises because 
$\mu$ does not necessarily equal $\tmu$. This makes the determination 
of $\mu$ more involved since the electroweak symmetry breaking
conditions only depend on its bosonic cousin $\tmu$; we will defer
the discussion of how $\mu$ and $\tmu$ are determined to the next
section.

\subsection{Higgs Boson Decoupling}\label{sec:higgsdec}
As discussed in Paper~I, the implementation of step-function decoupling
requires us to write the Lagrangian density with the fields in their
(approximate) mass eigenstate basis. This led us to rewrite the
interactions of the MSSM Higgs fields in terms of the spin-zero fields
$(\mathsf{h},\mathcal{H},H^\pm)$ defined \cite{rge1} in terms of the MSSM fields of
the scalar Higgs sector by,
\begin{eqnarray}
\label{eq:hrot}\left(\begin{array}{c}G^{+}\\\mathsf{h}\end{array}\right)&=\sn\left(\begin{array}{c}h^{+}_{u}\\[5pt]h^{0}_{u}\end{array}\right)+\cs\left(\begin{array}{c}h^{-*}_{d}\\[5pt]h^{0*}_{d}\end{array}\right)\\[5pt] 
\label{eq:Hrot}\left(\begin{array}{c}H^{+}\\\mathcal{H}\end{array}\right)&=\cs\left(\begin{array}{c}h^{+}_{u}\\[5pt]h^{0}_{u}\end{array}\right)-\sn\left(\begin{array}{c}h^{-*}_{d}\\[5pt]h^{0*}_{d}\end{array}\right)\;,
\end{eqnarray}
where $c=\cos\beta$ and $s=\sin\beta$.  For $m_H\gg M_Z$, the physical
Higgs bosons $h$, $H$, $A$ and $H^\pm$ are then approximately given by
\cite{rge1},\footnote{If $m_H$ is close to $M_Z$, all the bosons in the
Higgs sector have masses close to $M_Z$ and, as noted in Paper~I, the
resulting threshold corrections are small.}
\begin{eqnarray*}
\mathsf{h}&=\frac{h+iG^{0}}{\sqrt{2}}\\
\mathcal{H}&=\frac{-H+iA}{\sqrt{2}}\;.
\end{eqnarray*}
Here $G^+$ and $G^0$ are the would-be-Goldstone bosons that get
dynamically rearranged to become the longitudinal components of the
massive gauge bosons by the Higgs mechanism.  If $CP$ is violated in the
Higgs boson sector, the neutral spin-zero states $H$ and $A$ will
further mix with one another. When we eliminate the charged and neutral
components of the scalar doublets $H_u$ and $H_d$ and rewrite the
Lagrangian density in terms of $(\mathsf{h},\mathcal{H},H^\pm)$, new
operator structures, and concomitantly new couplings, appear in the
theory. For instance, the couplings of $h$ to fermions are given by the
SM Yukawa coupling matrices $\boldl_{u,d}$, and after decoupling of the
heavy Higgs scalars, the matrices $\bdf_{u,d}$ disappear altogether
(though, depending on the spectrum, their tilde cousins may remain).

For $Q>m_H$ where all Higgs fields are active in the RGEs, the rotation
to the Higgs field ``mass basis'' makes no difference as it is just a
field redefinition. For $Q<m_H$, however, this rotation is crucial
since it determines the particular combinations of fields that decouple from the RGEs. In this
case, only specific combinations of the parameters in the original theory
remain upon decoupling, while other combinations generally become irrelevant.
For example, when the two trilinear terms which couple $\tilde{u}_{L}$
and $\tilde{u}_{R}$ to the Higgs bosons are rotated into the
$\left(\mathsf{h},\mathcal{H}\right)$ basis, we see that
\begin{equation}\begin{split}
\mathcal{L}&\ni\tilde{u}^\dagger_{Rk}(\mathbf{a}^T_u)_{kl}\tilde{u}_{Ll}h^0_u-\tilde{u}^{\dagger}_{Rk}\mtsfuhut_{kl}\tilde{u}_{Ll}h^{0*}_{d}\\
&=\tilde{u}^\dagger_{Rk}(\mathbf{a}^T_u)_{kl}\tilde{u}_{Ll}\left(\sn\mathsf{h}+\cs\mathcal{H}\right)-\tilde{u}^{\dagger}_{Rk}\mtsfuhut_{kl}\tilde{u}_{Ll}\left(\cs\mathsf{h}-
\sn\mathcal{H}\right)\;.
\end{split}\end{equation}
Upon decoupling the field $\mathcal{H}$, this term becomes
\begin{equation}\label{eq:trileg}
\mathcal{L}\ni\tilde{u}^\dagger_{Rk}\left[\sn(\mathbf{a}^T_u)_{kl}-\cs\mtsfuhut_{kl}\right]\tilde{u}_{Ll}\mathsf{h}\;.
\end{equation}
Although in the complete MSSM case we can sensibly talk about the
evolution of the constituent pieces $(\mathbf{a}^T_u)_{kl}$ and 
$\mu$ separately, for $Q<m_H$, 
it is no longer possible to write RGEs for both
$\ba_{u}$ and $\tilde{\mu}^{*}\bdf^{h_{u}}_{u}$. We must instead talk
only about the single
 combination 
$\left[\sn(\mathbf{a}^T_u)_{kl}-\cs\mtsfuhut_{kl}\right]$
 that remains in the effective  theory below $Q=m_H$. In the same vein,
 we should mention that the quartic scalar operators with $\bdLm'_{abcd}$-type
 couplings in (\ref{eq:fourlag}) are absent above all SUSY thresholds
 in the $R$-parity conserving MSSM, 
 since in this case the number of daggered and undaggered fields is
 always the same (remember that these originate in the absolute square of a
 quadratic operator). We will see below that these $\bdLm'$-type
 couplings arise because of the way we define specific linear
 combinations of fields to take into account 
 threshold effects.

\subsection{Squark Decoupling}\label{subsec:sqdec}

Squarks of different flavours and types ($L$ or $R$) can mix as long as
they have the same electric charge. In the up-squark sector, squark mass
eigenstates are, therefore, combinations of $\tu_{L,R}$, $\tilde{c}_{L,R}$ and
$\tst_{L,R}$  squarks, and likewise in the down-squark sector. It is
these mass eigenstate fields that need to be decoupled for the evaluation of
threshold effects. 

The reader may have noticed that in Paper~I, as well as in this paper,
we have maintained manifest electroweak gauge invariance throughout, in
that we decouple entire multiplets together.  In keeping with this, we
have specified the boundary conditions for the $(3\times 3)$ Yukawa
coupling matrices at the scale $Q= m_t$, rather than at $Q=M_Z$ where the
top quark would have already been decoupled.\footnote{Specifying the
  boundary conditions at $Q=m_t$ is really
only necessary if we insist on decoupling every particle at the scale
equal to its mass. Such a procedure is advantageous in many situations,
but not compulsory. Indeed, as we will see in the next section, while we
decouple all sparticles and heavy Higgs bosons at their mass scale, we
decouple the top quark at $Q=M_Z$, below which the effective theory is
an $SU(3)_C\times U(1)_{\rm em}$ gauge theory.}
 To avoid any complications that may arise from splitting a multiplet,
for the purposes of evaluating the squark thresholds, {\it and for this
alone}, we will assume that left-right squark mixing effects which can
arise only from $SU(2)_L\times U(1)_Y$ breaking are not
important.\footnote{To be specific, we will focus our discussion on
squark thresholds, but exactly the same procedure applies for slepton
thresholds.}  This is clearly a sensible approximation when squark SSB
parameters are larger than the weak scale. Remembering that the
calculation is only logarithmically sensitive to the actual location of
the threshold, we can see that this approximation breaks down
significantly {\it only when the off-diagonal entries in the squark mass
matrices cause large cancellations in the calculation of the lighter
squark eigenvalue, leading to a physical mass squared much smaller than
either diagonal entry.} For pathological parameter values where such a
cancellation is operative and $\tq_L-\tq_R$ mixing indeed causes one of
the squarks to be much lighter than all other squarks of the same
charge, our approximation will not apply. Hereafter, we will assume that
any hierarchy in the squark mass spectrum has its origin in the values
of squark SSB parameters, and {\it is not the result of an accidental
cancellation} between the diagonal and off-diagonal parameters in the
squark-mass matrix. We will, therefore, neglect left-right mixing, for,
and only for, the purpose of determining the locations of the squark
thresholds.

Squark thresholds are then determined only by the SSB mass squared
matrices for the squarks. As with all other particles, we decouple any
squark below the scale equal to its mass. Since (for the purpose of
locating the squark mass thresholds) we are working in the approximation
that $SU(2)\times U(1)_Y$ breaking terms are neglected, the various left- and
right-squark mass thresholds are determined by the eigenvalues of the
corresponding SSB mass squared matrices. The associated technicalities
are best explained by describing the step-wise procedure that we use.
\begin{enumerate}
\item We evolve the gauge and Yukawa couplings to the high scale
  (usually taken to be $M_{\rm GUT}$) where the boundary conditions for
  the SSB parameters are specified. Toward this end, we first eliminate
  the quark masses in favour of the diagonal SM Yukawa coupling matrices
  in the quark mass basis at $Q=M_Z$ ($Q=m_t$ for the top quark), and
  rotate to a current basis (related to the quark mass basis by Eq.~(38)
  of Paper~I) at $Q=m_t$.\footnote{In our iterative procedure to solve
  the RGEs described below, it is important to remember that we retain
  the top quark all the way to $Q=M_Z$ so that electroweak gauge
  invariance is preserved.}  We morph the SM coupling
  matrices $\bm{\lambda}_{u,d}$ to the MSSM Yukawa coupling matrices
  $\bdf_{u,d}$ at $Q=m_H$, and include SUSY radiative corrections
  \cite{pierce} as given by ISAJET \cite{isajet} ({\it i.e} without
  flavour mixing among squarks) at $Q=M_{\rm SUSY}$, before continuing
  to evolve to $M_{\rm GUT}$.

\item All the dimensionless couplings and MSSM SSB parameters can now be
  evolved back to the low scale with Higgs bosons, higgsino and gaugino
  thresholds implemented as in Paper~I. We must, however, be careful in
  implementing the squark thresholds. At each step in the integration,
  we find the eigenvalues of the SSB squark mass squared matrices, and
  carry on the evolution in the original current basis if the scale
  $Q^2$ is larger than the highest of these eigenvalues, and none of the
  squarks are decoupled.

\item At some value of the scale, $Q_0$, $Q^2$ just crosses the largest
  eigenvalue of one of the SSB squark $\bdm^2$ matrices, evaluated at
  the scale $Q_0$. We decouple that particular squark at this scale, but
  retain all other squarks in the evolution. The up and down type
  squarks in the same doublet decouple together of course, so that
  $SU(2)$ is never broken by this procedure.  To decouple the squark, we
  rotate to the basis (denoted here by the superscript $M$) where the
  corresponding squark SSB matrix is diagonal at $Q=Q_0$ using the
  appropriate one of, %
\begin{subequations}\begin{eqnarray}
\left(\begin{array}{c}\tilde{u}_{L}\\\tilde{d}_{L}\end{array}\right)&=&\mathbf{R}_{Q}\left(\begin{array}{c}\tilde{u}^{M}_{L}\\\tilde{d}^{M}_{L}\end{array}
\right)\;,\\ \tilde{u}_{R}&=&\mathbf{R}_{u}\tilde{u}^{M}_{R}\;,\\
\tilde{d}_{R}&=&\mathbf{R}_{d}\tilde{d}^{M}_{R}\;, \ \ \ {\rm or\ 
one \ of}\\
\left(\begin{array}{c}\tilde{e}_{L}\\\tilde{\nu}_{L}\end{array}\right)&=&\mathbf{R}_{L}\left(\begin{array}{c}\tilde{e}^{M}_{L}\\\tilde{\nu}^{M}_{L}\end{array}\right)\;,\\
\tilde{e}_{R}&=&\mathbf{R}_{e}\tilde{e}^{M}_{R} \ \ \ {\rm when\ 
decoupling \ a \ slepton.}
\end{eqnarray}\end{subequations}
The  unitary rotation matrices, $\mathbf{R}_\bullet$, are chosen to diagonalize the
Hermitian SSB squark (or slepton) mass matrices, for example:
\begin{equation}
\left(\mathbf{R}^{\dagger}_{Q}\mathbf{m}^{2}_{Q}\mathbf{R}_{Q}\right)_{ij}
=\left(\mathbf{m}^{2}_{Q}\right)^{\mathrm{diag}}_{ij}
\equiv(\mathbf{m}^{2}_{Q})^{\mathrm{diag}}_{ii}\bm{\delta}_{ij}\;. 
\end{equation}
  Decoupling this squark is now straightforward --- as with all
  sparticles, we simply introduce a $\theta_{\tq_k}$ for the decoupled
  squark $\tq_k^M$ (where $k$ is the mass basis index, and $\tq=\tQ_L,\
  \tu_R$ or $\td_R$) into the RGEs, since we are in the ``mass basis''
  for this squark at least within the approximation that we have
  discussed.\footnote{From now on, we will no longer write the
  superscript $M$ for the squark mass eigenstates, but it will
  presumably be clear from the context whether we are referring to these,
  or to the states in the original basis.}
  Just for clarification, in any other basis these $\theta_{\tq_k}$
  would be matrices.  For instance, the left-handed squark theta,
  $\theta_{{\tilde{Q}_{k}}}$, can be written in the original current
  basis as,
\begin{equation}\label{eq:Thetadef}
(\bm{\Theta}_{Q})_{ij}=(\mathbf{R}_{Q}\bm{\Theta}^{\mathrm{diag}}_{Q}\mathbf{R}^{\dagger}_{Q})_{ij}=\theta_{{\tilde{Q}_{k}}}(\mathbf{R}_{Q})_{ik}\bm{\delta}_{kl}(\mathbf{R}^{\dagger}_{Q})_{lj}\;.
\end{equation}
We have checked that this same $\bm{\Theta}_Q$ matrix works for all
terms involving doublet squark thresholds, and likewise for the other
squarks and sleptons. We always write the RGEs in a current basis where
the SSB squark mass matrices are diagonal (see the item immediately
following this). This is why
we have $\theta$'s rather than the matrices $\bm{\Theta}$ appearing. We
rotate back to our original current basis when we present numerical
results.

\item To evolve to lower values of $Q$, we must be in {\it some current
  basis}. We thus also rotate the corresponding quarks (or leptons for
  slepton decoupling) by the same amount as the decoupled squark
  (slepton), and continue the evolution in this new current basis. The
  ``mass'' and the ``eigenvector'' of the decoupled squark is frozen at
  its decoupled value at $Q=Q_0$. As a technical aside, we note that the
  same RGE (with $(3\times 3)$ mass squared matrices) can be used to
  continue the evolution in this new current basis. The
  $\theta_{\tq_k}$'s that we introduce into the RGEs ensure that the
  decoupled squark does not contribute to the evolution of the remaining
  squarks. In other words, in the new mass basis the decoupled squark
  {\it never} contributes to the evolution of the SSB matrix elements
  for the $(2\times 2)$ sub-matrix in the sub-space orthogonal to the
  decoupled squark state.\footnote{Rather than re-code new RGEs where we
  retain only the two active squarks for the subsequent evolution, we
  continue the evolution with all the squarks. Below the scale $Q_0$,
  just the elements of the $(2\times 2)$ sub-matrix in the vector space
  orthogonal to the decoupled squark are physical. The remaining
  elements, though calculated are never used, and their value does not
  affect the calculation.} We use the RGEs
  to continue the evolution to lower scales, now obtaining the
  eigenvalues of the $(2\times 2)$ sub-matrix in the orthogonal sub-space
  just mentioned, and once again decoupling the squark when the larger
  of the two eigenvalues just crosses $Q^2$. The corresponding
  eigenvector of this sub-matrix gives us the second squark state with
  the same gauge quantum numbers to be decoupled (remember that we are
  in the new current basis). As before we freeze its mass and wave
  function upon decoupling. Notice that, by construction, the two
  decoupled states are orthogonal to one another and to the single
  remaining active state, as they should be. The single remaining state
  may be decoupled in a straightforward way. Since we know the relation
  between the new current basis and our original current basis, it is
  straightforward to rotate the SSB parameters back to the latter basis.

\item We have described squark decoupling for any one set of squarks, {\it
  i.e.} doublet squarks, the up-type singlet squarks or the down-type
  singlet squarks, but it should be clear that the procedure that we
  have described works for all the squarks as well as for sleptons. 

\item We must, likewise, decouple squark mass eigenstates obtained using
  our procedure during the evolution of all other parameters. However, 
  for the scalar trilinear coupling parameters, ${\bf a}$, there is the added
  complication that we must decouple the heavy Higgs bosons at the scale
  $Q=m_H$. In this case particular combinations of parameters that
  include ${\bf a}_{ij}$ need to be evolved,  as we have already discussed. 

\end{enumerate}

\section{Application to the MSSM}\label{sec:mssm}

We now apply the general results obtained in Sec.~\ref{sec:formal} to
derive the RGEs for the 
dimensionful couplings of the $R$-parity conserving
MSSM. As in Paper~I, our strategy will be to read off the couplings in
(\ref{eq:fourlag}) from the MSSM couplings 
detailed in Eq.~(16)-(22) of Paper~I, and from (\ref{eq:hinomass})-(\ref{eq:quarteg})
of this paper, and substituting these into the
general RGEs obtained in Sec.~\ref{sec:formal}. The quartic scalar
couplings not listed here may be derived from (\ref{eq:quartlag}) above,
and are also found in Ref.~\cite{wss}. Once the couplings are all
correctly identified, the derivation of the RGEs using (\ref{eq:RGEmx}),
(\ref{eq:RGEmxp}), (\ref{eq:4compH}) and (\ref{eq:4compmb}) is tedious
but straightforward.


We proceed by outlining how to go about deriving the RGEs for each group
of dimensionful couplings: gaugino and higgsino mass parameters,
trilinear couplings, and scalar soft mass parameters and their bilinear
cousins. Our purpose is to guide the reader interested in deriving the
RGEs as to where the various terms come from, and to point out potential
pitfalls that may be encountered on the way. The complete set of RGEs
for the dimensionful parameters is listed in Appendix~\ref{sec:rgeapp}. 

\subsection{Gaugino and Higgsino Mass Parameters}\label{subsec:GHP}

The application to the MSSM of the general RGEs in (\ref{eq:RGEmx}) and
(\ref{eq:RGEmxp}) is straightforward. The matrices $\mathbf{m}_{X}^{(\prime)}$ are
constructed from the $M^{(\prime)}_{1,2,3}$, the SSB gaugino mass terms, and
$\mu$ which appears via
the higgsino mass terms in (\ref{eq:hinomass}). The parameter
$\tilde{\mu}$ that appears in the scalar Higgs sector should not be 
confused with the corresponding fermion mass parameter $\mu$.


As previously noted, the trace terms in (\ref{eq:RGEmx}) and (\ref{eq:RGEmxp})
vanish so that we need only sum over the scalar index $b$ and
multiply out the terms. When the higgsinos are rotated as in Eq.~(14)
of Paper~I, $\mathbf{m}^{(\prime)}_{X}$ becomes diagonal and the derivation
simplifies further. We find, for example, that the RGE for $M_{2}$ is
\begin{equation}\label{eq:M2rgeill}\begin{split}
{\left(4\pi\right)}^2\frac{dM_2}{dt}=&M_2\swi\left[3\sqk\bgtq_{kl}\bgtq^\dagger_{lk}+\slk\bgtl_{kl}\bgtl^\dagger_{lk}+\sh\mgthusq\left(\sn^2\h+\cs^2\Hh\right)\right.\\
&\left.\qquad\quad+\sh\mgthdsq\left(\cs^2\h+\sn^2\Hh\right)\right]\\[5pt]
&+2\sn\cs\left(-\h+\Hh\right)\sh\left[\gthd\mu^*\gthu+(\gthd)^{*}\mu(\gthu)^{*}\right]-12\swi M_2g^2_2\;,
\end{split}\end{equation}
and draw the reader's attention to the following points:
\begin{enumerate}
\item The first term, which is $M_{2}$ multiplied by a number of gaugino
coupling terms, arises from the terms in (\ref{eq:RGEmx}) that have
$\left(\bdm_{X}\pm i\bdm'_{X}\right)$ on the extreme left or on the
extreme right. When
in this position the $\bdm^{(\prime)}_{X}$ is connected to an external
gaugino, and since $\bdm^{(\prime)}_{X}$ is diagonal it contributes a gaugino
mass term.  This term includes a sum over scalar fields and
corresponding fermion fields (that enter the RGE via 1-loop Feynman
diagrams), with concomitant couplings such as $\tilde{\mathbf{g}}^{Q}$
and $\gthu$ of fields that couple to the wino.
This sum over \textit{all} scalar fields includes all active sfermion
flavours at the scale $Q$, {\it i.e.} sfermions with masses larger than
$Q$ are decoupled. As in Paper~I, the decoupling of any particle
is introduced into the RGEs via the $\theta_\mathcal{P} = 1$ if $Q$ is larger than
the mass of $\mathcal{P}$, with $\theta_\mathcal{P} = 0$ otherwise. Thus, for instance, in the 
first term $\sqk\bgtq_{kl}\bgtq^\dagger_{lk}$, the summation over $k$
includes all {\it active} left-type squarks at the scale $Q$ (as
emphasized in the last section, this RGE is written in the basis where
the corresponding squark SSB matrix is diagonal), while the $l$ sums
over the quarks (that are all always assumed to be present in the effective
theory we use to compute the RGE\footnote{In this paper, we always assume
  that $Q\ge M_Z$, and retain all SM particles in the effective
  theory, but write an explicit $\theta_h$ only for the light scalar
  doublet that includes the would-be-Goldstone bosons.})  which is why we do
not include the corresponding $\theta$'s.  
Below the highest sfermion threshold the sfermion
gaugino coupling terms are, therefore, {\it truncated} traces. 

\item In the RGEs, we write $\theta_\mathcal{P}$'s for, and only for,
  particles $\mathcal{P}$ that enter via the internal lines of loop
  diagrams that contribute in the RGEs. Sometimes, this internal
  particle is the same as the particle on the external leg of the
  corresponding diagram, as illustrated by the appearance of
  $\theta_{\tw}$ in the first and third terms of
  (\ref{eq:M2rgeill}). Furthermore, since at any scale $Q$, we retain
  only those (non-SM) particles with masses larger than $Q$ in the effective
  Lagrangian that we use to derive the RGEs, we stop evolving the
  coefficient of any (composite) field operator --- these coefficients
  are just the Lagrangian parameters --- when the scale $Q$ falls below
  the mass of any of the fields that enter the operator. Thus, in the
  case of $M_2(Q)$, we freeze its evolution below the scale $Q_0$ where
  $Q_0=M_2(Q_0)$.

\item The reader may be struck by the second term on the
  right-hand-side, proportional to $\mu$, whose appearance seems odd at
  first sight.  It originates in the terms in (\ref{eq:RGEmx}) where
  $\bdm^{(\prime)}_{X}$ is sandwiched between two $\bx^{1,2}_{b}$
  matrices. Since $\bx^{1,2}_{b}$ connects gauginos to higgsinos, and
  the external fields are gauginos, the $\bdm^{(\prime)}_{X}$ is
  necessarily a higgsino mass term. Notice that this term is a threshold
  effect: it does not vanish only if $|\mu|<Q<m_{H}$. The light and
  heavy Higgs boson doublets in (\ref{eq:hrot}) and (\ref{eq:Hrot}),
  respectively, make equal but opposite $\mu$-dependent contributions to
  the evolution of $M_2$ which indeed cancel above all thresholds.  The
  appearance of $\mu$ is a general feature of the electroweak gaugino
  mass RGEs \mbox{---} listed in full in Appendix~\ref{sec:rgeapp},
  (\ref{app:rgem1})-(\ref{app:rgem2p}). For much the same reasons, for
  appropriate mass ordering, the RGE for the complex parameter $\mu$,
  (\ref{app:rgemu}), develops a dependence on the electroweak gaugino
  mass parameters, $M^{(\prime)}_{1}$ and $M^{(\prime)}_{2}$.

\item Despite the appearance of the complex $\mu$-dependent terms just
  discussed in the RGEs for the gaugino mass parameters, the reality of
  $M_i$ and $M_i^{'}$  under renormalization group evolution is preserved,
  as it must. 

\item Finally, notice that above all thresholds, where the $SU(2)$ gaugino
  coupling matrices
  $\bgt^\bullet$  reduce to $g_2$ times the unit matrix in the flavour
  space (see Paper~I) and 
   all $\theta_{i}=1$, we
recover the MSSM result upon summing over all flavours \cite{martv}.  

\end{enumerate}

\subsection{Trilinear Couplings}\label{subsec:tri}

For the $R$-parity conserving MSSM, the trilinear scalar couplings
$\mathbf{H}_{abc}$ involve couplings between one of the Higgs boson
fields and two sfermion fields. In order to facilitate decoupling of the
Higgs scalars, we saw that it is necessary to rotate these Higgs fields
to their (approximate) mass basis given by (\ref{eq:hrot}) and
(\ref{eq:Hrot}), so that $a,b$ and $c$ run over the {\it complex}
fields,
$\left\{\mathsf{h},\mathcal{H},G^{+},H^{+},\tilde{u}_{Li},\tilde{d}_{Li},\tilde{e}_{Li},\tilde{\nu}_{Li},\tilde{u}_{Ri},\tilde{d}_{Ri},\tilde{e}_{Ri}\right\}$,
and $i$ runs over all three flavours. Unlike the charged
would-be-Goldstone fields $G^\pm$ that appear explicitly, the neutral
would-be-Goldstone boson $G^0$ is contained in the complex field
$\mathsf{h}$.  When fully expanded out in flavour space,
$\mathbf{H}_{abc}$ is a $\left(25\times25\times25\right)$ array, whose
entries are mostly all zero, and with many non-zero entries related by
the $SU(2)$ gauge symmetry of the interactions, that we can easily read
off. For instance, from (\ref{eq:trileg}) we immediately see that,
\begin{equation}\label{eq:Hurhul}
\mathbf{H}_{\tilde{u}_{Rk},\mathsf{h},\tilde{u}_{Ll}}=\left[\sn(\mathbf{a}^T_u)_{kl}-\cs\mtsfuhut_{kl}\right]\;.
\end{equation}

As we mentioned below (\ref{eq:4compH}), we can without loss of
generality in the RGEs, always choose the external index $a$ to be a
sfermion, since then several terms drop out in the derivation of the
corresponding RGE for the trilinear parameter. Of course, we also need
to work out the $\bdLm$ and $\bdLm'$ matrices. These are
$\left(25\times25\times25\times25\right)$ arrays in scalar field space,
again with mostly zero entries and with most non-zero entries related by
the gauge invariance of the interactions. As noted previously, these
can be worked out from (\ref{eq:quartlag}). The extraction of the
$\bdLm$ terms is entirely straightforward, but care must be taken with
the sfermion flavours and squark colours, which are implicitly summed
over in (\ref{eq:quartlag}). As alluded to at the end of
Sec.~\ref{sec:higgsdec}, we note here that $\bdLm'=\bm{0}$ when the MSSM
Lagrangian is written in terms of $h_{u}$ and $h_{d}$ Higgs
fields. These couplings arise only when the Higgs boson doublets are
rotated to their mass basis using (\ref{eq:hrot}) and (\ref{eq:Hrot})
since only the conjugate fields $h_d^{-*}$ and $h_d^{0*}$ appear in the
linear combinations with the unstarred fields $h_u^+$ and $h_u^0$,
respectively.
In order to get only one daggered scalar field, lepton and baryon number
 conservation for the dimension four operators imply that we must have
 Higgs fields in the interactions. Moreover, 
 we must have
both an up-type and a down-type Higgs field {\it before} the Higgs field
 rotation to the mass basis, and that the
down-type Higgs fields must be $h^{(0,-)*}_{d}$ while the up-type fields
 must be $h_u^{(+,0)}$. 
This only occurs in quartics in (\ref{eq:quartlag}) that derive from
differentiating the superpotential with respect to the superfields
${\hat{u}}_L$ or ${\hat{d}}_L$, and so
have a $\tilde{u}_{iR}$ and a
$\tilde{d}_{jR}$ in the interaction. 
As a result, there are very few non-vanishing $\bdLm'$-type quartic
interactions. Indeed, these are completely given by,
\begin{equation}
{\mathcal L} \ni \left({\bf f}_u^T{\bf f}_d^*\right)_{kl}\left(\mathsf{h}H^+ -
G^+\mathcal{H}\right) \tu_{Rk}^\dagger\td_{Rl} + {\rm h.c.}
\end{equation}

In deriving the RGEs, we must remember to include the contributions
from the would-be-Goldstone fields in the sum over {\it complex} scalar
fields. The neutral Goldstone boson is automatically present in the field
$\mathsf{h}$ along with the (almost) SM-like Higgs boson, but the
charged fields $G^{\pm}$ must explicitly be included in the sum. For
example, the RGE for
$\mathbf{H}_{\tilde{u}_{Ri},\mathsf{h},\tilde{u}_{Lj}}$ includes a term
$\bdLm_{e,f,\mathsf{h},\tilde{u}_{Lj}}\mathbf{H}_{\tilde{u}_{Ri},e,f}$
and we must sum $e$ and $f$ over \textit{all} scalars in the theory for
which
$\bdLm_{e,f,\mathsf{h},\tilde{u}_{Lj}}\mathbf{H}_{\tilde{u}_{Ri},e,f}\ne0$.

Using Eq.~(\ref{eq:4compH}) to derive the RGE for the operator in
(\ref{eq:Hurhul}), writing all internal threshold $\theta_\mathcal{P}$'s
explicitly, we obtain:\\
${\left(4\pi\right)}^2\frac{d\left[\sn\au_{ij}-\cs\mtsfuhu_{ij}\right]}{dt}=$
\begin{equation}\begin{split}\label{eq:RGEtriu}
&\suk\left\{\h\left[\sn\au_{ik}-\cs\mtsfuhu_{ik}\right]\left[\frac{2\glp^{2}}{3}\left(\cs^{2}-\sn^{2}\right)\delta_{kj}+2\sn^{2}\left[\fuul^{\dagger}\fuul\right]_{kj}\right]\right.\\
&\qquad\left.+\Hh\sn\cs\left[\cs\au_{ik}+\sn\mtsfuhu_{ik}\right]\left[-\frac{4\glp^{2}}{3}\delta_{kj}+2\left[\fuul^{\dagger}\fuul\right]_{kj}\right]\right\}\\
&+\sul\sqk\left[-2\left(\frac{\glp^{2}}{9}+\frac{4\gthl^{2}}{3}\right)\delta_{ik}\delta_{lj}+6\fuhu_{ij}\fuhu^{\dagger}_{lk}\right]\left[\sn\au_{kl}-\cs\mtsfuhu_{kl}\right]\\
&+2\sqk\left\{\h\left[\left(\frac{\glp^{2}}{12}-\frac{3\gtwl^{2}}{4}\right)\left(\sn^{2}-\cs^{2}\right)\delta_{ik}+2\sn^{2}\left[\fuur\fuur^{\dagger}\right]_{ik}-\cs^{2}\left[\fddr\fddr^{\dagger}\right]_{ik}\right]\right.\\
&\qquad\qquad\times\left[\sn\au_{kj}-\cs\mtsfuhu_{kj}\right]+\Hh\sn\cs\left[\left(\frac{\glp^{2}}{6}-\frac{3\gtwl^{2}}{2}\right)\delta_{ik}+2\left[\fuur\fuur^{\dagger}\right]_{ik}\right.\\
&\qquad\qquad\left.\left.+\left[\fddr\fddr^{\dagger}\right]_{ik}\right]\times\left[\cs\au_{kj}+\sn\mtsfuhu_{kj}\right]\right\}\\
&+2\Hh\sdk\left[\sn\ad_{ik}+\cs\mtsfdhd_{ik}\right]\left[\fddl^{\dagger}\fudl\right]_{kj}\\
&+\frac{2}{3}\sbi\sn\left(M_{1}-iM'_{1}\right)\left(\sh(\gtphu)^{*}\bgtpq^{*}_{ik}\bftur_{kj}-\frac{4}{3}\bgtpq^{*}_{ik}(\bdf_{u})_{kl}\bgtpur^{*}_{kj}\right.\\
&\left.-4\sh\bftuq_{ik}\bgtpur^{*}_{kj}(\gtphu)^{*}\right)-\frac{32}{3}\sgl\sn\left(M_{3}-iM'_{3}\right)\bgtsq^{*}_{ik}(\bdf_{u})_{kl}\bgtsur^{*}_{lj}\\
&-6\swi\sh\sn\left(M_{2}-iM'_{2}\right)(\gthu)^{*}\bgtq^{*}_{ik}\bftur_{kj}+\frac{2}{3}\sbi\sh\cs\mu^{*}\gtphd\left(4\bftuq_{ik}\bgtpur^{*}_{kj}-\bgtpq^{*}_{ik}\bftur_{kj}\right)\\
&+6\sh\swi\cs\mu^{*}\gthd\bgtq^{*}_{ik}\bftur_{kj}-4\sh\cs\mu^{*}\bftdq_{ik}(\bdf^{\dagger}_{d})_{kl}\bftur_{lj}\\
&+\suk\left[\sn\au_{ik}-\cs\mtsfuhu_{ik}\right]\left[\frac{8}{9}\sbi\bgtpur^{T}_{kl}\bgtpur^{*}_{lj}+\frac{8}{3}\sgl\bgtsur^{T}_{kl}\bgtsur^{*}_{lj}+2\sh\bftur^{\dagger}_{kl}\bftur_{lj}\right]\\
&+\h\left[3\sn^{2}(\bdf^{\dagger}_{u})_{kl}(\bdf_{u})_{lk}+\cs^{2}\left\{3(\bdf^{\dagger}_{d})_{kl}(\bdf_{d})_{lk}+(\bdf^{\dagger}_{e})_{kl}(\bdf_{e})_{lk}\right\}\right]\left[\sn\au_{ij}-\cs\mtsfuhu_{ij}\right]\\
&+\Hh\sn\cs\left[3(\bdf^{\dagger}_{u})_{kl}(\bdf_{u})_{lk}-\left\{3(\bdf^{\dagger}_{d})_{kl}(\bdf_{d})_{lk}+(\bdf^{\dagger}_{e})_{kl}(\bdf_{e})_{lk}\right\}\right]\left[\cs\au_{ij}+\sn\mtsfuhu_{ij}\right]\\
&+\frac{1}{2}\sh\left\{\h\left[\cs^{2}\left(\sbi\mgtphdsq+3\swi\mgthdsq\right)+\sn^{2}\left(\sbi\mgtphusq+3\swi\mgthusq\right)\right]\right.\\
&\qquad\quad\;\;\times\left[\sn\au_{ij}-\cs\mtsfuhu_{ij}\right]+\Hh\sn\cs\left[-\left(\sbi\mgtphdsq+3\swi\mgthdsq\right)\right.\\
&\qquad\quad\;\;\left.\left.+\left(\sbi\mgtphusq+3\swi\mgthusq\right)\right]\times\left[\cs\au_{ij}+\sn\mtsfuhu_{ij}\right]\right\}\\
&+\sql\left[\sh\bftuq_{ik}\bftuq^{\dagger}_{kl}+\sh\bftdq_{ik}\bftdq^{\dagger}_{kl}+\frac{1}{18}\sbi\bgtpq^{*}_{ik}\bgtpq^{T}_{kl}+\frac{3}{2}\swi\bgtq^{*}_{ik}\bgtq^{T}_{kl}\right.\\
&\left.\qquad\quad+\frac{8}{3}\sgl\bgtsq^{*}_{ik}\bgtsq^{T}_{kl}\right]\left[\sn\au_{lj}-\cs\mtsfuhu_{lj}\right]\\
&-3\left\{\left(\frac{1}{36}\sqi+\frac{4}{9}\suj+\frac{1}{4}\h\right)g'^{2}+\frac{3}{4}\left(\sqi+\h\right)g^{2}_{2}+\frac{4}{3}\left(\sqi+\suj\right)g^{2}_{3}\right\}\left[\sn\au-\cs\mtsfuhu\right]_{ij}
\end{split}\end{equation}
The first set of terms before the appearance of $(M_{1}-iM'_1)$ arises
from the $\bdLm$ and $\bdLm'$ terms in ($\ref{eq:4compH}$).
All $\bdLm$ terms have a non-zero contribution
but since most $\bdLm'$ entries are zero, only
$\bdLm'_{abef}\mathbf{H}^{*}_{cef}$ contributes.

The second set of entries, containing gaugino mass terms, originate in the
non-vanishing traces in ($\ref{eq:4compH}$) where $(\bdm_{X}-i\bdm'_{X})$
is on the extreme left. For our case, $\Phi_a =\tu_{Rj}$, so that
$\bv_a$ has a quark/gaugino as the first/second index, while
$\bw_a$ has the higgsino/quark as the first/second index. Remembering
that taking the transpose, or the dagger, flip the order of these indices,
it is quite straightforward to see that the first set of trace terms
where  $(\bdm_{X}-i\bdm'_{X})$
is located on the extreme left necessarily give contributions
proportional to SSB gaugino masses. For this, we must keep in mind that
the matrices $\bx^{1,2}_{\bullet}$ connect gauginos to higgsinos, and vice-versa.  
A similar analysis of the next set of trace terms in (\ref{eq:4compH}),
with $(\bdm_{X}+i\bdm'_{X})$ enclosed by other fermion matrices, shows
that the altered location of the Majorana  fermion mass matrix
now results in the appearance of the ``higgsino mass'',
$\mu^*$ (without any tilde). 

The remaining terms, except for the very last one which obviously arises
from
$C_{2}(S)\mathbf{H}_{abc}$, are from the traces with primed scalar indices.
The important thing to note is that the trace in ($\ref{eq:4compH}$)
denotes  a sum over fermion types, which sometimes, but not necessarily,
becomes a trace over fermion flavours, since now a scalar
may carry a flavour index. In the present case,
the trace over the product of two $\bu^{1,2}_{\bullet}$ matrices does
result in a trace over fermion flavours, since then the scalar
index on the $\bu^{1,2}_{\bullet}$ necessarily corresponds to a Higgs field which
does not carry any flavour. We refer the interested reader to the
discussion below Eqs.~(36) and (37) of Paper~I where a completely
analogous situation is discussed in more detail.

We have illustrated the derivation of the RGE for the trilinear scalar
coupling of squarks to the lighter of the two Higgs
doublets. The RGE for the corresponding coupling,
$\left[\cs\au_{ij}+\sn\mtsfuhu_{ij}\right]$, to the heavier doublet can
be obtained in the same manner. By taking linear combinations of these
RGEs, we can obtain the separate RGEs for ${\ba}_u$ and
$\tmu^*\bdf_u^{h_u}$.\footnote{The $s$ and $c$ can be taken out of the
derivatives on the left-hand-side as noted in Paper~I. For $Q>m_H$, when
both doublets are in the theory, the rotation is irrelevant. Otherwise,
the input value of $\tan\beta$ corresponds to the ratio of VEVs at the
scale $Q=m_H$ when the two doublet model reduces to the one-doublet
model. We may think about this as evolving the couplings $\ba_u$ and
$\bdf_u^{h_u}$ from the high scale down to the scale $Q=m_H$, at which
we {\it must} do the Higgs rotations to reduce to the one-doublet
model.}
Since the coupling $\tmu^*\bdf_u^{h_u}$ always occurs as a
product, it is not possible to obtain the RGEs for the individual
factors. Of course, above all thresholds, we must have $\tmu=\mu$ and
$\bdf_u^{h_u}=\bdf_u$.  We have checked that with these
replacements, our RGEs reduce to the MSSM RGEs \cite{martv} if we put
all $\theta_i=1$ and take care to sum over all internal flavours.




\subsection{Soft Masses}

Finally, we turn to the RGEs for the SSB mass parameters and their
bilinear cousins $\bm{\mathcal{B}}$; we have almost all the matrices
necessary to use (\ref{eq:4compmb}) to find the RGEs. The majority of
the missing $\mathbf{m}^{2}$ and $\bm{\mathcal{B}}$ terms appear in the
SSB Lagrangian and so the required matrices can be written down
directly. Note that when we write the Lagrangian in the rotated Higgs
basis, there are no $\bm{\mathcal{B}}$ terms so that
$\bm{\mathcal{B}}_{ef}=0$. In this case, several terms drop out of the
RGE in (\ref{eq:4compmb}).

\subsubsection{Higgs Mass Terms}

Using (\ref{eq:4compmb}) to derive the RGE for the coefficient 
of the $\mathsf{h}^\dagger\mathsf{h}$ term in the Lagrangian we find that\\
${\left(4\pi\right)}^{2}\frac{d\left[\sn^{2}M^{2}_{H_{u}}+\cs^{2}M^{2}_{H_{d}}-\sn\cs\left(b+b^{*}\right)\right]}{dt}$
\begin{equation}\begin{split}
=&\frac{3}{2}\h\left[\glp^{2}+\gtwl^{2}\right]\left(\cs^{2}-\sn^{2}\right)^{2}\left[\sn^{2}M^{2}_{H_{u}}+\cs^{2}M^{2}_{H_{d}}-\sn\cs\left(b+b^{*}\right)\right]\\
&+\Hh\left[\glp^{2}\left(-\cs^{4}+4\sn^{2}\cs^{2}-\sn^{4}\right)+6\sn^{2}\cs^{2}\gtwl^{2}\right]\left[\cs^{2}M^{2}_{H_{u}}+\sn^{2}M^{2}_{H_{d}}+\sn\cs\left(b+b^{*}\right)\right]\\
&-6\h\Hh\left[\glp^{2}+\gtwl^{2}\right]\sn\cs\left(\cs^{2}-\sn^{2}\right)\left[\sn\cs\left\{M^{2}_{H_{u}}-M^{2}_{H_{d}}\right\}-\frac{1}{2}\left(\cs^{2}-\sn^{2}\right)\left(b+b^{*}\right)\right]\\
&+\suk\sul\left[-2\glp^{2}\left(\sn^{2}-\cs^{2}\right)\delta_{lk}+6\sn^{2}\left[\fuult\fuuls\right]_{lk}\right]\musq_{kl}\\
&+\sqk\sql\left[\glp^{2}\left(\sn^{2}-\cs^{2}\right)\delta_{lk}+6\sn^{2}\left[\fuurs\fuurt\right]_{lk}+6\cs^{2}\left[\fddrs\fddrt\right]_{lk}\right]\mqsq_{kl}\\
&+\sdk\sdl\left[\glp^{2}\left(\sn^{2}-\cs^{2}\right)\delta_{lk}+6\cs^{2}\left[\fddlt\fddls\right]_{lk}\right]\mdsq_{kl}\\
&+\slk\sll\left[-\glp^{2}\left(\sn^{2}-\cs^{2}\right)\delta_{lk}+2\cs^{2}\left[\feers\feert\right]_{lk}\right]\mlsq_{kl}\\
&+\sek\sel\left[\glp^{2}\left(\sn^{2}-\cs^{2}\right)\delta_{lk}+2\cs^{2}\left[\feelt\feels\right]_{lk}\right]\mesq_{kl}\\
&+6\suk\sql\left[\sn\au_{lk}-\cs\mtsfuhu_{lk}\right]\left[\sn\au^{\dagger}_{kl}-\cs\mtsfuhu^{\dagger}_{kl}\right]\\
&+6\sqk\sdl\left[\cs\ad_{lk}-\sn\mtsfdhd_{lk}\right]\left[\cs\ad^{\dagger}_{kl}-\sn\mtsfdhd^{\dagger}_{kl}\right]\\
&+2\slk\sel\left[\cs\ae_{lk}-\sn\mtsfehd_{lk}\right]\left[\cs\ae^{\dagger}_{kl}-\sn\mtsfehd^{\dagger}_{kl}\right]\\
&-2\sh\left|\mu\right|^{2}\left\{\sbi\left[\sn^{2}\mgtphusq+\cs^{2}\mgtphdsq\right]+3\swi\left[\sn^{2}\mgthusq+\cs^{2}\mgthdsq\right]\right\}\\
&-2\sh\left\{\sbi\left(M^{2}_{1}+M'^{2}_{1}\right)\left[\sn^{2}\mgtphusq+\cs^{2}\mgtphdsq\right]+3\swi\left(M^{2}_{2}+M'^{2}_{2}\right)\left[\sn^{2}\mgthusq+\cs^{2}\mgthdsq\right]\right\}\\
&-\frac{1}{2}\left\{-4\sh\sbi\sn\cs\mu^{*}\gtphu\gtphd\left(M_{1}+iM'_{1}\right)-12\sh\swi\sn\cs\mu^{*}\gthu\gthd\left(M_{2}+iM'_{2}\right)\right\}\\
&-\frac{1}{2}\left\{-4\sh\sbi\sn\cs\mu(\gtphu)^{*}(\gtphd)^{*}\left(M_{1}-iM'_{1}\right)-12\sh\swi\sn\cs\mu(\gthu)^{*}(\gthd)^{*}\left(M_{2}-iM'_{2}\right)\right\}\\
&-\h\left(\frac{3g'^{2}}{2}+\frac{9g^{2}_{2}}{2}\right)\left[\sn^{2}M^{2}_{H_{u}}+\cs^{2}M^{2}_{H_{d}}-\sn\cs\left(b+b^{*}\right)\right]\\
&+\h\left\{\sn^{2}\left[6\bdf^{*}_{u}\bdf^{T}_{u}\right]_{kk}+\cs^{2}\left[6\bdf^{*}_{d}\bdf^{T}_{d}+2\bdf^{*}_{e}\bdf^{T}_{e}\right]_{kk}+\sbi\sh\left[\sn^{2}\mgtphusq+\cs^{2}\mgtphdsq\right]\right.\\
&\left.\qquad\;\;+3\swi\sh\left[\sn^{2}\mgthusq+\cs^{2}\mgthdsq\right]\right\}\left[\sn^{2}M^{2}_{H_{u}}+\cs^{2}M^{2}_{H_{d}}-\sn\cs\left(b+b^{*}\right)\right]\\
&+\Hh\sn\cs\left\{\left[6\bdf^{*}_{u}\bdf^{T}_{u}\right]_{kk}-\left[6\bdf^{*}_{d}\bdf^{T}_{d}+2\bdf^{*}_{e}\bdf^{T}_{e}\right]_{kk}+\sbi\sh\left[\mgtphusq-\mgtphdsq\right]\right.\\
&\left.\qquad\qquad+3\swi\sh\left[\mgthusq-\mgthdsq\right]\right\}\left[\sn\cs\left\{M^{2}_{H_{u}}-M^{2}_{H_{d}}\right\}-\frac{1}{2}\left(\cs^{2}-\sn^{2}\right)\left(b+b^{*}\right)\right]\;,
\end{split}\end{equation}
where $M^{2}_{H_{u}}\equiv\left(\mhusq+\mtsq\right)$ and $M^{2}_{H_{d}}\equiv\left(\mhdsq+\mtsq\right)$.

All terms up to the trilinear scalar 
couplings derive from the single term,  $\bdLm_{afbe}\bdm^{2}_{ef}$,
in (\ref{eq:4compmb}). All other quartic terms are zero, either because
$\bdLm'$ vanishes when $a=b$, or because
$\bm{\mathcal{B}}_{ef}=0$. Since this operator has no flavour indices, all
flavours are internal and, therefore, summed over. Extra care should be
applied when dealing with sums over squarks since there are additional
factors of $3$ arising from a sum over colours. Once the entries of the
$\bdLm$ matrix have been worked out, this contribution should not
pose any special difficulty. The so-called $S$-term is also contained in
the terms proportional to $g^{\prime2}$ from this contribution. 

The terms containing the trilinear scalar couplings obviously come from
the $\left(\mathbf{H}^{*}_{eaf}+\mathbf{H}_{fae}\right)\mathbf{H}_{ebf}$
term in (\ref{eq:4compmb}). Since we have chosen the first index of
$\mathbf{H}_{abc}$ to always be a sfermion, $\mathbf{H}_{aef}=0$ for
$a=\mathsf{h}$. For the contributions from the squark-Higgs scalar
trilinear interactions the factor $2$ changes to a factor $6$ due
to a colour sum.

Next, we turn to the contributions from the Majorana fermion mass terms
from the $\bdm_X^{(\prime)}$ matrices in (\ref{eq:4compmb}).  When the two
factors of $\left(\bdm_{X}\pm i\bdm'_{X}\right)$ are next to one
another, it should be clear that we should obtain either a
$\left|\mu\right|^{2}$ or an $\left(M^{2}_{1,2}+M'^{2}_{1,2}\right)$
term. On the
other hand, when the factors of $\left(\bdm_{X}\pm i\bdm'_{X}\right)$
are separated by an $\bx^{1,2}_{b}$-matrix, we obtain a product of
gaugino and higgsino mass parameters because $\bx^{1,2}_{b}$ only
connects gauginos with higgsinos.  The trace, of course, is a sum over
all gauginos and higgsinos in the theory.

The terms that derive from the next term,
$\mathbf{C}_{2}(S)\bdm^{2}_{ab}$, should be obvious. 
Following these, we have the set of traces with
primed scalar indices, $a'$ and $b'$. As discussed earlier, many of
these terms are zero when $a$ and $b$ are Higgs scalars. We are left
with traces over only $\bu^{1,2}_{\bullet}$ and
$\bx^{1,2}_{\bullet}$. The trace over the $\bu^{1,2}_{\bullet}$ matrices
leads to a trace over matter fermion flavours, of which the quark traces
acquire an additional colour factor of $3$.

The Higgs scalar mass term for which we just obtained the RGE written
above is the only combination which remains in the effective theory for
$Q<m_{H}$. The complete set of mass terms in the rotated Higgs basis
are, 
\begin{equation}\begin{split}\label{eq:higgsterms}
\mathcal{L}\ni&-\left[\sn^{2}\left(m^{2}_{H_{u}}+\left|\tilde{\mu}\right|^{2}\right)+\cs^{2}\left(m^{2}_{H_{d}}+\left|\tilde{\mu}\right|^{2}\right)-\sn\cs\left(b+b^{*}\right)\right]\mathsf{h}^\dagger\mathsf{h}\\
&-\left[\cs^{2}\left(m^{2}_{H_{u}}+\left|\tilde{\mu}\right|^{2}\right)+\sn^{2}\left(m^{2}_{H_{d}}+\left|\tilde{\mu}\right|^{2}\right)+\sn\cs\left(b+b^{*}\right)\right]\mathcal{H}^\dagger\mathcal{H}\\
&-\left[\sn\cs \left(m^{2}_{H_{u}}+\left|\tilde{\mu}\right|^{2}\right)-\sn\cs
\left(m^{2}_{H_{d}}+\left|\tilde{\mu}\right|^{2}\right)-\cs^{2}b+\sn^{2}b^{*}\right]\mathsf{h}^\dagger\mathcal{H}\\
&-\left[\sn\cs \left(m^{2}_{H_{u}}+\left|\tilde{\mu}\right|^{2}\right)-\sn\cs
\left(m^{2}_{H_{d}}+\left|\tilde{\mu}\right|^{2}\right)+\sn^{2}b-\cs^{2}b^{*}\right]\mathcal{H}^\dagger\mathsf{h}\;,
\end{split}\end{equation}
and the RGEs for these coefficients can be obtained in a similar manner.  We
can then obtain the separate RGEs for the real parameters $\left(m^{2}_{H_{u}}+\left|\tilde{\mu}\right|^{2}\right)$,
$\left(m^{2}_{H_{d}}+\left|\tilde{\mu}\right|^{2}\right)$ and the complex parameter $b$, valid for $Q>m_H$, by taking
appropriate linear combinations.

It is important to note that since the terms $\mhusq$ and $\mtsq$ only
appear in the Lagrangian in the combination $\left(\mhusq+\mtsq\right)$,
we cannot derive an RGE for them separately.  Of course, above all
thresholds, supersymmetry requires $\mu=\tmu$, so that we can use the
RGE for $\mu$, (\ref{app:rgemu}), to extract the RGE for the soft mass
parameters $\mhusq$ and $\mhdsq$, but these will cease to be valid once
any one particle is decoupled from the theory. We have checked that
these RGEs reduce to the standard ones \cite{martv} once all $\theta_i$
are set equal to unity, and that our RGE for $m_{\mathsf{h}}^2$ reduces
to the RGE for the SM Higgs boson mass parameter \cite{luoprl} if we
set the quartic scalar coupling in the SM to be the appropriate
combination of gauge couplings.

\subsubsection{Sfermion Mass Terms}

The RGEs for sfermion soft mass terms are somewhat simpler on account of
the fact that they have no Higgs fields in their operator in the
Lagrangian. Otherwise, the derivation follows in a similar fashion to
the Higgs mass terms, with the obvious differences in the terms in
(\ref{eq:4compmb}) that contribute to the RGE. There is still only one
$\bdLm$ term (because again the $\bdLm'$ and $\bm{\mathcal{B}}$ are
zero), and this contributes a large proportion of the RGE including
terms with, and terms without, traces over sfermion mass matrices. These
terms include the $S$-term as for the Higgs mass RGE above. When $e$ and
$f$ are both Higgs fields, this gives a $b$-parameter dependence in the
RGEs for the sfermions listed in Appendix~\ref{sec:rgeapp} for $Q<m_{H}$.
Since the external
fields are sfermions, both trilinear terms in (\ref{eq:4compmb}) now
contribute, and in the trace terms, only $\bv_{\bullet}$ and
$\bw_{\bullet}$ terms are non-zero. These terms yield the contributions
that depend on the square of the gaugino mass parameters, and also on
$|\mu|^2$. Above all thresholds, these latter contributions cancel with
$\tmu^2$ contributions arising from Higgs boson loops.

There is one point about the RGEs for the sfermion SSB mass parameters
that we ought to draw attention to. For $Q<m_H$, where we only have the
light Higgs doublet in the theory, we would expect that the couplings
${\bf f}_u$ can occur only in the combination $s{\bf f}_u$. A look at
the term just before the trilinear coupling terms in
(\ref{app:mQ2rgelow}) shows, however, that this is not the case. (There
are analogous terms in the RGEs for ${\bf m}_{U,D}^2$, and also for
${\bf m}_{L,E}^2$.) The ``product of Yukawa couplings'' that appears in
these terms (without accompanying $s^2$ or $c^2$ factors) is really a
$\bm{\Lambda}_{iklj}$-type coupling for the quartic interaction of
squarks which, as we have already explained, we have approximated by its
supersymmetric limit and set equal to the product of the corresponding
elements of the Yukawa coupling matrix, each frozen at the scale
$Q=m_H$; {\it e.g.} $\bm{\Lambda}_{\tQ_i\tu_{Rk}\tu_{Rl}\tQ_j}\sim
\left({\bf f}_u\right)_{ik}^* \left({\bf f}_u\right)_{lj}^T$ in the
first of such terms in (\ref{app:mQ2rgelow}).

We have checked that, except for terms involving couplings and mass
parameters of Higgs boson fields, the RGEs that we obtain agree with
those in Ref.~\cite{sakis}.  Indeed the RGEs in Ref.~\cite{sakis} do
reduce to the MSSM RGEs if we set all the $\theta$'s to be one. However,
since the RGEs of Ref.~\cite{sakis} appear to have been written
\textit{without any rotation} of the Higgs boson fields, we found it
impossible to compare contributions involving thresholds for Higgs boson
fields. For these same reasons, we are unable to see how their RGEs for
Yukawa couplings reduce to the corresponding RGEs for the SM
\cite{rge1} when all new particles are decoupled. Likewise, we are not able to
obtain the RGE for the SM Higgs boson mass parameter using the RGEs for
the Higgs scalar SSB parameters as given in Ref.~\cite{sakis}.

\section{Solutions to the RGEs and Flavour-Violating SSB
  Parameters}\label{sec:flav}
\setcounter{subsubsection}{0}

While the {\it quark mass basis}, where both up and down quark Yukawa
matrices are diagonal (at a chosen scale), may be the most physical basis
to work in, as we saw in Paper~I, it is more convenient to work in a
{\it current basis} where either the up or the down, but not both,
Yukawa matrices are diagonal. In this case, we rotate the entire quark
doublet (in the flavour space) thereby preserving the $SU(2)$ gauge
symmetry. To similarly preserve the underlying supersymmetry, we should
also rotate the squark multiplets the same way that we rotate the
quarks; {\it i.e.} we should rotate the quark (and also
lepton) superfields. Of course, just as the Yukawa coupling matrices
transform under this change of basis (see (38) and (39) of Paper~I to
set the notation), the SSB mass and ${\bf a}$-parameter matrices in a
arbitrary basis are related to the corresponding matrices in the
basis where the up (or the down) quark Yukawa coupling matrix is diagonal at
$Q=m_t$ (the matrices in this special  basis are denoted with the
superscript $M$) according to,
\begin{align}
{({\ba}_{u,d})}^T & = {{\mathbf{V}}_R}(u,d){{\left({\ba}^{M}_{u,d}\right)}}^{T}{{\mathbf{V}}^\dagger_L}(q)\;,\\
\label{eq:mqmix}
{\bdm}^{2}_Q & = {\mathbf{V}}_L(q){\left({\bdm}^{2}_Q\right)}^M{{\mathbf{V}}^{\dagger}_L}(q)\;,\\
{\bdm}^{2}_{U,D} & =
{\mathbf{V}}_R(u,d){\left({\bdm}^{2}_{U,D}\right)}^M{{\mathbf{V}}^{\dagger}_R}(u,d)\;,
\end{align}
where we set $q=u(d)$ in ${\bf V}_L(q)$ in the basis where up (down)
type quark Yukawa coupling matrices are diagonal at $Q=m_t$.  The
matrices ${\bf V}_{L,R}(u,d)$, defined via (38) of Paper~I, are the
linear transformations that connect any current basis to the quark mass
basis. The SSB matrices ${\bf a}_{u,d}^M$
 and $\left(m_\bullet^2\right)^M$ in the basis where up quark Yukawa
 couplings are diagonal are related to the corresponding matrices in
 which the down quark Yukawa couplings are diagonal by,
\begin{align}
{({\ba}_{u,d}^M)}^T(u) & = {({\ba}_{u,d}^M)}^T(d){\bf K}^\dagger,
\label{eq:tranuda}\\ 
({\bdm}^{2}_Q)^M(u) & = {\bf K}({\bdm}^{2}_Q)^M(d) {\bf K}^\dagger, 
\label{eq:tranudmQ}\\
({\bdm}^{2}_{U,D})^M(u) & = ({\bdm}^{2}_{U,D})^M(d).
\label{eq:tranudmR}
\end{align}
Here $u$ and $d$ in the parenthesis denote whether the up or down type
Yukawa coupling matrices, respectively, are diagonal at $Q=m_t$, after
the complete transformation (38) of Paper~I; {\it i.e.} when we go to
the basis where the up-type Yukawa matrix is diagonal, we also
transform the singlet down sector by the matrix ${\bf V}_R(d)$,
and likewise if we transform to the basis where the down-type Yukawa
coupling matrix is diagonal. The matrix ${\bf K}$ that appears in
(\ref{eq:tranuda}) and (\ref{eq:tranudmQ})
is the Kobayashi-Maskawa matrix \cite{km} given by,
\begin{equation}\label{eq:KM}
\mathbf{K}=\mathbf{V}^{\dagger}_{L}(u)\mathbf{V}_{L}(d)\;.
\end{equation}
%

There is no {\it a priori} preference for the one or the other
basis. However, if in a process there are only up, or only down, type
quarks in the initial and final states, it makes sense to use the mass
basis for these quarks. In the next section, we will apply the
considerations of this paper to the rate for the decay of the lightest
up-type squark to the final state $c\tz_1$. For this reason, other than
where explicitly stated, we will work in a basis where the up-type
quark Yukawa couplings are diagonal at $Q=m_{t}$, {\it i.e.}  we will
take $\mathbf{V}_{L,R}(u)=\dblone$. We then have $\mathbf{V}_{L}(d)={\bf
K}$. Finally, to fully specify a current basis we must also choose
$\mathbf{V}_{R}(d)$ \mbox{---} the matrix for the rotation of the right-handed
down-type quarks.\footnote{Within the SM, physics is independent of our choice
of the matrices $\mathbf{V}_{R}(u)$ and $\mathbf{V}_{R}(d)$, and depends
on $\mathbf{V}_{L}(u,d)$ only through the KM matrix. This is also
the case for the MSSM with mSUGRA
boundary conditions, but more generally, physics also depends on
the right-handed quark rotation matrices.} Except when we
illustrate results for extensions of the mSUGRA model, we also set
$\mathbf{V}_{R}(d)=\dblone$.

Since the boundary conditions for the dimensionless couplings of SM
particles are specified at the weak scale, while those for the
dimensionful parameters of the MSSM are usually given at a high scale,
we solve the RGEs using an iterative procedure. This is, of course,
quite standard, but the incorporation of threshold effects entails some
new complications. We begin by evolving the measured gauge and Yukawa
couplings of quarks and leptons, specified in a chosen current basis,
from $Q=M_Z$ to the high scale (usually $Q=M_{\rm GUT}$) where the SSB
parameters of the MSSM are specified.\footnote{Although the gauge and
Yukawa coupling RGEs along with their tilde cousins form a closed
system, the solution to these RGEs may be sensitive to sparticle
thresholds through particle decoupling effects. We will see that these
effects can be especially important in non-universal models where squark
mass matrices are non-diagonal (in the basis where the corresponding
quark Yukawa couplings are diagonal) and the squark mass eigenvalues
have substantial splitting.} We can now evolve {\it all} the MSSM
parameters down to the weak scale, decoupling the particles one-by-one
as described in Sec.~\ref{sec:dec}. In the course of this downward
evolution, we must split the SM couplings $g_i$ and ${\bf f}_{u,d,e}$
from their SUSY cousins ${\bf \tg}^\bullet_i$ and ${\bf
\tf}_{u,d,e}^\bullet$ as discussed in Paper~I, and also incorporate the
distinction between $\mu$ and $\tmu$. As we will see shortly, this
complicates the introduction of the electroweak symmetry breaking
conditions which are incorporated at $M_{\rm SUSY}$. Finally, below the
scale $m_H$ where the two Higgs doublet MSSM transitions to the one
Higgs doublet model, we switch from the MSSM quark and lepton Yukawa
couplings ${\bf f}_\bullet$ to the corresponding SM couplings
$\bm{\lambda}_\bullet$, which we reset along with the gauge couplings,
after evolving these to $Q=M_Z$ ($Q=m_t$ for the top quark Yukawa
coupling). We then evolve back to the GUT scale where we re-set
$\mu=\tmu$ as discussed below, and iterate the solution to the system of
RGEs until convergence is obtained to the specified precision.

\subsubsection{Weak scale boundary conditions}
For the gauge sector, 
we take as our input at the weak scale the current PDG values \cite{pdg}
$\alpha_{em}$, $\alpha_{s}$ and $\sin^{2}{\theta_{W}}$, which are
\begin{gather*}
\alpha^{-1}_{em}(M_{Z})=127.925\pm0.016\;;\;\alpha_{s}(M_{Z},\msb)=0.1176\pm0.002\;;\\
\sin^{2}{\theta_{W}}(M_{Z},\msb)=0.23119\pm0.00014. 
\end{gather*}
These are the couplings extracted using the effective theory with the
electroweak gauge bosons as well as the top quark integrated out at
$Q=M_Z$. In order to use the SM for evolution for $Q>M_Z$, we must match
these couplings to those of the full SM, which, to two-loop accuracy
implies that the SM gauge couplings in the $\msb$ scheme are given by
\cite{shifts,ramond1},
%
\begin{subequations}\begin{align}
\frac{1}{\alpha_{1}(M_{Z})}=&\frac{3}{5}\left[\frac{1-\sin^{2}{\theta_{W}}(M_{Z})}{\alpha_{em}(M_{Z})}\right]+\frac{3}{5}\left[1-\sin^{2}{\theta_{W}}(M_{Z})\right]4\pi\Omega(M_{Z})\;,\\
\frac{1}{\alpha_{2}(M_{Z})}=&\frac{\sin^{2}{\theta_{W}}(M_{Z})}{\alpha_{em}(M_{Z})}+\sin^{2}{\theta_{W}}(M_{Z})4\pi\Omega(M_{Z})\;,\\
\frac{1}{\alpha_{3}(M_{Z})}=&\frac{1}{\alpha_{s}(M_{Z})}+4\pi\Omega_{3}(M_{Z})\;,
\end{align}\end{subequations}
where
\begin{subequations}\begin{align}
\Omega(\mu)=&\frac{1}{24\pi^{2}}\left[1-21\ln{\left(\frac{M_{W}}{\mu}\right)}\right]+\frac{2}{9\pi^{2}}\ln{\left(\frac{m_t}{\mu}\right)}\;,\\
\Omega_{3}(\mu)=&\frac{2}{24\pi^{2}}\ln{\left(\frac{m_{t}}{\mu}\right)}\;.
\end{align}\end{subequations}
Notice that in order to preserve the $SU(2)$ symmetry of the effective
theory down to $Q=M_Z$, we have, as mentioned earlier, integrated out
the top quark at $Q=M_Z$ rather than at its mass as we do for all other
particles.  This is the origin of the $\ln(m_t/\mu)$ terms in the
matching conditions for the gauge couplings above. We emphasize that we
decouple all SUSY particles as well as the additional Higgs bosons at
the scale of their mass: as a result, we do not get corresponding jumps
in the gauge couplings as these decouple. Our method --- which is also
used in ISAJET --- has an important advantage in that it ``sums the logs of
the ratio of any large mass to $M_Z$'', in contrast to the frequently
used procedure that uses MSSM evolution down to $M_Z$, and then corrects
for this via a ``single step evolution'' (between the heavy scale and
$M_Z$) to take into account the difference between the running in the
MSSM and in the SM.

Next, we convert the values of these gauge couplings in the $\msb$ scheme
to their corresponding values in the
$\drb$ scheme using the relations~\cite{martv2}:
\begin{subequations}\begin{align}
\frac{1}{\alpha_{1}(\drb)}=&\frac{1}{\alpha_{1}(\msb)}\;,\\
\frac{1}{\alpha_{2}(\drb)}=&\frac{1}{\alpha_{2}(\msb)}-\frac{1}{6\pi}\;,\\
\frac{1}{\alpha_{3}(\drb)}=&\frac{1}{\alpha_{3}(\msb)}-\frac{1}{4\pi},
\end{align}\end{subequations}
and use the results as 
boundary conditions at $Q=M_Z$ when solving the RGEs.

For the Yukawas, we begin with the quark masses at $Q=M_{Z}$ (the masses
of the light quarks and leptons at $M_{Z}$ can be found in
Ref.~\cite{fusaoka}), and convert to SM Yukawas using
$v_{SM}=248.6/\sqrt{2}$ as in Ref.~\cite{pierce}. The masses of the
first two generations of quarks have substantial error, which leads to a
corresponding error in their Yukawa couplings. The third generation
quark masses are more precisely known \mbox{---} we take the top pole
mass $m_{t}=172$~GeV, and $m_b(M_Z)=2.83$~GeV \cite{javier}.  We rotate the diagonal Yukawa couplings, which
are in the ``quark mass basis'' to a current basis
using (39) of Paper~I at $Q=m_t$. 
We include SUSY radiative corrections
\cite{pierce} (from ISAJET), with inter-generation quark mixing
neglected, at $Q=M_{\rm SUSY}$.
Finally, for the numerical results that we
present in the rest of the paper, we parametrize the Kobayashi-Maskawa
matrix as in (\ref{eq:rotmat}) below, and take $\sin\alpha=0.2243$,
$\sin\beta=0.0037$ and $\sin\gamma=0.0413$, $\delta_{\beta}=60^\circ$
and $\delta_{\alpha}=\delta_\gamma=0$.

\subsubsection{Electroweak symmetry breaking}

It is traditional to use the observed value of $M_Z^2$
to determine the value of $\mu^2$ from the minimization conditions for
the scalar potential in the Higgs sector. The inclusion of threshold
effects causes additional complications for this program. Recall that
the electroweak symmetry breaking conditions are imposed at
$M_{\mathrm{SUSY}}=\sqrt{m_{\tilde{t}_{L}}m_{\tilde{t}_{R}}}$ which is
always smaller than the mass of the heaviest SUSY particle. As a result,
$\mu^2$ is now conceptually and numerically different from $|\tmu|^2$,
which is of course the parameter that enters in the Higgs boson
potential. Moreover, the Higgs potential depends only on
$M^{2}_{H_{u}}\equiv\left(\mhusq+\mtsq\right)$ and
$M^{2}_{H_{d}}\equiv\left(\mhdsq+\mtsq\right)$, so that it is not
possible to separate $\tmu^2$ from the SSB parameters
$m_{H_u}^2$ and $m_{H_d}^2$ that are specified at the high scale.
%
%
Using the combinations,
\begin{gather*}
\left(M^{2}_{H_{u}}+M^{2}_{H_{d}}\right)= m^{2}_{H_{u}}+m^{2}_{H_{d}}+2\left|\tilde{\mu}\right|^{2}\quad\mathrm{and}\\
\left(M^{2}_{H_{d}}-M^{2}_{H_{u}}\right)= m^{2}_{H_{d}}-m^{2}_{H_{u}}\;,
\end{gather*}
the tree level minimisation conditions of the Higgs potential can be
written as,
\begin{gather}\begin{split}
\left(M^{2}_{H_{u}}+M^{2}_{H_{d}}\right)=-\frac{1}{\cos{2\beta}}\left(M^{2}_{H_{d}}-M^{2}_{H_{u}}\right)-\frac{1}{2}\left(g'^{2}+g^{2}\right)\left(v^{2}_{u}+v^{2}_{d}\right)\;,\label{eq:EWSB}
\end{split}\\
b=\sn\cs\left(M^{2}_{H_{u}}+M^{2}_{H_{d}}\right)\;.\label{eq:EWSB2}
\end{gather}
The first of these fixes the sum $(M^{2}_{H_{u}}+M^{2}_{H_{d}})$ in terms
of the difference $(M^{2}_{H_{u}}-M^{2}_{H_{d}})$.  Since we know the
difference at the GUT
scale, we can evolve this down to $M_{\rm SUSY}$ (along with other SSB
parameters) during  the iterative process that we use to solve the RGEs.
At $Q=M_{\mathrm{SUSY}}$
we use (\ref{eq:EWSB}) to solve for $M^{2}_{H_{u}}+M^{2}_{H_{d}}$, which
can be evolved back to the GUT scale. 
%
%
%
We then use the sum to fix $\tilde{\mu}=\mu$ at the GUT scale, reset the
difference to its input value, and iterate. The value of the higgsino
parameter $\mu$ can then be obtained at all scales using
(\ref{app:rgemu}).  The $b$-parameter can, as usual, be eliminated in
favour of $\tan\beta$ using (\ref{eq:EWSB2}).\footnote{The alert reader
will notice that something is amiss in (\ref{eq:EWSB2}): the
$b$-parameter is complex, while the right-hand-side is manifestly
real. The point is that at any one scale, chosen here to be $M_{\rm
SUSY}$, the $b$-parameter can always be made real. Indeed, the very fact
that we have written positive values for the VEVs $v_u$ and $v_d$
mandates that $b$ is real and positive at $Q=M_{\rm SUSY}$. To be
specific, we can always make a gauge transformation such that just the
lower component of the scalar doublet $H_u$ has a VEV, and that this VEV is real and
positive. Then the minimization of the scalar potential in the Higgs
sector requires that the VEV of $H_d$ is aligned; {\it i.e.} it is also only in
its lower component. This alignment is a result of the dynamics. Finally, we
can redefine the phase of the doublet superfield ${\hat{H}}_d$ so that
$v_d$ is real and positive. This is not compulsory, but is the customary
practice that allows us to define $\tan\beta$ to be real and
positive. If instead we write the VEV of the down type Higgs field as
$|v_d|\exp{(i\theta_d)}$ and $\tan\beta=\frac{v_u}{|v_d|}$, the
left-hand-side of (\ref{eq:EWSB2}) would have to be amended to
$b\exp{i\theta_b}$. So, although we choose real VEVs and a concomitantly
real $b$-parameter at $Q=M_{\rm SUSY}$, we retain the complex $b$ in the
RGEs, since $b$ will not remain real at other scales.}

In our discussion up to this point we have ignored another potential
complication that arises if $m_H>M_{\rm SUSY}$. In this case, the
heavy particles of the Higgs sector decouple, and for $M_{\rm
SUSY}<Q<m_H$, we only have the light doublet in the effective theory
that we use to calculate the RGEs. In this case, the heavy Higgs doublet
mass term
$\left[\cs^{2}\left(m^{2}_{H_{u}}+\left|\tilde{\mu}\right|^{2}\right)
+\sn^{2}\left(m^{2}_{H_{d}}+\left|\tilde{\mu}\right|^{2}\right)+
\sn\cs\left(b+b^{*}\right)\right]$ and the Higgs mixing term,
$\left[\sn\cs
\left(m^{2}_{H_{u}}+\left|\tilde{\mu}\right|^{2}\right)-\sn\cs
\left(m^{2}_{H_{d}}+\left|\tilde{\mu}\right|^{2}\right)+\sn^{2}b
-\cs^{2}b^{*}\right]$ (and its complex conjugate) in
Eq.~(\ref{eq:higgsterms}), together with $\tan\beta$, are frozen at their
values at $Q=m_H$, while the light doublet mass parameter,
$\left[\sn^{2}\left(m^{2}_{H_{u}}+\left|\tilde{\mu}\right|^{2}\right)
+\cs^{2}\left(m^{2}_{H_{d}}+\left|\tilde{\mu}\right|^{2}\right)
-\sn\cs\left(b+b^{*}\right)\right]$,  along with
$v_{\rm SM}=\sqrt{v_u^2+v_d^2}$, continues to evolve to $M_{\rm SUSY}$.
The three frozen coefficients together with the evolved mass term for
the light doublet can now be used to solve for
$\left(m^{2}_{H_{d}}+\left|\tilde{\mu}\right|^{2}\right)$,
$\left(m^{2}_{H_{u}}+\left|\tilde{\mu}\right|^{2}\right)$ and the
complex $b$-parameter. We can now find an iterative solution in the same
manner as  for $M_{\rm SUSY}>m_H$.

Before closing this section, we should add that although we have
discussed EWSB conditions only at tree-level, in practice, we minimize
the one-loop effective potential including effects of third generation
Yukawa couplings, but ignoring all flavour-mixing effects in this
computation. These corrections, which effectively shift the Higgs boson
SSB mass squared parameters by $\Sigma_u$ and $\Sigma_d$, respectively,
are evaluated by replacing $f_{t,b,\tau}$ in the relations by the (3,3)
element of the corresponding Yukawa matrices, and with the dimensionful
parameters also replaced by the (3,3) element of the corresponding
matrix (or the appropriate frozen value).

\subsubsection{SSB parameters at the high scale}

Since our purpose in these papers is to address flavour physics of
sparticles in as general a way as possible, subject to experimental
constraints that seem to suggest that flavour physics is largely
restricted by the structure of the Yukawa coupling matrices, we thought
it would be useful to first seek a general parametrization for SSB
parameters that does not introduce a new source of flavour-violation,
but allows for non-universality of model parameters. Additional, but
uncontrolled, flavour-violation can easily be incorporated by allowing
for other contributions to the SSB mass and trilinear
parameter matrices. We use the (s)quark sector to illustrate our
arguments, but almost identical considerations will apply to (s)leptons, 
except that in this case we would also have to include additional
lepton number and lepton-flavour violating matrices in the singlet
(s)neutrino sector.

Within the framework of the $R$-parity conserving MSSM, the SSB matrices
$\bdm^2_{U,D,Q}$ and ${\bf a}_{u,d}$ potentially include new sources of
flavour-violation, not included in the superpotential Yukawa
couplings. In order not to introduce a new source of flavour-violation,
these SSB matrices must be diagonal in the same superfield basis (where
the SM fermions and their scalar superpartners are rotated by the same
matrices) that the superpotential Yukawa interactions (renormalized at
the same high scale as the SSB parameters) are diagonal.\footnote{This
is {\it not equivalent} to the requirement that the Yukawa couplng
matrix commute with the corresponding ${\bf a}$-parameter matrix because
these non-Hermitean matrices are diagonalized by bi-unitary rather than by
unitary transformations.}  Of course, it is impossible to simultaneously
diagonalize ${\bf f}_u$ and ${\bf f}_d$, but what we mean is that the
SSB mass matrices that describe the mixing of both left- and right-up
type squarks, and their trilinear couplings must be diagonal in the
basis that the matrix ${\bf f}_u$ is diagonal, and likewise for the down
sector. However, since $SU(2)$ symmetry dictates that the ${\bf
m}^{2}_{\tu_L}$ and ${\bf m}^{2}_{\td_L}$ SSB matrices must be the
identical, this doublet squark mass-squared matrix must be proportional
to the unit matrix, $\dblone$, in order to remain diagonal, both when
the up- or the down-type Yukawa coupling matrix is diagonal. In
contrast, the matrices ${\bf m}_U^2$ and ${\bf a}_u$ (${\bf m}_D^2$ and
${\bf a}_d$) can be functions of the up (down) type Yukawa coupling
matrices (and their Hermitian adjoints) chosen in such a way that these
matrices are simulaneously diagonal when we transform these to the basis
where the corrsponding superpotential Yukawa coupling matrix is
diagonal.

To find the most general parametrization of the ${\bf m}_{U,D}^2$ and
${\bf a}_{u,d}$ matrices of the type that we are looking for, we first
note that these, respectively, transform in the same way as the matrices
$\left({\bf f}^T_{u,d}{\bf f}^*_{u,d}\right)^n$ and ${\bf
f}_{u,d}\left({\bf f}_{u,d}^\dagger {\bf f}_{u,d}\right)^n$, where $n$
is any integer. Thus any linear combination of these matrices (with
$n=0,1,2\cdots$) is guaranteed to be diagonal in the basis that ${\bf
f}_{u.d}$ is diagonal (at the high scale at which we input the SSB
parameters). The only question, then, is just how many terms we need to
allow in the linear combinations that make up ${\bf m}^2_{U,D}$ or ${\bf
  a}_{u,d}$, to guarantee the most general form for
these SSB matrices, so that flavour violation enters only through the
superpotential Yukawa coupling matrices? This is easiest to see in the
diagonal basis for the Yukawa couplings. The SSB matrices are also
diagonal in this basis, and so are completely specified by $n_g$
diagonal elements, where $n_g$ is the number of
generations. Transforming to a general basis does not alter the number
of parameters that we need: we thus know that we must have $n_g$ terms
in each of the linear combinations for ${\bf m}^2_{U,D}$ and for ${\bf
a}_{u,d}$ that we discussed above. For the MSSM with $n_g=3$
generations, we thus parametrize the SSB sfermion mass and ${\bf
a}$-parameter matrices at the high scale as,
\begin{subequations}\label{eq:GUTbound}
\begin{align}
\bdm^2_{Q,L}&=m^2_{\{Q,L\}0}\dblone+\mathbf{T}_{Q,L}\;, \\
\bdm^2_{U,D,E}&=m^2_{\{U,D,E\}0}[c_{U,D,E}\dblone+R_{U,D,E}\bdf^T_{u,d,e}
\bdf^*_{u,d,e}+S_{U,D,E}(\bdf^T_{u,d,e}\bdf^*_{u,d,e})^2]+\mathbf{T}_{U,D,E}\label{eq:GUTboundsferm}\;,\\
\ba_{u,d,e}&=\bdf_{u,d,e}[A_{\{u,d,e\}0}\dblone+W_{u,d,e}\bdf^\dagger_{u,d,e}\bdf_{u,d,e}+X_{u,d,e}(\bdf^\dagger_{u,d,e}\bdf_{u,d,e})^2]+\mathbf{Z}_{u,d,e}\;,
\label{eq:GUTboundtri}
\end{align}
\end{subequations}
where ${\bf f}_{u,d,e}$ are the superpotential Yukawa coupling matrices
{\it in an arbitrary current basis} at the same scale at which the SSB
parameters of the model are specified. Here, $c_{U,D,E} =0$ or 1 is
introduced only to allow the facility to ``switch off'' the universal
term if desired. The matrices $\mathbf{T}_{Q,L,U,D,E}$ and
$\mathbf{Z}_{u,d,e}$ have been introduced only to allow for additional
sources of flavour-violation not contained in the Yukawa
couplings. Setting $\mathbf{T}_{Q,L,U,D,E}=\mathbf{Z}_{u,d,e}=\bm{0}$
gives us the most generation parametrization of the three-generation
$R$-parity conserving MSSM where the superpotential Yukawa interactions
are the sole source of flavour violation. In other words, any top-down
theory of flavour, will specify the form of the matrices
$\mathbf{T}_{Q,L,U,D,E}$ and $\mathbf{Z}_{u,d,e}$ along with the
coefficients $m^2_{\bullet0}$, $A_{\bullet 0}$, $c_{\bullet}$, 
$R_{\bullet}$, $S_{\bullet}$, $W_{\bullet}$ and $X_{\bullet}$ that
appear above.

The physics behind our anstaz (\ref{eq:GUTbound}) for SSB parameters,
with $\mathbf{T}_\bullet=\mathbf{Z}_\bullet=\bm{0}$, is that we assume
that the SUSY breaking mechanism, for reasons that are not understood
today, does not introduce a new source of flavour
violation.\footnote{For a complementary approach, where supersymmetry
breaking is the origin of fermion flavour, see Ref.~\cite{javier2}, and
the extensive bibliography therein.} We stress that this differs from
the minimal flavour violation \cite{mfv} assumption, where the new physics
may introduce an additional flavour violation in a controlled way. Our
ans\"atz is thus a special case of the minimal flavour violation
scenario. Of course, the form of (\ref{eq:GUTbound}) is not invariant
under renormalization group evolution, and at the weak scale we will end
up with the same form for the SSB parameters as the minimal flavour
violation ansatz (suitably generalized from the first paper of
Ref.~\cite{mfv} to include higher order terms in the Yukawa couplings),
but with appropriate relations between the coefficients of the various
terms in the minimal flavour violation formulae for SSB parameters. We
will return to the difference between minimal flavour violation and
(\ref{eq:GUTbound}) in Sec.~\ref{sec:stop}.

We have created a code (to be incorporated into ISAJET) that allows the
reader to choose the arbitrary values for the various coefficients as
well as
the matrices $\mathbf{T}_{Q,L,U,D,E}$ and $\mathbf{Z}_{u,d,e}$ that
appear in
 (\ref{eq:GUTbound}) as  boundary conditions for the RGEs for the SSB
parameters of the MSSM. This code provides a tool for a study of
(s)quark flavour
violation in sparticle processes in a arbitrary theory that reduces to
the MSSM at the scale where the SSB parameters are specified. We can
evolve the gauge and Yukawa couplings from the weak scale to this high scale, 
use the boundary conditions (\ref{eq:GUTbound}) as inputs for the SSB
parameters, and proceed to determine the iterative solution to the RGEs
as described above.
The familiar universal mSUGRA boundary conditions
are reproduced by setting $c_{U,D,E}=1$;
$m^{2}_{\{Q,L\}0}=m^{2}_{\{U,D,E\}0}=m^{2}_{0}$; $A_{\{u,d,e\}0}=A_{0}$;
$R_{U,D,E}=S_{U,D,E}=W_{u,d,e}=X_{u,d,e}=0$;
$\mathbf{T}_{Q,L}=\mathbf{T}_{U,D,E}=\mathbf{Z}_{u,d,e}=\bm{0}$ in
(\ref{eq:GUTbound}). 

\subsection{Quark Yukawa Couplings}\label{sec:quarkyuk}

\begin{figure}\begin{centering}
\includegraphics[width=.7\textwidth]{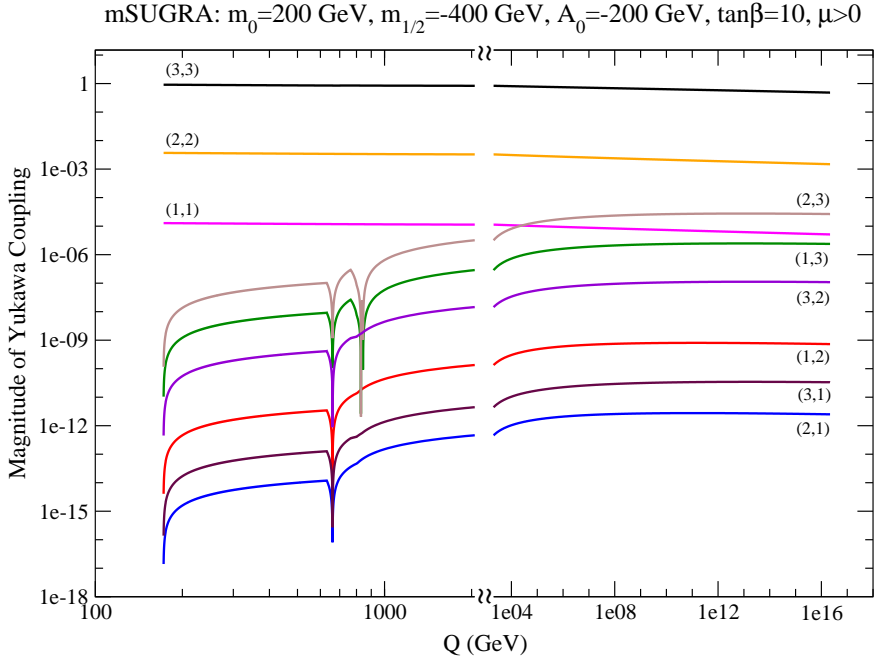}
\caption{Evolution of the magnitudes of the complex elements of the
up-quark Yukawa coupling matrix of the mSUGRA model with $m_0=200$~GeV,
$m_{1/2}=-400$~GeV, $A_0=-200$~GeV, $\tan\beta=10$ and $\mu >0$ in the
basis where this matrix is diagonal at $Q=m_t$.  For
$Q>m_{H}$ $(\simeq 631~\mathrm{GeV})$ we plot $\left|(\bdf_{u})_{ij}\right|$
whereas for $Q<m_{H}$, where the effective theory includes just one
scalar Higgs doublet, we plot
$\left|(\bm{\lambda}_{u})_{ij}\right|/\sin{\beta}$ which is equal to
$\left|(\bdf_{u})_{ij}\right|$ at $Q=m_{H}$. In all the figures we
take $m_{t}=172$~GeV.}
\label{fig:SUGRAfu}
\end{centering}
\end{figure}

We begin the discussion of our numerical results by showing the
magnitude of the complex elements of the up quark Yukawa coupling matrix
${\bf f}_u$ above $Q=m_H$, and $\boldl_u/\sin\beta$ below $Q=m_{H}$, for the mSUGRA model
with $m_0=200$~GeV, $m_{1/2}=-400$~GeV, $A_0=-200$~GeV, $\tan\beta=10$
and $\mu >0$ in Fig.~\ref{fig:SUGRAfu}.\footnote{The reader may well
wonder why we choose the gaugino mass to be negative and of the opposite
sign to that of the $\mu$ parameter, a relative sign that is ``apparently
disfavoured'' by the Brookhaven $g_\mu-2$ measurement. The reason is
that our convention for the sign of $\mu$ \cite{wss} (as well as the one
used in ISAJET) is opposite the one that is usually used. To compensate
for this, ISAJET internally (and unknown to the user) flips the sign of
the gaugino masses relative to what the user uses as an input. Since the
physics depends only on the relative sign between the gaugino masses and
$\mu$, this sign flip is equivalent to a flipped sign for $\mu$
(together with sign flips for $A_0$ and $b$), making it appear that the
ISAJET output is in accord with the usual convention for the sign of
$\mu$. We should also mention that here and in Ref.~\cite{wss}, the
convention for the sign of the ${\bf a}$ is opposite that to the one in
Ref.~\cite{martv} and of ISAJET. To reproduce the results with the sign
conventions for mSUGRA parameters that we use in this paper (and in
Ref.~\cite{wss}) using ISAJET, we should merely switch the sign of the
gaugino masses in the ISAJET inputs. Our results (except for
inter-generation mixing effects) can thus be obtained by running ISAJET
with $m_0=200$~GeV, $m_{1/2}=400$~GeV, $A_0=-200$~GeV $\tan\beta=10$ and
$\mu>0$.} The approximate spectrum for this illustrative scenario is
shown in Table~\ref{tab:SUGRAthresh}. Fig.~\ref{fig:SUGRAfu} should be 
compared with Fig.~1 of Paper~I, where we had shown the evolution of the
same Yukawa coupling, but in a
simplified scenario where the SUSY thresholds were clustered in two
regions, one at $600~\GeV$ and the other at $2~\mathrm{TeV}$. 
\begin{table}
\begin{tabular*}{.7\textwidth}{@{\extracolsep{\fill}}ll}
$M_{\rm SUSY}$ & 703 GeV\\
Higginos ($\mu$)&538 GeV\\
Gluinos ($m_{\tilde{g}}$)&941 GeV\\
$\mathcal{H}$, $H^{\pm}$ ($m_{H}$)&631 GeV\\
Bino ($|M_{1}|$)&166 GeV\\
Winos ($|M_{2}|$)&315 GeV\\
$(\tilde{u}_{L},\tilde{d}_{L}),\;(\tilde{c}_{L},\tilde{s}_{L}),\;(\tilde{t}_{L},\tilde{b}_{L})$&837
GeV, 837 GeV, 763 GeV\\
$\tilde{u}_{R},\;\tilde{c}_{R},\;\tilde{t}_{R}$&809 GeV, 809 GeV, 645
GeV\\
$\tilde{d}_{R},\;\tilde{s}_{R},\;\tilde{b}_{R}$&806 GeV, 806 GeV,
801 GeV\\
$(\tilde{\nu}_{eL},\tilde{e}_{L}),\;(\tilde{\nu}_{\mu
L},\tilde{\mu}_{L}),\;(\tilde{\nu}_{\tau L},\tilde{\tau}_{L})$&331 GeV,
331 GeV, 329 GeV\\
$\tilde{e}_{R},\;\tilde{\mu}_{R},\;\tilde{\tau}_{R}$&249 GeV, 249 GeV,
245 GeV\\
\end{tabular*}
\caption{The location of the thresholds for our canonical mSUGRA case
in Fig.~\ref{fig:SUGRAfu} and in several subsequent figures.}
\label{tab:SUGRAthresh}
\end{table}

The most striking feature of the figure are the dips in the off-diagonal
elements of the ${\bf f}_u$ matrix. The aligned dips at $Q\sim 650$~GeV,
common to {\it all} the off-diagonal elements, occur because of the
change in the sign of the coefficient of the ${\bf f}_d{\bf
f}_d^\dagger$-type terms that drive the growth of these off-diagonal
elements from zero at $Q=m_t$. This has been discussed in detail in
Paper~I, where we have also explained why the magnitudes of all the
off-diagonal Yukawa couplings vanish at essentially a common value of
$Q$.  The presence of the second zero at the higher value of $Q$, just
in the (2,3) and (1,3) elements, is accidental. It occurs because of
conspiracies between terms in the corresponding $\beta$-function as the
left-type squarks are decoupled. Notice that the lowest four curves,
though they do not have this additional dip, show kinks at these same
values of $Q$, corresponding to the decoupling of these squarks. For
several other mSUGRA cases, we have checked that while squark decoupling
causes kinks in the curves, the coupling does not drop to even close to
zero for a second time, in contrast to the behaviour in our illustrative
example in the figure.

\begin{figure}
\begin{centering}
\includegraphics[width=.7\textwidth]{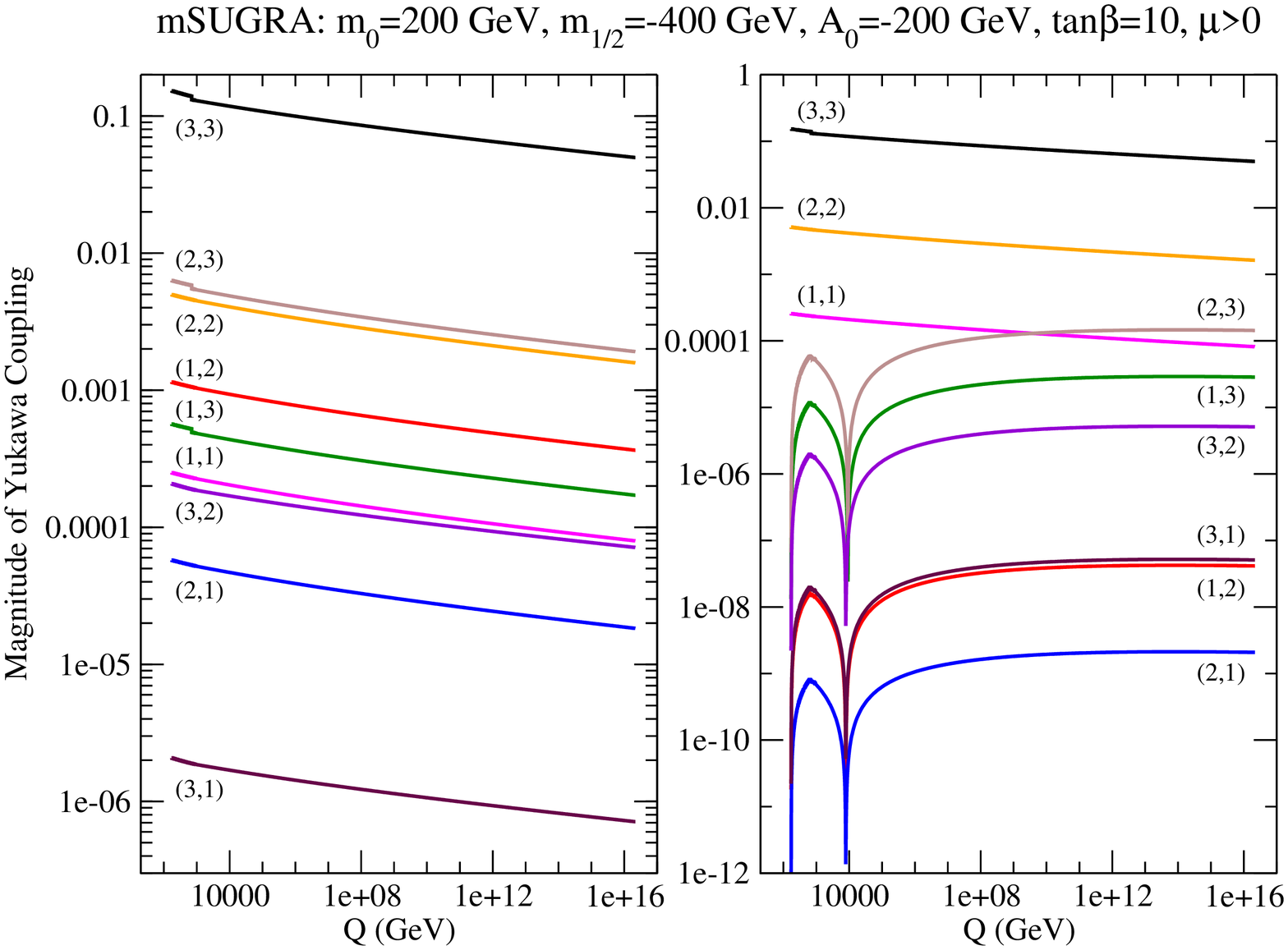} \caption{ Evolution
of the magnitudes of the complex elements of the down-quark Yukawa
coupling matrix of the mSUGRA model with $m_0=200$~GeV,
$m_{1/2}=-400$~GeV, $A_0=-200$~GeV, $\tan\beta=10$ and $\mu >0$.  In the
left-frame, we show the elements of the matrix in the same basis as in
Fig.~\ref{fig:SUGRAfu}, where $\bm{\lambda}_u$ is diagonal at $Q=m_t$,
whereas in the right frame, we show these elements in the basis where
the $\bm{\lambda}_d$ is diagonal at $Q=m_t$. For $Q>m_{H}$ $(\simeq 631~
\mathrm{GeV})$ we plot $\left|(\bdf_{d})_{ij}\right|$ whereas for
$Q<m_{H}$, where the effective theory includes just one scalar Higgs
doublet, we plot $\left|(\bm{\lambda}_{d})_{ij}\right|/\cos{\beta}$
which is equal to $|\left(\bdf_{d}\right)_{ij}|$ at $Q=m_{H}$.}
\label{fig:SUGRAfd} \end{centering} \end{figure}

The evolution of the down-quark Yukawa coupling matrix for the same 
mSUGRA point is shown in Fig.~\ref{fig:SUGRAfd}.  
We have shown these
couplings both in what we will subsequently refer to as our ``standard''
current basis where the up-type Yukawa couplings are diagonal at
$Q=m_t$ in the left frame, and in the basis where the down-type Yukawa
coupling matrix is diagonal at $Q=m_t$ in the right frame.  The matrices
in the two bases are connected by the KM matrix, in an analogous way to
(\ref{eq:tranuda}). The curves in the left frame are all smooth (except
for the small kink in the curves for $|\left(\bdf_{d}\right)_{i3}|$ that
occurs because of the SUSY correction to the bottom quark Yukawa
coupling \cite{pierce}), and do not show the dip to zero that appeared
in the previous figure. This is not surprising because underlying the
explanation of this dip was the fact that the off-diagonal elements
evolved from zero at $Q=m_t$ \cite{rge1}.  Notice also that in this
frame the off-diagonal elements are not necessarily smaller than the diagonal
elements even for $Q\alt 1$~TeV. The magnitudes of the off-diagonal
matrix elements in the frame on the right, which do start at zero at
$Q=m_t$ show the anticipated aligned dips, except that the location of
the dip is shifted considerably to the right, relative to
Fig.~\ref{fig:SUGRAfu}. This shift is not difficult to understand. The
large top quark Yukawa coupling in the SM governs the evolution of the
off-diagonal elements of the down-type Yukawa couplings, causing them to
evolve much more rapidly from zero in the right-hand frame of 
Fig.~\ref{fig:SUGRAfd}, so that
in order to evolve back to zero after the sign flip in the
$\beta$-function due to the additional Higgs and SUSY particles, a
longer evolution distance is needed in the present
case. Furthermore, beyond the Higgs boson threshold the off-diagonal
elements of $\bdf_{u}$ in Fig.~\ref{fig:SUGRAfu} are accelerated to
zero on account of the fact that the down-type Yukawa couplings
$\bdf_{d}$, that enter in the evolution of these elements, are enhanced
by a factor $\sim1/\cos{\beta}$, pushing the dip in this figure to a low
value of $Q$.

We now turn to briefly discuss what happens when we allow
non-universality of GUT scale squark mass parameters, so as to split the
squarks more than in mSUGRA. Recall that if the squarks all decouple
together, all that happens is a change in the slope of the
$\beta$-function. If, however, the squark masses are not all the same,
our decoupling procedure entails an additional rotation to the squark
mass basis. If this basis differs significantly from our ``standard'' basis
in which the quark Yukawa coupling matrix is diagonal (at $Q=m_t$), one
may expect considerable deviation in the evolution of the off-diagonal
elements from the mSUGRA case for $Q$ values in between the highest and
lowest squark thresholds. The introduction of non-universal squark mass
parameters via non-vanishing values of $R_\bullet$ and $S_\bullet$ in
(\ref{eq:GUTbound}) never leads to significant effects because the
rotation from the ``standard'' basis to the squark mass basis is small by
construction. 

The question then is whether we can have large deviations
from Fig.~\ref{fig:SUGRAfu} and~\ref{fig:SUGRAfd} via the ${\bf T}_{\bullet}$ (or
${\bf Z}_{\bullet}$) matrices. To examine this, 
we set all GUT scale inputs to be the same as the mSUGRA case in
Fig.~\ref{fig:SUGRAfu} except  that we now take, 
\begin{subequations}\label{eq:T2bc}\begin{gather} m^{2}_{\{U,D\}0}=0\;,
    \ \ \ \ \ {\rm and} \\
\mathbf{T}_{U,D}=\mathit{diag}\left\{10000,40000,90000\right\} {\rm
  GeV}^2\;.
\end{gather}\end{subequations}
We need, of course, to specify the basis in which the squark mass matrix
is diagonal. If this is the ``standard'' current basis, we have checked that
although the right squark masses are now significantly split relative to
the mSUGRA case, except for detailed changes in the evolution for $Q$
values in between the squark thresholds ({\it e.g.} the dip at the
higher value of $Q$ in the upper curves in Fig.~\ref{fig:SUGRAfu}
develops more structure, and there are jumps in the (1,2) and (2,1)
elements), the evolution is not altered in any important way at large
values of $Q$. This is because the rotation between the ``standard'' basis
and the squark mass basis is again small.

If we take instead the right-squark matrices in (\ref{eq:T2bc})
to be diagonal in a completely different basis
specified by unitary matrices ${\bf V}_L(u)$, ${\bf V}_R(u)$ and ${\bf
  V}_R(d)$ that take us to our ``standard'' basis, and specified to be of the form,
\begin{equation}\label{eq:rotmat}
\mathbf{V}=\left(\begin{array}{ccc}c_{\alpha}c_{\beta}&s_{\alpha}c_{\beta}e^{-i\delta_{\alpha}}&s_{\beta}e^{-i\delta_{\beta}}\\
-s_{\alpha}c_{\gamma}e^{i\delta_{\alpha}}-c_{\alpha}s_{\beta}s_{\gamma}e^{i\left(\delta_{\beta}-\delta_{\gamma}\right)}&c_{\alpha}c_{\gamma}-s_{\alpha}s_{\beta}s_{\gamma}e^{i\left(-\delta_{\alpha}+\delta_{\beta}-\delta_{\gamma}\right)}&c_{\beta}s_{\gamma}e^{-i\delta_{\gamma}}\\
s_{\alpha}s_{\gamma}e^{i\left(\delta_{\alpha}+\delta_{\gamma}\right)}-c_{\alpha}s_{\beta}c_{\gamma}e^{i\delta_{\beta}}&-c_{\alpha}s_{\gamma}e^{i\delta_{\gamma}}-s_{\alpha}s_{\beta}c_{\gamma}e^{i\left(-\delta_{\alpha}+\delta_{\beta}\right)}&c_{\beta}c_{\gamma}\end{array}\right)\;,
\end{equation}
(where $s_{\alpha}=\sin{\alpha}$, $c_{\alpha}=\cos{\alpha}$, {\it etc.})
with randomly chosen values of $\alpha$, $\beta$, $\gamma$ and
$\delta_{\beta}$ (we take $\delta_{\alpha}=\delta_{\gamma}=0$ for
simplicity), we may expect the evolution of Yukawa couplings to depart
from the corresponding evolution in mSUGRA.\footnote{The matrices ${\bf
V}_{L,R}(u)$ should be numerically unitary to high accuracy. Otherwise,
numerical errors from inverting the up Yukawa couplings from this new
current basis to our ``standard'' basis will leave residual off-diagonal
elements rather than zero even at $Q=m_t$. If the size of these elements
is comparable to the values of the smallest off-diagonal elements at
values of $Q$ substantially away from $m_t$, it is clear that our
solutions will be dominated by the error from the non-unitarity of the
${\bf V}_{L,R}(u)$ matrices. A similar consideration applies to ${\bf
V}_R(d)$. } We emphasize that this scenario (which is only of
pedagogical interest) where ${\bf m}_{U,D}^2$ are diagonal in the basis
where the corresponding Yukawa coupling matrices have large off-diagonal
elements includes potentially very large flavour-violation in the
singlet squark SSB mass matrices, and is likely excluded.

\begin{figure}\begin{centering}
\includegraphics[width=.7\textwidth]{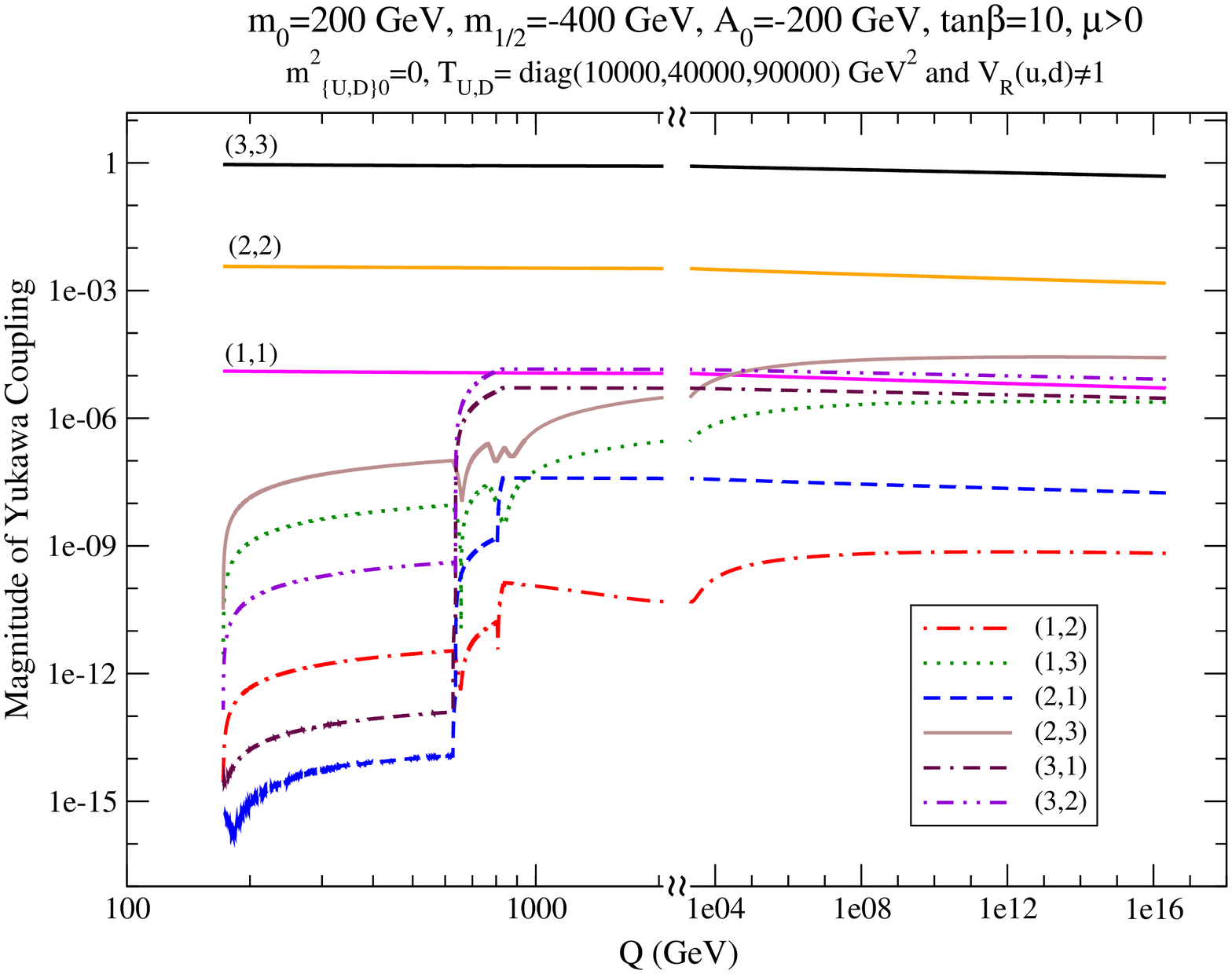}
\caption{The same as Fig.~\ref{fig:SUGRAfu} except that the GUT scale
  right-squark mass matrices are now given by (\ref{eq:T2bc}) in
  the basis specified by (\ref{eq:genrot}) of the text. Note, however,
  that these matrix
  elements are shown in the ``standard'' current basis where the matrix is
  diagonal at $Q=m_t$.}
\label{fig:T2fu}
\end{centering}
\end{figure}

In Fig.~\ref{fig:T2fu} we illustrate the evolution of the magnitudes of
the elements of the up-quark Yukawa coupling matrix with,
\begin{subequations}\label{eq:genrot}\begin{align}
\mathbf{V}_{L}(u):&\;\alpha=2.053,\;\beta=0.254,\;\gamma=2.030,\;\delta_{\beta}=0.4829\;,\\
\mathbf{V}_{R}(u):&\;\alpha=1.188,\;\beta=2.218,\;\gamma=0.763,\;\delta_{\beta}=0.87\;,\\
\mathbf{V}_{R}(d):&\;\alpha=1.904,\;\beta=2.947,\;\gamma=1.847,\;\delta_{\beta}=1.14\;.
\end{align}\end{subequations}
Note that,
just as in Fig.~\ref{fig:SUGRAfu}, we have plotted these elements in our
``standard'' basis where the Yukawa coupling matrix is diagonal at
$Q=m_t$. The GUT scale matrices ${\bf m}_{U}^2$ and ${\bf m}_D^2$ will
be random-looking Hermitian matrices in our ``standard'' basis. 
%
We see that the evolution of the diagonal elements is not significantly
altered from mSUGRA. This is because although there is a large mismatch
between the squark mass basis and our ``standard'' basis, this mismatch is
operative over the small range of $Q$ between the highest and lowest 
thresholds and so has little impact on the largest elements. 
The two largest  off-diagonal entries (i.e. $(\bdf_{u})_{23}$
and $(\bdf_{u})_{13}$) similarly do not change significantly from
Fig.~\ref{fig:SUGRAfu}, but all other entries are greatly altered.
While it may seem that the values of these matrix elements at a large
scale is quite irrelevant phenomenologically, the altered form of the
Yukawa coupling matrix at the high scale could be of relevance to
model-builders.

We have seen that the magnitudes of the off-diagonal elements in
Fig.~\ref{fig:T2fu} remain small because the splitting in the squark
spectrum is limited to ${\cal O}(100-1000)$~GeV. The natural question
then is whether we can get these to be larger by choosing extreme
intra-generational squark splitting. Even putting aside potentially
unacceptable flavour-changing effects that might result, this is not
easy. In general, such a GUT scale splitting also has a large value for
$S=m_{H_u}^2-m_{H_d}^2 +Tr\left[{\bf m}^2_Q-{\bf m}^2_L-2{\bf
m}^2_U+{\bf m}^2_D+{\bf m}^2_E\right]$, which pulls the other squarks
also to large masses, so the squark mass splitting is reduced by RGE
effects. It may be possible to obtain split squarks by adjusting $S$ to
be zero ($S$ is then invariant under renormalization group evolution), but we
have not investigated this here.

\subsection{Gaugino Mass Parameters in Split SUSY Models }

We have seen in Sec.~\ref{subsec:GHP} that the evolution of the
electroweak gaugino mass parameters acquires a dependence on the
$\mu$-parameter and {\it vice-versa}, once threshold effects from
splitting in the Higgs boson sector are included. In models where $m_H
\gg |\mu|, |M_{1,2}|, |M_{1,2}^{\prime}|$, the effect of the term explicitly
dependent on $\mu$ in (\ref{app:rgem1})-(\ref{app:rgem2p}) (and the
corresponding terms dependent on the gaugino mass parameters in the RGE
for $\mu$) may be significant, so that the relation
$M_2/M_1=\alpha_2/\alpha_1$ expected in many models is modified.  We
should keep in mind that two loop terms will, in general, also alter
this relation. Our point is that we should expect threshold corrections,
from the $\mu$ term in the RGE, as well as from the decoupling of
sfermions, to be comparable to (or even larger than) the two loop
modifications, and so need to be included in a quantitative analysis.
Within the mSUGRA context, we have small values of $|\mu|$, and hence,
$m_H \gg |\mu|$ in the hyperbolic branch/focus point (HB/FP) region
\cite{focus} which occurs for large values of $m_0$, and is one of the
regions selected out \cite{relic} by the relic-density measurement
\cite{wmap}. We mention here that {\em the location of this HB/FP region
is significantly altered by the inclusion of the threshold corrections.}

These considerations led us to examine the evolution of the gauge
couplings and the electroweak gaugino parameters for a
relic-density-consistent mSUGRA model point in the HB/FP region
with $m_0=3075$~GeV, $m_{1/2}=-600$~GeV, $A_0=0$, $\tan\beta=10$ and
$\mu>0$, for which ($\mu, M_1, M_2) \simeq (306, -258, -496)$~GeV at the
weak scale.\footnote{Without threshold corrections, ISAJET gives a
similar spectrum for $m_0\simeq 3660$~GeV.} We found that at the two
loop level with all threshold effects included, $M_2/M_1=1.919$ to be
compared to $M_2/M_1=1.881$ obtained without including threshold
effects. Thus, although threshold effects actually bring us closer to
${\alpha_2/\alpha_1}=1.964$, their inclusion is clearly necessary for
a quantitative analysis of mass parameters that may be extracted at an
$e^+e^-$ linear collider, where a precision of better than 1\% will be
possible if charginos are kinematically accessible.

The threshold corrections to gaugino mass parameters can be much larger
in the so-called split SUSY model \cite{split} that has received
considerable attention in the recent literature. In this scenario, 
 the naturalness of the scalar Higgs sector
(which we view as one of the primary motivations for weak scale SUSY) is
abandoned, while  gauge coupling unification and the neutralino dark matter
candidate of $R$-parity violating models are preserved.
Gaugino mass parameters and $|\mu|$ are assumed to be at the weak scale,
while scalar mass parameters are set to be at an intermediate scale. This
means that sfermion masses as well as $m_H$ are very large (with the SM Higgs
doublet fine-tuned to be light), so that charginos and
neutralinos are the only new particles (other than a SM Higgs boson) at
the weak scale.

\begin{figure}\begin{centering}
\includegraphics[width=.7\textwidth]{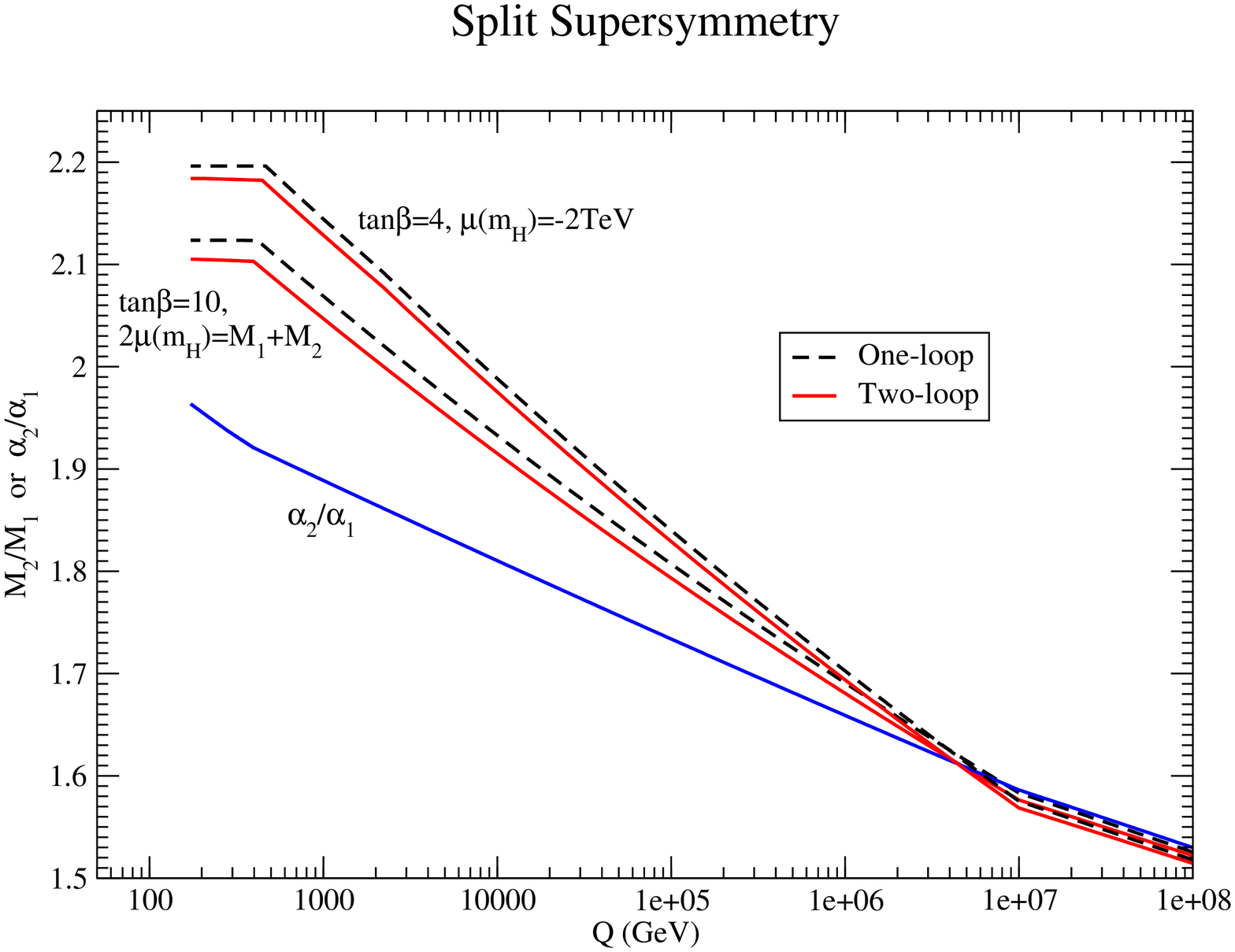}
\caption{The evolution of the gaugino mass ratio $M_2/M_1$ at the one
  (dashed) and two (solid) loop levels, along with the two-loop evolution of 
  $\alpha_{2}/\alpha_{1}$, for a split SUSY model where
  scalar masses are around $10^7$~GeV, with gauginos and higgsinos at
  the weak scale. Gaugino mass unification is assumed and as everywhere
  else in this paper, the gaugino mass parameters are negative. The other
  parameters are as mentioned in the text.}
\label{fig:split}
\end{centering}
\end{figure}

As an illustration, in Fig.~\ref{fig:split} we plot the variation of the
ratio $M_{2}/M_{1}$ at the one-loop level (dashed), as well as at the
two-loop level (solid), with the renormalization scale $Q$, along with
the two-loop value of $\alpha_{2}/\alpha_{1}$. We show results, first
where the value of $|\mu|$ is set exactly in between $|M_1|$ and $|M_2|$
so that the lightest neutralino acquires a significant higgsino
component, to qualitatively mimic mixed higgsino dark
matter.\footnote{In the absence of a real theory of split SUSY, we
should view this figure only as a qualitative illustration of
potentially large threshold effects. Here, we take the sfermion mass
parameters to be $10^7$~GeV at $Q=M_{\rm GUT}$, $m_{1/2}=-350$~GeV and
$A_0=0$. Since it is not possible to satisfy the EWSB conditions except
when $\tan\beta$ is hierarchically large --- this would cause down-type
Yukawa couplings to become non-perturbative --- we treat $\mu$ and
$\tan\beta$ as phenomenological parameters, and fix $m_H$ to be
$10^7$~GeV in this figure. The parameters $m_{H_u}^2$ and $m_{H_d}^2$
(indeed all scalar mass parameters) are never needed since the RGEs for
gaugino masses, $\mu$ and the $\ba$-parameters, and the dimensionless
couplings, form a closed set even at the two-loop level. Sfermion masses
only enter via the location of thresholds.} We have checked that the
$M_2/M_1$ values do not change in a significant way for yet larger
values of $\tan\beta$.  It is clear that the relation
$\frac{M_2}{M_1}=\frac{\alpha_2}{\alpha_1}$ is violated at the several
percent level by the threshold corrections, without which the $M_2/M_1$
lines would have continued with the same slope that they have above
$10^7$~GeV to low values.  For this point, the $2sc\times \mu$ term that
explicitly appears in the RGE is very small so that the result is
independent of the sign of $\mu$: most of the difference is an effect of
the sfermion loop contributions being switched off below $10^7$~GeV. To
gain some idea of how large the effect of this $\mu$ term might be, we
have also shown $M_2/M_1$ for $\mu=-2$~TeV, with $sign(\mu M_2)>0$ and
$\tan\beta=4$.  Since such a large value of $\mu$ would be totally
incompatible with the measured relic density and small values of
$\tan\beta$ are unnatural in these models without some modification to
the EWSB sector, the reader should view these curves only as a guide to
how much the gaugino mass ratio may deviate from its ``unification
value''. The difference between the two cases is almost entirely due to the
different choice of $\mu$.
We see that the gaugino mass unification condition will, in
this case, be violated by $\sim 10$\%. If instead we choose the opposite
sign for $M_2\mu$, but keep $|\mu|=2$~TeV, this
``large $|\mu|$ line'' would be {\it lower} than the corresponding
line with $\mu = (M_1+M_2)/2$ by about the same amount that it is
higher than this line  in the figure.  Clearly, increasing the splitting
between the scalar and the gaugino/higgsino sector of the theory will cause even
further violation of the unification condition, and $\sim 20$\% effects
appear to be plausible if the scalars are instead at the $10^{11}$~GeV
scale.

\subsection{Trilinear Couplings}

\begin{figure}\begin{centering}
\includegraphics[width=.7\textwidth]{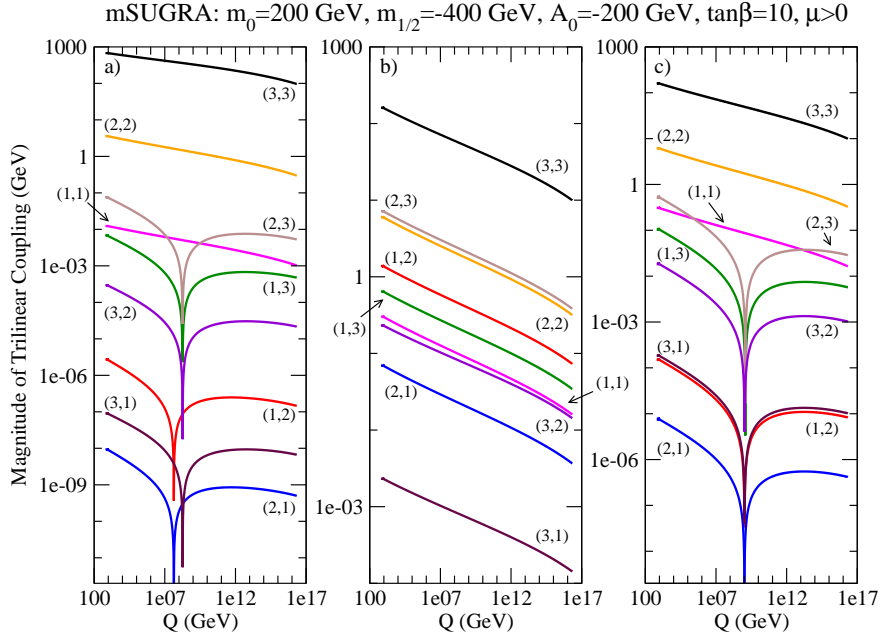}
\caption{The magnitude of the elements of the trilinear coupling matrix
  ${\bf a}_\bullet$ for the mSUGRA model in Fig.~\ref{fig:SUGRAfu}. We
  show ({\it a})~$|\left({\bf a}_u\right)_{ij}|$, ({\it b})~$|\left({\bf
  a}_d\right)_{ij}|$, in the basis where up-quark
  Yukawa couplings are diagonal at $Q=m_t$, and ({\it c})~$|\left({\bf
  a}_d\right)_{ij}|$  in the basis where down-quark
  Yukawa couplings are diagonal at $Q=m_t$. The curves extend between
  $Q=m_H$ and $Q=M_{\rm GUT}$.}
\label{fig:SUGRAtri}
\end{centering}
\end{figure}

We begin our discussion of the trilinear couplings by returning to the
mSUGRA case considered in Figs.~\ref{fig:SUGRAfu}
and~\ref{fig:SUGRAfd}. The magnitudes of the individual entries of the
trilinear coupling matrices ${\bf a}_u$ and ${\bf a}_d$ are shown in
Fig.~\ref{fig:SUGRAtri}, where we plot ({\it a})~$|\left({\bf
a}_u\right)_{ij}|$ and ({\it b})~$|\left({\bf a}_d\right)_{ij}|$ in our
``standard'' current basis where the up-type quark Yukawa coupling matrix is
diagonal at $Q=m_t$, along with $|\left({\bf a}_d\right)_{ij}|$ in the
basis where the down-type quark Yukawa coupling matrix is diagonal at
$Q=m_t$ in frame ({\it c}). We terminate the curves at $Q=m_H$ since
below this scale we have a single Higgs scalar doublet model, and the
trilinear couplings ${\bf a}_{u,d}$ evolve only as part of a linear
combination with $\tilde{\mu}^{*}\bdf^{h_{u,d}}_{u,d}$ as discussed in Sec.~\ref{subsec:tri}.
We see that the curves in frame ({\it a}) show a simple dip structure
indicating that the real and imaginary parts of ${\bf a}_{u,d}$ are
really monotonic functions of $Q$ that pass through zero together, in a
manner similar to the elements of the Yukawa coupling matrix
\cite{rge1}. The actual location of the zero is somewhat harder to
analyse because even though ${\ba_{u}}$ obtains off-diagonal components
only because the down-quark Yukawa matrix is not diagonal at $Q=m_t$,
the matrix ${\bf a}_u$ is off-diagonal even at the GUT scale. The
off-diagonal elements of ${\bf a}_d$ in frame ({\it b}) start off with a
much larger magnitude in the ``standard'' Yukawa basis at $Q=M_{\rm
GUT}$ because the corresponding Yukawa coupling matrix has large
off-diagonal pieces. In this case, the evolution of these off-diagonal
elements receives significant contributions from {\it all} entries in
the RGE (unlike the evolution of the off-diagonal elements in frame
({\it a}) or of the off-diagonal Yukawa couplings discussed at length in
Paper~I where contributions from the off-diagonal down-type Yukawa
matrices govern the evolution), and never go through zero; the situation
is similar to that in the first frame of Fig.~\ref{fig:SUGRAfd}. We see
from frame ({\it c}) that the magnitudes of $\left({\bf
a}_d\right)_{ij}$, in the basis that the down-type Yukawa coupling
matrix is diagonal at $Q=m_t$, again show the characteristic dip
structure indicating that the off-diagonal elements increase in
magnitude from their value at $Q=M_{\rm GUT}$ to some maximum magnitude
at an intermediate scale, but then smoothly reverse direction and
thereafter evolve monotonically through zero to the low scale. The
elements in frames ({\it b}) and ({\it c}) are, of course related by,
(\ref{eq:tranuda}). We note that because the off-diagonal elements in
frame ({\it c}), are driven by the larger (and significantly
off-diagonal) ``up-type'' Yukawa couplings, these are bigger than those in
frame ({\it a}) where it is the down-type Yukawa coupling matrix that
has the significant off-diagonal elements, and so largely determines the
entries.

%
%
%

%
%
%
%

\begin{figure}\begin{centering}
\includegraphics[width=.7\textwidth]{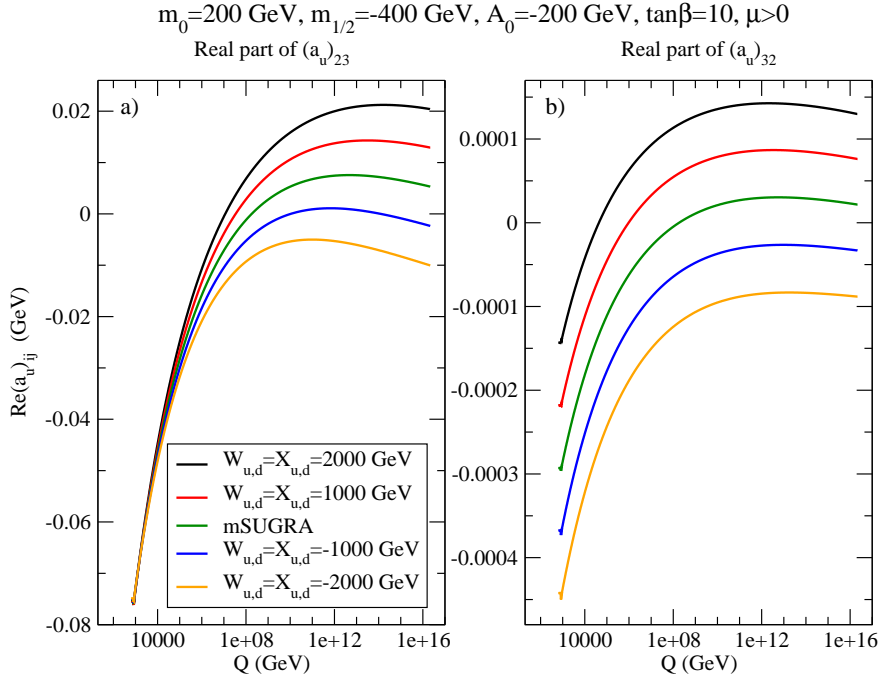}
\caption{The evolution of ({\it a})~Re$\left({\bf a}_u\right)_{23}$ and
({\it b})~Re$\left({\bf a}_u\right)_{32}$ for a model with the GUT scale
values of ${\bf a}$-parameters set as in (\ref{eq:GUTboundtri}) for
several values of $W$ and $X$ shown in the legend (with 
$A_{\{u,d\}0}=A_{0}$). The imaginary parts
of the matrix elements are about four orders of magnitide smaller. The
curves are in the same order as in the legend.  The GUT scale SSB scalar
and gaugino mass parameters are assumed to be universal, with the same
value as in Fig.~\ref{fig:SUGRAtri}. We terminate the curves at $Q=m_H$
at the lower end.}
\label{fig:au23wx}
\end{centering}
\end{figure}

Finally, in Fig.~\ref{fig:au23wx} we consider a model with non-universal
values of ${\bf a}$-parameters, but where these are not a new source of
flavour violation.  Specifically, we consider a model with the same
values for mSUGRA parameters as in Fig.~\ref{fig:SUGRAtri}, but with
non-zero values for  $W$ and $X$ in (\ref{eq:GUTboundtri}) 
(with $A_{\{u,d\}0}=A_{0}$), and illustrate
the evolution of ({\it a})~Re$\left({\bf a}_u\right)_{23}$ and 
({\it b})~Re$\left({\bf a}_u\right)_{32}$. We have checked that the
imaginary parts of these matrix elements are about four orders of
magnitude smaller.    
The striking feature of frame~({\it a}) is that the various curves which
 start with very different values of Re$\left({\bf a}_u\right)_{23}$ at
 $Q=M_{\rm GUT}$, appear to focus to a common value at the low scale. We
 have checked, however, that although they all cross at $Q\simeq
 1.5$~TeV, they do not all converge at precisely the same value of
 $Q$. This apparent convergence, which persists for other values of
 mSUGRA parameters, is sensitively dependent on the special GUT scale
 boundary conditions for ${\bf a}_u$ that we have used. We have checked
 that if instead we use a general matrix ${\bf Z}_u$ in
 (\ref{eq:GUTboundtri}), the corresponding evolution is completely
 different.  We do not have a good explanation for the seeming
 convergence in frame ({\it a}), and only note that it is not generic to
 {\it all} elements of ${\bf a}_u$ as evidenced, for example, by the
 corresponding evolution of Re$\left({\bf a}_u\right)_{32}$ in
 frame~({\it b}) of the figure.

\subsection{Soft Masses}

\begin{figure}\begin{centering}
\includegraphics[width=.7\textwidth]{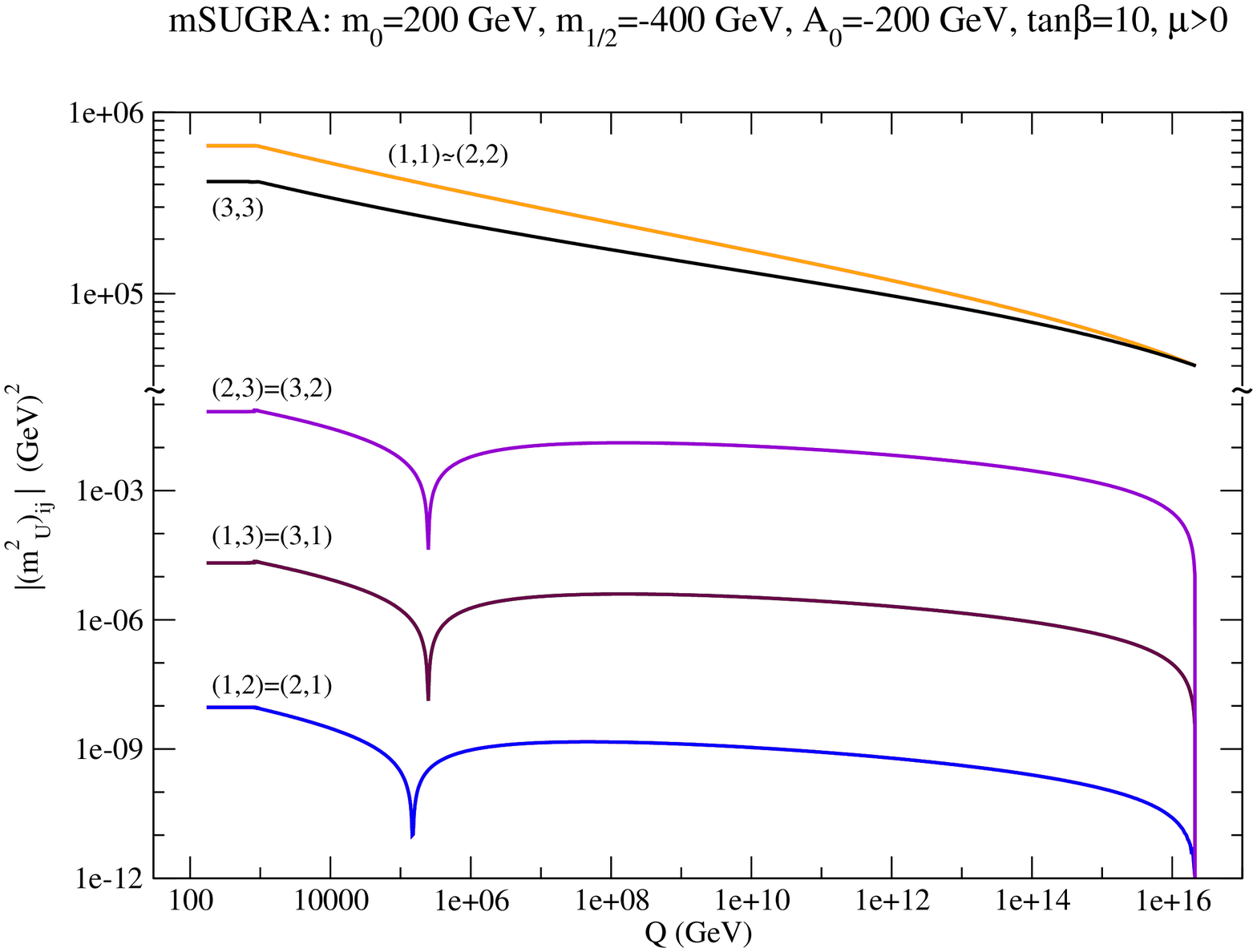}
\caption{The scale dependence of the magnitudes of the entries of the
  matrix ${\bf m}_U^2$ for the mSUGRA model with parameters as in
  Fig.~\ref{fig:SUGRAfu} in the basis where the up-type quark Yukawa coupling
  matrix is diagonal at $Q=m_t$. Note the we have broken the vertical
  scale at 0.1~GeV$^2$ and also used a different scale above this to
  better show the splitting of the (3,3) entry from the other diagonal
  entries.}
\label{fig:SUGRAmup}
\end{centering}
\end{figure}

We begin our discussion of the evolution of the scalar mass SSB
parameters by showing in Fig.~\ref{fig:SUGRAmup}
the evolution of the magnitudes of 
${\bf m}_U^2$ in our ``standard'' current basis for the mSUGRA model with
the same parameters as in Fig.~\ref{fig:SUGRAfu}. 
The diagonal matrix elements start from a common value $m_0$ and
increase as we go to the weak scale because of gauge (and gaugino)
interactions. The splitting between the (1,1) and (2,2) elements that
occurs because of the Yukawa couplings is too small to be visible in the
figure. The (3,3) element is, however, reduced significantly on account
of the large (3,3) entry in ${\bf f}_u$. Notice that the curves become
flat once the squarks are all decoupled. The magnitudes of the three
independent off-diagonal elements of the Hermitian matrix ${\bf m}_U^2$
start from zero at $Q=M_{\rm GUT}$, and rapidly rise because ${\bf f}_u$
has off-diagonal entries at $Q=M_{\rm GUT}$: the much more off-diagonal
${\bf f}_d$ matrix affects the evolution of ${\bf m}_U^2$ only at the
two-loop level. Note the break in the
vertical scale in the figure. The ordering of the magnitudes of
the off-diagonal elements of ${\bf m}_U^2$ can be simply gauged from the
up-type Yukawa coupling matrix. These off-diagonal elements  start from
zero at $Q=M_{\rm GUT}$, evolve to a maximum magnitude, then smoothly
reverse direction at an intermediate scale and then continue to evolve
monotonically all the way to the weak scale.  The dips in the figure
occur where the real, and simultaneously the imaginary, part of
$\left({\bf m}_U^2\right)_{ij}$ changes
sign during the course of its evolution to $Q=m_t$ where we terminate
the plot.

Before proceeding further we draw the reader's attention to a technical
point that is important for the numerical solution of the RGEs, once
squark decoupling is included as described in
Sec.~\ref{subsec:sqdec}. As we have already explained, in order to
correctly implement the decoupling procedure for values of $Q$ below the
highest sfermion threshold, at each step we need to rotate to the basis
where the SSB squark mass squared matrices are diagonal, which, in turn,
requires us to obtain the unitary matrix ${\bf R}$ that relates our
``standard'' basis to this ``squark mass basis''. After evaluating the
right-hand side of the RGEs, we always rotate back to our ``standard''
basis using ${\bf R}^\dagger$. While this is straightforward in
principle, the practical problem is that when two squark eigenvalues
become very close --- this is always the case in mSUGRA because the up
and charm squark mass parameters only evolve differently because of
effects of the small first and second generation quark Yukawa couplings
--- ${\bf R}{\bf R}^\dagger$ develops non-zero off-diagonal matrix
elements due to numerical noise at the 10$^{-10}$ level (The noise level
depends on the computer system.). This then ruins the delicate
cancellations that are necessary to obtain the tiny magnitude of
$\left({\bf m}_U^2\right)_{12}$ seen in the bottom curve of
Fig.~\ref{fig:SUGRAmup}. Our procedure for dealing with this is detailed
in Appendix~\ref{sec:orthfix}.  It is for essentially the same reason
that we had to use the manifestly unitary form (as in (\ref{eq:rotmat}))
for the matrices ${\bf V}_L(u)$, ${\bf V}_R(u)$ and ${\bf V}_R(d)$ when
we discussed the evolution of Yukawa couplings when the matrices ${\bf
m}^2_{U,D}$ were diagonal in a general current basis in
Fig.~\ref{fig:T2fu}; see the footnote just after Eq.~\eqref{eq:rotmat}.

\begin{figure}\begin{centering}
\includegraphics[width=.7\textwidth]{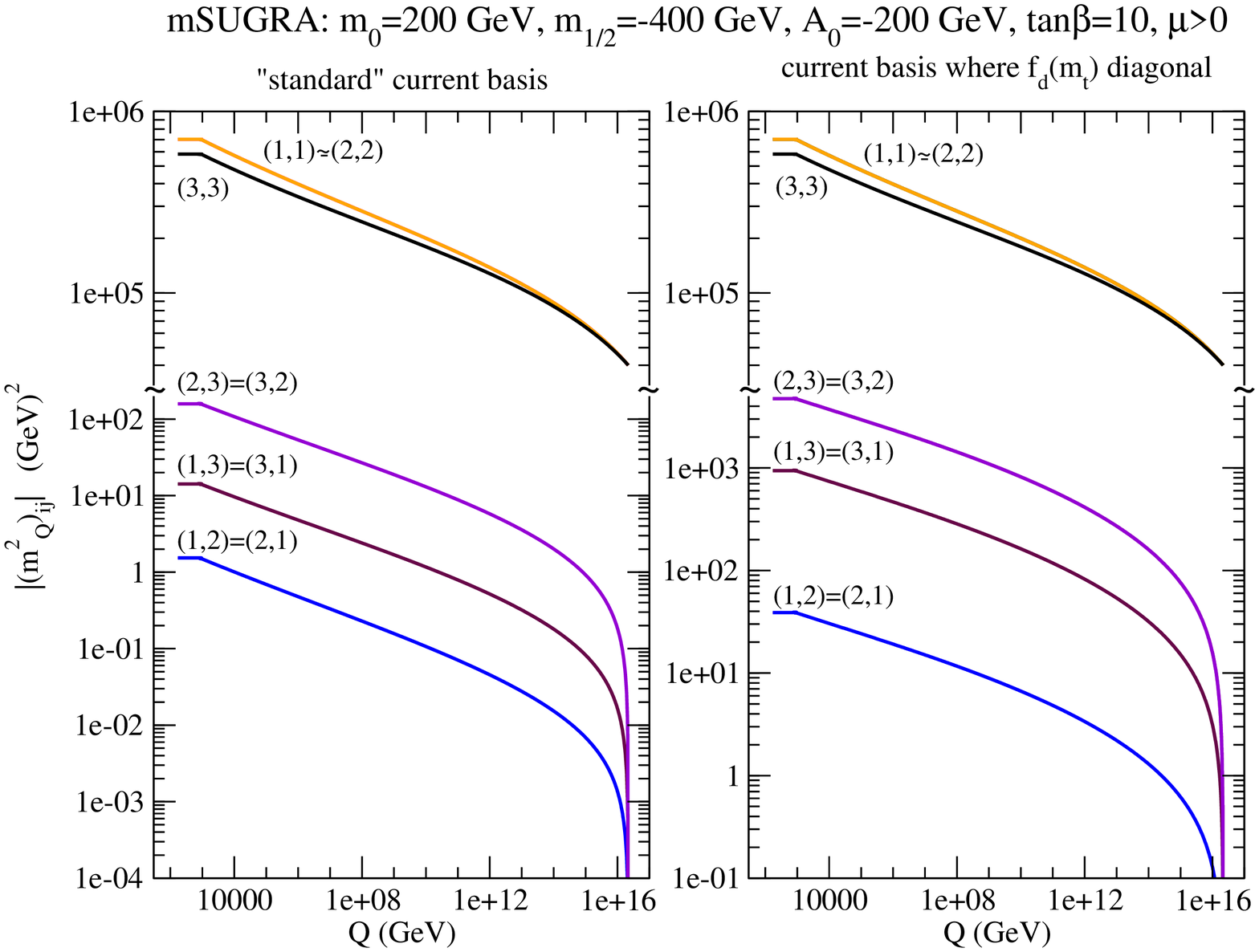}
\caption{The scale dependence of the magnitudes of the entries of the
  matrix ${\bf m}_Q^2$ for the mSUGRA model with parameters as in
  Fig.~\ref{fig:SUGRAfu} in the basis where the up-type (left frame), or
  the down-type (right frame),
  quark Yukawa coupling matrix is diagonal at $Q=m_t$. Note the we have broken
  the vertical scales to better display the matrix elements.}
\label{fig:SUGRAmq}
\end{centering}
\end{figure}

Fig.~\ref{fig:SUGRAmq} shows the evolution of $\bdm^{2}_{Q}$ for the
same mSUGRA point.
In the left frame, we show the magnitudes of the elements in our
``standard'' basis, where the up quark Yukawa coupling matrix is diagonal at
$Q=m_{t}$. The frame on the right shows the magnitudes of the elements
of this same matrix, but in the basis where the down-type quark Yukawa
coupling matrix is diagonal at $m_{t}$. We have checked that the
large difference in the size of the off-diagonal elements in the two
frames is indeed accounted for by the fact that the corresponding
matrices are related by (\ref{eq:tranudmQ}). Unlike in
Fig.~\ref{fig:SUGRAmup}, there is no dip in the magnitudes of the
off-diagonal elements because they evolve monotonically from zero at the GUT
scale, until the squarks are all decoupled. 

\begin{figure}\begin{centering}
\includegraphics[width=.7\textwidth]{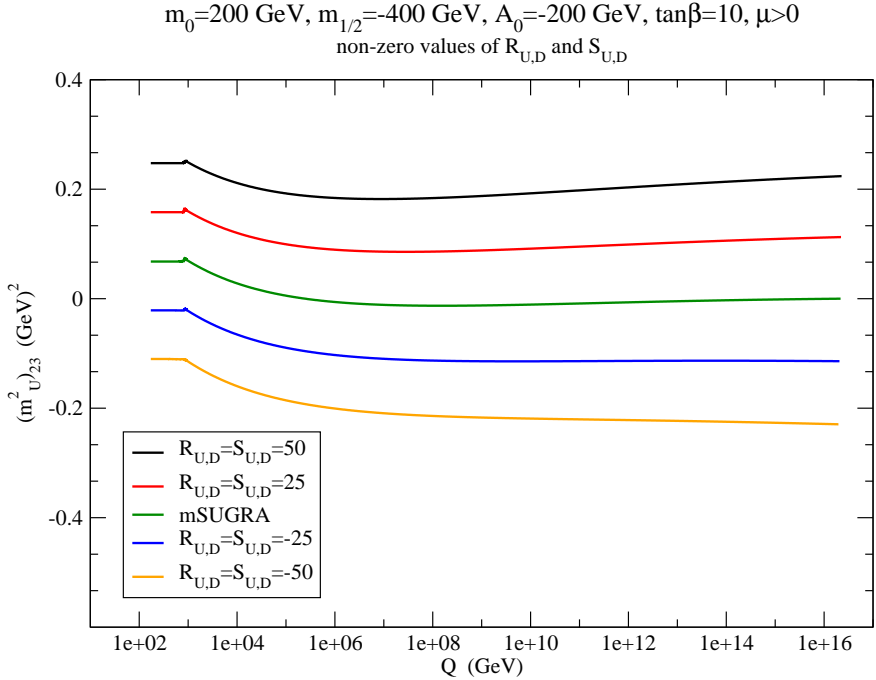}
\caption{The scale dependence of $(\bdm^{2}_{U})_{23}$ for several sets
of values of $R_{U,D}$ and $S_{U,D}$ that appear in
(\ref{eq:GUTboundsferm}) for the model with non-universal GUT scale SSB
squark mass parameters, but $c_{U,D}=1$. The curves are in the same order as the legend, 
and all the other parameters are set as in
Fig.~\ref{fig:SUGRAmup}. }
\label{fig:rsmu}
\end{centering}
\end{figure}

To understand how the non-universal boundary conditions 
in (\ref{eq:GUTboundsferm}) impact the evolution of the squark mass
matrices we examine a model with non-zero values of $R_{U,D}$ and
$S_{U,D}$, but with universal gaugino masses and ${\bf a}$-parameters. 
As an illustration we show in
 Fig.~\ref{fig:rsmu} the value of 
$(\bdm^{2}_{U})_{23}$ for the set of values of $R_{U,D}$ and $S_{U,D}$
shown in the figure, with all other parameters set as in
Fig.~\ref{fig:SUGRAmup} (including $c_{U,D}=1$) so that $R_{U,D}=S_{U,D}=0$ corresponds to
mSUGRA model in this figure. 
We see that with non-zero values of $R_{U,D}$ and $S_{U,D}$, $\left({\bf
  m}_U^2\right)_{23}$ already starts off with a substantial value
  (positive or negative) at $Q=M_{\rm GUT}$, and evolves slowly with
  $Q$. This situation is qualitatively similar to that in
  Fig.~\ref{fig:SUGRAmup}, once $\left({\bf m}_U^2\right)_{ij}$ has
  evolved away from its value at $M_{\rm GUT}$, and
  has had a chance to grow from zero. However, because the curves start
  of with rather large values at the GUT scale (except for the middle
  mSUGRA curve) the evolution does not take them through zero for any
  value of $Q>m_t$. As a result, the dip which was the most prominent
  feature of Fig.~\ref{fig:SUGRAmup} is absent, except in the middle
  curve which does cross zero for $Q\sim 2.5\times 10^5$~GeV.

\begin{figure}\begin{centering}
\includegraphics[width=.7\textwidth]{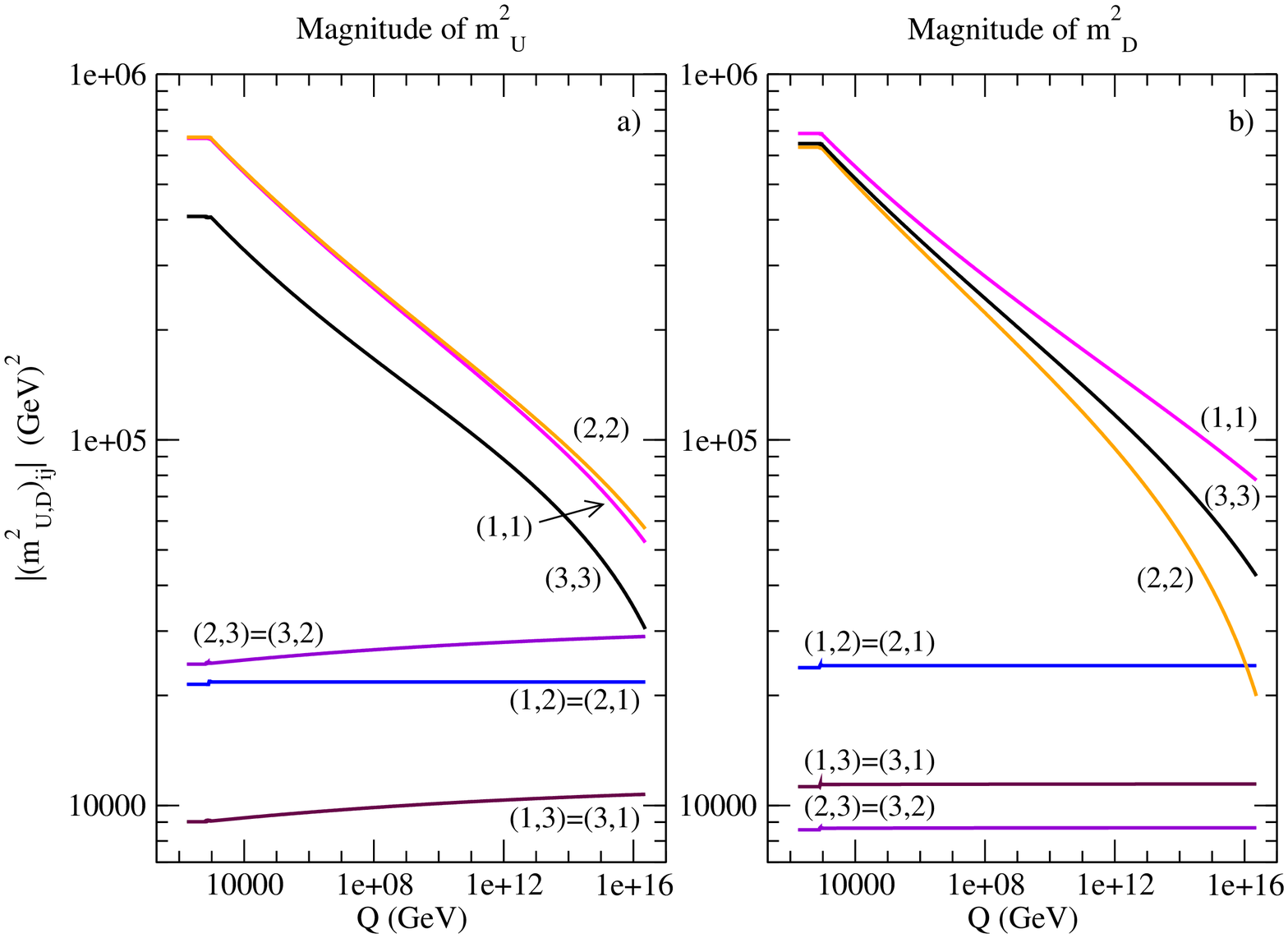}
\caption{({\it a})~The same as Fig.~\ref{fig:SUGRAmup} except for the
  non-universal model considered in Fig.~\ref{fig:T2fu}, and ({\it
  b})~the magnitude of the corresponding elements of ${\bf m}_D^2$ for
  the same scenario. The elements are plotted in our ``standard'' current
  basis but are exactly the same in the basis where the down quark Yukawa
  couplings are diagonal.}
\label{fig:T2mu}
\end{centering}
\end{figure}

Finally, in Fig.~\ref{fig:T2mu} we return to the non-mSUGRA case that we
considered in Fig.~\ref{fig:T2fu}.
We emphasize again that in this case ${\bf m}_U^2$ and ${\bf m}_D^2$ are
diagonal only in the basis where the superpotential Yukawa coupling
matrices have large off-diagonal elements so that we would expect that
this model includes new and potentially large sources of
flavour-violation in the singlet squark SSB mass matrices that may well
be excluded by data.  As in Fig.~\ref{fig:SUGRAmup}, we show
$|\left(\bdm^{2}_{U}\right)_{ij}|$ in the basis where the up-quark
Yukawa coupling matrix is diagonal at $Q=m_t$.  We see from the figure
that the matrix ${\bf m}_U^2$ has large off-diagonal entries at
$Q=M_{GUT}$. As expected, the gauge (and gaugino) interactions cause the
diagonal entries to rapidly increase, whereas the off-diagonal elements
which do not ``feel these terms'' evolve much more slowly with $Q$. Note
also that because
$\left(\bdm^{2}_{U}\right)_{11}<\left(\bdm^{2}_{U}\right)_{22}$ at
$Q=M_{\rm GUT}$, Yukawa coupling effects draw them closer as we go to
low scales. We also remark that the negative GUT scale value of $S$
tends to reduce the diagonal elements of $\bdm^{2}_{U}$ as we go to low
scales but pulls up the corresponding elements of $\bdm^{2}_{D}$.
Although the renormalization group evolution increases the gap between
the off-diagonal and diagonal elements at the low scale, notice that the
off-diagonal elements are separated by just one order of magnitude from
the {\it difference between the diagonal elements}, so that we may
expect large flavour mixing between the $SU(2)$ singlet up squarks. This
mixing, if anything, is even larger in the down squark sector as can be
seen in frame ({\it b}). A careful evaluation of inter-generation squark
mixing is clearly necessary for any discussion of flavour-violation in
squark decays, the subject of the next section.

\section{Flavour-Changing Squark Decay: An Illustrative Application
}\label{sec:stop}

We are now ready to apply the considerations of our analysis to flavour
violating decays of squarks. In models with additional sources of
flavour violation in the SSB sector through non-vanishing values of the
matrices ${\bf T}_\bullet$ and/or ${\bf Z}_\bullet$ in
(\ref{eq:GUTbound}), at least some the SSB mass and ${\bf a}$-parameter
matrices will not be diagonal in the basis that quarks are diagonal, and
we will have flavour-violating decays of squarks at the ``tree-level'',
{\it i.e.} even without any renormalization group evolution of the SSB
parameters. If the elements of these matrices take arbitrary values
comparable to $M_{\rm SUSY}$, flavour-violating quark-squark-neutralino
and quark-squark-gluino couplings become comparable in magnitude to the
corresponding flavour-conserving vertices, resulting in the well-known
SUSY flavour problem. Minimal flavour violation \cite{mfv} has been
suggested as a way of introducing flavour violation into the
SSB parameters in a controlled way. We present formulae 
for flavour violating couplings and partial widths for
the corresponding decays of squarks,
and then go on to a numerical analysis of these decays. For
definiteness, we will focus on up-type squarks.

\subsection{Squark decay width}

In the current basis the up-type squark mass terms in the Lagrangian are, 
\begin{equation}
\mathcal{L} \ni -\left( {\tilde{u}}^{\dag}_{Ll}, {\tilde{u}}^{\dag}_{Rr}
\right) {\left( \bm{\mathcal{M}}^2_{\tilde{u}} \right)}_{(lr)(ms)}
\left( \begin{array}{c} \tilde{u}_{Lm} \\ \tilde{u}_{Rs} \end{array}
\right)\;, 
\end{equation}
where the $(6\times6)$ up-squark mass matrix takes the form,
\begin{equation}
\left(\bm{\mathcal{M}}^{2}_{\tilde{u}}\right)_{(lr)(ms)}\equiv\left(\begin{array}{cc}\left(\bm{\mathcal{M}}^{2}_{LL}\right)_{lm}&\left(\bm{\mathcal{M}}^{2}_{LR}\right)_{ls}\\\left(\bm{\mathcal{M}}^{2}_{LR}\right)^{\dagger}_{rm}&\left(\bm{\mathcal{M}}^{2}_{RR}\right)_{rs}\end{array}\right).
\label{eq:sqmassmat}
\end{equation}
Here, the indices $l,m$ label left-handed squarks
($\tilde{u}_{Ll}=\left(\tilde{u}_{L},\tilde{c}_{L},\tilde{t}_{L}\right)$),
while $r,s$ label right-handed squarks
($\tilde{u}_{Rr}=\left(\tilde{u}_{R},\tilde{c}_{R},\tilde{t}_{R}\right)$).
The elements of the squark mass matrix are given by
\begin{subequations}
\begin{align}
\left(\bm{\mathcal{M}}^{2}_{LL}\right)_{lm} = & {(\bdm^2_{Q})}_{lm}+ v_u^{2} {\left({\bdf^{*}_u}\bdf^{T}_u\right)}_{lm}+\left(\frac{\glp^{
2}}{12}-\frac{\gtwl^{2}}{4}\right)\left(v_{u}^{2}-v_{d}^{2}\right)\bm{\delta}_{lm}\;,\\
{(\bm{\mathcal{M}}^{2}_{RR})}_{rs} = & {(\bdm^2_{U})}_{rs}+ v_u^{2}{\left(\bdf^{T}_u{\bdf^{*}_u}\right)}_{rs}-\frac{\glp^{2}}{3}\left(v_{u}^{2}-v_{d}^{2}\right)\bm{\delta}_{rs}\;,\\
\left(\bm{\mathcal{M}}^{2}_{LR}\right)_{ls}=&-v_{u}\left(\ba_{u}\right)^{*}_{ls}+v_{d}\mtfuhus_{ls}\;.
\end{align}
\end{subequations}
Of course, if we are
evaluating $\left(\bm{\mathcal{M}}^{2}_{LR}\right)_{ls}$ for $Q<m_H$, 
it is the combination
$\left[-\sn\left(\ba_{u}\right)^{*}_{ls}+\cs\mtfuhus_{ls}\right]$ (for
which the RGE is \eqref{eq:triu} in Appendix~\ref{sec:rgeapp}) that
enters the mass matrix. 

The physical squarks ($\tilde{u}^M_\sigma$, $\sigma=1-6$) are
obtained by diagonalizing the squark matrix (\ref{eq:sqmassmat}) by a
unitary transformation $\bm{\mathcal{U}}$, and are given in terms of the current
squarks by,
\begin{equation}\label{eq:squarkrot}
\left(\begin{array}{c}\tilde{u}_{Ll}\\\tilde{u}_{Rr}\end{array}\right)=\left(\begin{array}{c}\left.\bm{\mathcal{U}}_{L}\right._{l\sigma}\\\left.\bm{\mathcal{U}}_{R}\right._{r\sigma}\end{array}\right)\tilde{u}^{M}_{\sigma}\;,
\end{equation}
where, for later convenience, we write the $(6\times 6)$ matrix $\bm{\mathcal{U}}$
in terms of two $(3\times 6)$ blocks $\bm{\mathcal{U}}_L$ and $\bm{\mathcal{U}}_R$.
We label the physical squarks in order of their mass, with
$\tilde{u}^M_1$ being the lightest, and $\tilde{u}^M_6$ being the
heaviest. Similar considerations hold for down-type squarks.
%
%


The couplings of the physical squarks to neutralinos can readily be
evaluated by inserting the squark mass eigenstate fields obtained by inverting
(\ref{eq:squarkrot}) into the obvious generalization,
\begin{subequations}\begin{align}
\mathcal{L}\left(\tilde{q}_{Lm}q_{a}\tz_{i}\right) =& {\tilde{q}_{Lm}}^{\dag}\overline{\tz_{i}}\left[i\left(\mathbf{A}^{q}_{\tz_{i}}\right)_{ma}P_L-{(i)}^{\theta_i}\bftqq^{*}_{ma}v^{(i)}_{q}P_R\right]q_{a} + \mathrm{h.c.} \\
\mathcal{L}\left(\tilde{q}_{Rs}q_{a}\tz_{i}\right) =&{\tilde{q}_{Rs}}^{\dag}\overline{\tz_{i}}\left[i\left(\mathbf{B}^{q}_{\tz_{i}}\right)_{sa}P_R-{\left(-i\right)}^{\theta_i}\bftqr^{T}_{sa}v^{(i)}_{q}P_L\right]q_{a} + \mathrm{h.c.} 
\end{align}\end{subequations}
of the quark-squark-neutralino couplings in Ref.~\cite{wss} that
incorporates the fact that the matrix couplings $\tilde{\mathbf{g}}$ and $\tilde{\mathbf{f}}$ are
generally different from the usual gauge and Yukawa couplings.  Here, we
have used $\bftqr_{sa}$, $\bftqq_{ma}$ and $v^{(i)}_{q}$ to signify
$\bftur_{sa}$, $\bftuq_{ma}$ and $v^{(i)}_{1}$ for $q=u$. We have the
same form for the interactions of down quarks and squarks, but we must
then remember to use $v^{(i)}_q=v^{(i)}_2$. The matrices ${\bf
A}^q_{\tz_i}$ and ${\bf B}^q_{\tz_i}$ that appear in the couplings
generalize to,
\begin{subequations}\label{eq:AmBs}\begin{align}
\left(\mathbf{A}^{u}_{\tz_{i}}\right)_{ma}\equiv&\frac{{\left(-i\right)}
^{\theta_i-1}}{\sqrt{2}}\left[\bgtq_{ma}v^{(i)}_3
+\frac{\bgtpq_{ma}}{3}v^{(i)}_4\right]\;,\\
\left(\mathbf{A}^{d}_{\tz_{i}}\right)_{ma}\equiv&\frac{{\left(-i\right)}^{\theta_i-1}}{\sqrt{2}}\left[-\bgtq_{ma}v^{(i)}_3+
\frac{\bgtpq_{ma}}{3}v^{(i)}_4\right]\;,\\
\left(\mathbf{B}^{u}_{\tz_{i}}\right)_{sa}\equiv&\frac{4}{3\sqrt{2}}
\bgtpurd_{sa}{\left(i\right)}^{\theta_i-1}v^{(i)}_4\;,\\
\left(\mathbf{B}^{d}_{\tz_{i}}\right)_{sa}\equiv&-\frac{2}{3\sqrt{2}}\bgtpdrd_{sa}{\left(i\right)}^{\theta_i-1}v^{(i)}_4\;,
\end{align}\end{subequations}
in direct correspondence with (8.87) of Ref.~\cite{wss}. The Lagrangian
for physical squarks then takes the form,
\begin{equation}
\mathcal{L}\left(\tilde{q}^{M}_\sigma q_{a}\tz_{i}\right) \ni {\tilde{q}^{M\dagger}_\sigma}\overline{\tz_{i}}\left[\left(\bm{\alpha}^{q}_{\tz_{i}}\right)_{\sigma a} P_L+\left(\bm{\beta}^{q}_{\tz_{i}}\right)_{\sigma a} P_R\right]q_{a}+\mathrm{h.c.},
\end{equation}
with
\begin{subequations}\label{eq:albet}\begin{align}
\left(\bm{\alpha}^{u}_{\tz_{i}}\right)_{\sigma a}\equiv&i(\bm{\mathcal{U}}_{L})^{\dagger}_{\sigma m}\left(\mathbf{A}^{u}_{\tz_{i}}\right)_{ma}-{\left(-i\right)}^{\theta_i}v^{(i)}_1(\bm{\mathcal{U}}_{R})^{\dagger}_{\sigma s}\bftur^{T}_{sa}\;,\\
\left(\bm{\beta}^{u}_{\tz_{i}}\right)_{\sigma
  a}\equiv&i(\bm{\mathcal{U}}_{R})^{\dagger}_{\sigma
  s}\left(\mathbf{B}^{u}_{\tz_{i}}\right)_{sa}-{\left(i\right)}^{\theta_i}v^{(i)}_1(\bm{\mathcal{U}}_{L})^{\dagger}_{\sigma m}\bftuq^{*}_{ma}\;,\\
\left(\bm{\alpha}^{d}_{\tz_{i}}\right)_{\sigma a}\equiv&i(\bm{\mathcal{D}}_{L})^{\dagger}_{\sigma m}\left(\mathbf{A}^{d}_{\tz_{i}}\right)_{ma}-{\left(-i\right)}^{\theta_i}v^{(i)}_2(\bm{\mathcal{D}}_{R})^{\dagger}_{\sigma s}\bftdr^{T}_{sa}\;,\\
\left(\bm{\beta}^{d}_{\tz_{i}}\right)_{\sigma a}\equiv&i(\bm{\mathcal{D}}_{R})^{\dagger}_{\sigma s}\left(\mathbf{B}^{d}_{\tz_{i}}\right)_{sa}-{\left(i\right)}^{\theta_i}v^{(i)}_2(\bm{\mathcal{D}}_{L})^{\dagger}_{\sigma m}\bftdq^{*}_{ma}\;.
\end{align}\end{subequations}

Here, the $(3\times 6)$ matrices $\bm{\mathcal{D}}_{L}$ and $\bm{\mathcal{D}}_{R}$,
which enter via the diagonalization of the $(6\times 6)$ down squark mass
matrix, are the exact analogues of the matrices $\bm{\mathcal{U}}_{L}$ and
$\bm{\mathcal{U}}_{R}$ in (\ref{eq:squarkrot}). 
The partial width for the  $\tilde{q}^{M}_\sigma \to q_{a}
\tilde{Z}_{i}$ can then be written as,
\begin{equation} \begin{split} \label{eq:twobodgamma}
\Gamma(\mathit{\tilde{q}^{M}_{\sigma}\rightarrow q_{a}\tilde{Z}_{i}})=&\frac{1}{16\pi M^3_{\tilde{q}^{M}_\sigma}}\left\{\left({\left|\left(\bm{\alpha}^{q}_{\tz_{i}}\right)_{\sigma a}\right|}^2+{\left|\left(\bm{\beta}^{q}_{\tz_{i}}\right)_{\sigma a}\right|}^2\right)\left(M^2_{\tilde{q}^{M}_\sigma}-m^2_{q_{a}}-m^2_{\tilde{Z}_{i}}\right)\right.\\
&\qquad\qquad\quad\left.-2m_{q_{a}}m_{\tilde{Z}_{i}}\left[\left(\bm{\alpha}^{q}_{\tz_{i}}\right)_{\sigma a}\left(\bm{\beta}^{q}_{\tz_{i}}\right)^*_{\sigma a}+\left(\bm{\beta}^{q}_{\tz_{i}}\right)_{\sigma a}\left(\bm{\alpha}^{q}_{\tz_{i}}\right)^{*}_{\sigma a}\right]\right\}\\
&\qquad\qquad\qquad\qquad\qquad\qquad\qquad\qquad\qquad\quad\times\lambda^{1/2}\left(M^2_{\tilde{q}^{M}_\sigma},m^2_{q_{a}},m^2_{\tilde{Z}_{i}}\right)\;,
\end{split}\end{equation}
with 
\begin{equation}
\nonumber \lambda\left(x,y,z\right)=x^2+y^2+z^2-2xy-2xz-2yz.
\end{equation}

\subsection{Flavour-Violating Squark Decay}

Formula (\ref{eq:twobodgamma}) is very general and applies to two body
decays of squarks to neutralinos for an arbitrary flavour structure of
SSB parameters. We would expect that except, possibly, in the cases
where we have non-vanishing values of ${\bf T}_{Q,U,D}$ and/or ${\bf
Z}_{u,d}$ in (\ref{eq:GUTbound}), rates for flavour-conserving decays to
neutralinos will overwhelm the corresponding flavour-violating
decays. In many SUSY models, flavour-violating decays of squarks are
thus phenomenologically relevant only when all flavour-conserving
two-body decays are kinematically inaccessible. Because of the small
values of down-type squark masses, this is unlikely to be the case for
down-type squarks.  In other words, these flavour-violating decays are
likely to be most relevant for $\tu^M_1$, the lightest of the up-type
squarks. In many models --- certainly in models where the SSB parameters
are given by the ans\"atz (\ref{eq:GUTbound}) with modest values of
$R_\bullet$, $S_\bullet$, $W_\bullet$ and $X_\bullet$, and $c_\bullet=1$
--- the lightest charge $\frac{2}{3}$ squark is likely to be $\tst_1$,
the lighter of the two squarks with the greatest top-squark content, and
only small admixtures of $\tu_{L,R}$ and $\tc_{L,R}$.  In models where
the superpotential Yukawa interactions are the {\it sole source} of
flavour-violation, the decay $\tst_1 \to c\tz_1$ has a larger rate than
$\tst_1 \to u\tz_1$ because of the structure of the Kobayashi-Maskawa
matrix. It is for this reason that this decay has received considerable
attention in the literature \cite{hikasa,han,djouadi,wohr}. Although
other squark mass patterns are certainly possible, we will focus our
attention on the $\tst_1\to c\tz_1$ decay for the remainder of this
section, assuming that the lightest up-type squark $\tu^M_1=\tst_1$ and
that the decays $\tst_1 \to t\tz_1$ as well as $\tst_1 \to b\tw_1$ are
kinematically forbidden.

Before turning to numerical details, we qualitatively estimate the
various contributions to the effective flavour-changing coupling that
causes the decay $\tst_1\to c\tz_1$, beginning with inter-generational
squark mixing. For obvious reasons, we will work in the basis that
up-type quark Yukawas are diagonal at the weak scale. We see from the
RGEs in Appendix~\ref{sec:rgeapp} that (above all thresholds) the mixing
among singlet up-type squarks occurs only via the Yukawa coupling matrix
${\bf f}_u$, and so vanishes (except for two-loop RGE effects) if the
first and second generation up Yukawa couplings are neglected. Up-type
singlet squarks can still, however, mix with doublet up-type squarks via
the ${\bf a}$-parameters. In contrast, inter-generation up squark mixing
between doublet up squarks can occur via {\it down}-type Yukawa coupling
matrices (which are not diagonal) even if the first two generations have
vanishing Yukawa couplings. Thus, in models where, in the standard
current basis, up squark mass matrices are (essentially) diagonal at the high
scale, the dominant contribution to the mixing occurs via mixing between
$\tst_L$ or $\tst_R$ with $\tc_L$. Within mSUGRA (or similar models), if
we assume ${\bf a}_d \sim A_0{\bf f}_d$, with $A_0\sim m_0$, we can
estimate that the weak scale value of
$$\left|\left({\bf m}^2_Q\right)_{23}\right|\sim \frac{8}{16\pi^2}m_0^2
f_b^2|\mathbf{K}_{23}||\mathbf{K}_{33}|\log{(M_{\rm GUT}/M_{\rm
weak})}\;,$$ which leads to a ``mixing angle'' $\theta_{\tc_L\tst_L}\sim
\left|\left({\bf m}^2_Q\right)_{23}\right|/({\textit{few}\times
m_0^2})\sim \textit{few}\times10^{-4}$ \cite{hikasa}. With similar
assumptions, the $\tst_R-\tc_L$ mixing angle has a comparable magnitude
which, however, scales with the $A_d$-parameter. Indeed, the mixing
parameter that determines the decay rate is given to an excellent
approximation in models with small flavour mixings among squarks, by,
\begin{equation}
\epsilon=\frac{\left(\bm{\mathcal{M}}_{\tu}\right)_{23}
\left(\bm{\mathcal{U}}^\dagger_L\right)_{13}+
\left(\bm{\mathcal{M}}\right)_{26}\left(\bm{\mathcal{U}}^\dagger_R\right)_{13}}{m^2_{\tilde{t}_1
}-\bm{\mathcal{M}}_{22}},
\label{eq:epsilon}
\end{equation}
where the elements $\left(\bm{\mathcal{U}}^\dagger_L\right)_{13}$ and
$\left(\bm{\mathcal{U}}^\dagger_R\right)_{13}$ (which are just
$\cos\theta_t$ and $-\sin\theta_t$ \cite{wss} in the absence of
flavour-mixing) can be reliably estimated using the $(2\times 2)$
submatrix of the full up-squark mass matrix for the mixing between
$\tst_L$ and $\tst_R$.  More generally,
$\left|\epsilon\right|=|\left.(\bm{\mathcal{U}}^{\dagger}_{L})\right._{12}|$.

How does this ``mixing contribution'' compare to the ``direct
contribution'' from the induced ${\bgt}$-type couplings? A similar
analysis to that of the previous paragraph shows that the dominant
contribution once again arises via the $\bdf_{d}$ and 
$\tilde{\bdf}^{q}_{d}$-type
couplings.\footnote{Just to be sure there is no confusion, we reiterate
that in models with non-vanishing values for ${\bf T}$ and ${\bf Z}$
matrices, this need not be the case.}  The main difference is that
because off-diagonal elements of $\bgt$ arise only below the scale of
the heaviest sparticle, the large logarithm $\log(M_{\rm GUT}/M_{\rm
weak})$ is absent. Moreover, the combinatorial factor of 8 in the
previous paragraph is about unity so that these direct contributions are
typically two orders of magnitude smaller in models where all sparticles
are at the TeV scale. We can also check this from the magnitudes of the
Yukawa couplings in Fig.~\ref{fig:SUGRAfd}, since the difference between
these and the $\tilde{\bdf}^{q}_{d}$ couplings is small. We have also 
checked that the
value of the width calculated, ignoring the difference between the tilde
couplings and the true couplings, differs from the full calculation by a
few percent, as the reader may well expect.

\subsection{Single-step RGE integration and the stop decay rate}

The decay rate for the $\tst_1\to c\tz_1$ was first estimated by Hikasa
and Kobayashi \cite{hikasa} who were considering squarks of around 30~GeV
that might have been accessible at the TRISTAN collider. They used what
is essentially the equivalent of (\ref{eq:twobodgamma}), but estimated
the size of the off-diagonal elements of the up-squark mass matrix that
enter (\ref{eq:epsilon}) by a single step integration of the
RGEs. Working within the mSUGRA framework, and using the approximation
that $\tz_1\simeq \tilde{\gamma}$, they showed that if tree-level two
body decays of $\tst_1$ were kinematically forbidden, $\tst_1\to c\tz_1$
would be the dominant decay mode of $\tst_1$. While their
approximations\footnote{Effects of general mixing in the neutralino
  sector are simple to include \cite{han}.}
and analysis are certainly valid for $m_{\tst_1}$ values that they
considered twenty years ago, this is not the case for top squark masses
in the range of interest today. 

For top squarks in the TRISTAN range analysed in Ref.~\cite{hikasa}, the
competing tree-level decays are four-body decays (we assume here that
the sneutrinos are heavier than $\tst_1$) of $\tst_1$, which are both
higher order in the couplings and suppressed by four-body phase
space. However, for $m_{\tst_1}$ in the range of interest today, the
three body decay $\tst_1 \to bW\tz_1$ may well be kinematically
accessible and could compete with the flavour-changing two body
decay. It is with this in mind that we are led to re-visit the rate for
$\tst_1\to c\tz_1$, but this time integrating the RGEs numerically to
obtain the off-diagonal elements of the up-squark mass matrix, as
opposed to obtaining this via a single-step integration as is common
practice in the literature \cite{djouadi,wohr}.

To illustrate the need for integrating the RGEs we compare the rates for
the flavour-conserving three-body decay of $\tst_1$ with the
flavour-violating $\tst_1\to c\tz_1$ decay for the mSUGRA point
$m_0=250$~GeV, $m_{1/2}=-250$~GeV, $A_0=-930$~GeV $\tan\beta=20$ and
$\mu<0$ (where we have chosen a large value of $|A_0|$ to obtain a light
$\tst_1$) in Table~\ref{tab:msugcomp}.  For this point,
$m_{\tilde{t}_{1}}\simeq 181$~GeV, $m_{\tilde{Z}_{1}}=102$~GeV,
$m_{\tnu_\tau} =270$~GeV and $\tilde{W}_{1}=197$~GeV so that tree level
two-body decays of the $\tilde{t}_{1}$ as well as three-body decays,
$\tst_1 \to b\ell\tnu_\ell$, to sneutrinos are kinematically forbidden,
but both the two-body loop decay that we are considering and the three-body
decay to $\mathit{bW\tilde{Z}_{1}}$ \cite{wohr} are allowed.  We show the
two-body decay rate, both for the single-step estimate as well as with
the full integration of the RGEs.
\begin{table}
\centering
\begin{tabular}{llr}
&Method&Width (GeV)\\ \hline
$\Gamma(\mathit{\tst_1 \to bW\tilde{Z}_{1}})$&&$8.6\times10^{-8}$\\
$\Gamma(\tst_1 \to \mathit{c\tilde{Z}_{1}})$&``single-step'' approximation&$\sim41\times10^{-8}$\\
&full calculation&$3.3\times10^{-8}$\\
\end{tabular}
\caption{Partial widths for the two- and three-body decays of the
$\tilde{t}_{1}$ within the mSUGRA framework, with $m_0=250$~GeV,
$m_{1/2}=-250$~GeV, $A_0=-930$~GeV $\tan\beta=20$ and $\mu<0$, for which
the decays $\tst_1\to b\tw_1$ and $\tst_1\to t\tz_1$ are kinematically
forbidden. The result for the two-body decay is calculated using two
methods, the single-step integration of the RGE (see text) commonly used
in the literature, and the full integration of the
RGEs.}\label{tab:msugcomp}
\end{table}
We see from Table~\ref{tab:msugcomp} that the single-step approximation
over-estimates the decay rate by about a factor of 13.6,\footnote{Since
$\tst_1$ is mainly $\tst_R$, we evaluate the decay rate using parameters
at a scale equal to the lightest right-handed squark threshold. We have
checked that the partial width changes by $\sim 6$\% if instead we had
calculated it using parameters evaluated at the highest right-squark
threshold.} and would lead us to conclude that the branching ratio
$\mathrm{B}(\tst_1 \to \mathit{c\tilde{Z}_{1}})\simeq 0.83$, whereas the
full calculation shows that $\mathrm{B}(\mathit{\tst_1\to
bW\tilde{Z}_{1}})\simeq0.72$, completely changing the qualitative
picture of top squark decays!  Admittedly, this striking change is
because we are close to the kinematic boundary for the $\tst_1 \to
bW\tz_1$ decay. We have checked, however, that the single-step
approximation over-estimates $\Gamma(\tst_1 \to c\tz_1)$ by a factor of
$\sim 10-25$ in mSUGRA models where the decay $\tst_1\to b\tw_1$ is
kinematically forbidden, and where the LSP is the lightst
neutralino. This may make the four-body decay modes of $\tst_1$ more
competitive than previously thought.

\subsection{Model dependence of $\Gamma(\tst_1\to c\tz_1)$  }

The rate for flavour-violating squark decays will sensitively depend on
whether the SSB parameters include genuinely new sources of
flavour violation. To illustrate this, we return to our sample mSUGRA
point from the previous section, namely $m_0=200$~GeV,
$m_{1/2}=-400$~GeV, $A_0=-200$~GeV $\tan\beta=10$ and $\mu>0$, and
include non-zero values for $\mathbf{T}_{Q,U,D}$ to obtain a variety of
scenarios. We mention that the scenarios we examine are unrealistic from
the perspective of observing flavour-violating $\tst_1$ decays: indeed
for this sample mSUGRA scenario, tree-level decays of $\tst_1$ are
accessible. Our purpose here is to understand how $\Gamma(\tst_1 \to
c\tz_1)$ is altered (and also the quantities that it is sensitive to) as we
alter the scenarios, to systematically allow increased non-universality
and/or flavour violation in the SSB sector.  Our results are shown in
Table~\ref{tab:widthTcomp}, beginning with Scenario $(1)$ which is the
reference mSUGRA case.
\begin{table}
\centering
\begin{tabular*}{0.85\textwidth}{@{\extracolsep{\fill}}cllc}
\multicolumn{3}{l}{Scenario}&Width\\ \hline
$(1)$&\multicolumn{2}{l}{mSUGRA \mbox{---} no dependence on specific $\mathbf{V}_{L,R}(u,d)$}&$2.2\times10^{-9}$~GeV\\[5pt]
$(2a)$&$\mathbf{V}_{R}(u)=\mathbf{V}_{R}(d)=\mathbf{V}_{L}(u)=\dblone$&${\mathbf{T}}_{U,D}\neq\mathbf{0}$&$3.9\times10^{-9}$~GeV\\
$(2b)$&&${\bf T}_{Q}\neq\mathbf{0}$&$1.6\times10^{-9}$~GeV\\[5pt]
$(3a)$&$\mathbf{V}_{L}(u)\neq\dblone,\;
\mathbf{V}_{R}(u)=\mathbf{V}_{R}(d)=\dblone$&${\bf T}_{U,D}\neq\mathbf{0}$&$3.9\times10^{-9}$~GeV\\
$(3b)$&&${\bf T}_{Q}\neq\mathbf{0}$&$2.7\times10^{-5}$~GeV\\[5pt]
$(4a)$&$\mathbf{V}_{R}(d)\neq\dblone,\;
\mathbf{V}_{R}(u)=\mathbf{V}_{L}(u)=\dblone$&${\bf T}_{U,D}\neq\mathbf{0}$&$3.6\times10^{-9}$~GeV\\
$(4b)$&&${\bf T}_{Q}\neq\mathbf{0}$&$1.6\times10^{-9}$~GeV\\[5pt]
$(5a)$&$\mathbf{V}_{R}(u)\neq\dblone,\;
\mathbf{V}_{R}(d)=\mathbf{V}_{L}(u)=\dblone$&${\bf T}_{U,D}\neq\mathbf{0}$&$5.8\times10^{-3}$~GeV\\
$(5b)$&&${\bf T}_{Q}\neq\mathbf{0}$&$1.6\times10^{-9}$~GeV\\[5pt]
$(6a)$&$\mathbf{V}_{R}(u)\neq\dblone,\; \mathbf{V}_{R}(d)\neq\dblone,\;
\mathbf{V}_{L}(u)\neq\dblone$&${\bf T}_{U,D}\neq\mathbf{0}$&$5.8\times10^{-3}$~GeV\\
$(6b)$&&${\bf T}_{Q}\neq\mathbf{0}$&$2.7\times10^{-5}$~GeV\\[5pt]
\end{tabular*}
\caption{A comparison of the two-body loop decay widths for six
scenarios. For each scenario we list the basis used for our GUT scale
inputs, with rotation matrices as specified by \eqref{eq:genrot} in the
text. In case $(a)$ of each scenario, we take
$\mathbf{T}_{Q}=\mathbf{0}$ and
$\mathbf{T}_{U,D}=\mathit{diag}\{10000,40000,90000\}~\mathrm{GeV}^{2}$. In
case $(b)$, $\mathbf{T}_{U,D}=\mathbf{0}$ and
$\mathbf{T}_{Q}=\mathit{diag}\{10000,40000,90000\}~\mathrm{GeV}^{2}$. When
$\mathbf{T}_{Q,U,D}\neq\mathbf{0}$, the corresponding
$m^{2}_{\{Q,U,D\}0}=0$. Here, ${\bf V}_{L,R}(u,d)$ are the matrices
needed to transform from the current basis in which the matrices ${\bf
T}_U$, ${\bf T}_D$ or ${\bf T}_Q$ are diagonal at $Q=M_{\rm GUT}$ to the basis
where the up or the down quark Yukawa coupling matrices are diagonal at
the weak scale.  We have checked that $m_{\tst_1}$ is constant to within
about 5\% across the Table, so its contribution to the variation of the
partial width is small.}\label{tab:widthTcomp}
\end{table}
In mSUGRA, any dependence of the width on the matrices
$\mathbf{V}_{L,R}(u,d)$ enters only via the KM matrix.
\begin{enumerate}
\item In Scenario $(2)$, we allow ({\it a})~non-universality by allowing ${\bf
  T}_{U,D}\not=\mathbf{0}$ but diagonal in our standard current basis where
  up-type Yukawas are diagonal at $m_t$, while in ({\it b}) we allow
  ${\bf T}_Q\not=\mathbf{0}$ and diagonal. The up-type squark matrices are
  (approximately) aligned with the up-quark Yukawa couplings so we do
  not expect large flavour-violating effects in up-squark decays in this
  case. Nevertheless, we do see changes of ${\cal O}(1)$ from mSUGRA
  predictions. We emphasize that the situation would be quite different for
  flavour-violating decays of down-type squarks, since the down-squark
  matrix is now not aligned with the corresponding Yukawa coupling
  matrix. 

\item This is exemplified in Scenario $(3b)$ where the mass matrix for
  left type up squarks is completely unaligned with the up Yukawa
  couplings. Not surprisingly, this leads to a very large increase in
  the rate for the flavour-violating decay of $\tst_1$. In contrast, the
  width for this decay in Scenario $(3a)$ coincides with that in $(2a)$
  because $\bdm^2_U$ is unchanged when we transform to our standard
  basis, so that scenarios $(2a)$ and $(3a)$ are really identical.

\item In Scenario $(4)$, the rotation matrices used mean that the
boundary condition for Scenario $(4a)$, the
$\mathbf{T}_{U,D}\neq\mathbf{0}$ case, differs from scenario $(2a)$ only
in the boundary condition for $\bdm^{2}_{D}$ which, in turn, affects the
running of the $\bdm^{2}_{Q,U}$ matrices to a small extent through the
RGEs. As a result, we see that the width is different from the other
scenarios by a few percent. On the other hand, the boundary
condition on $\bdm^{2}_{Q}$ is not dependent on $\mathbf{V}_{R}(d)$ so
that scenarios $(2b)$ and $(4b)$ are identical, and
the width is, therefore, the same in the two cases.

\item In Scenario $(5a)$, the matrix $\bdm^2_U$ is unaligned with ${\bf
  f}_u$ resulting in a large flavour mixing among {\it singlet} up
  squarks in our standard basis, and a concomitantly large rate for the
  flavour-violating decay. It is only in this scenario that
  $\tst_R-\tc_R$ mixing is the dominant source of the flavour-violation,
  since in all the other scenarios that we have considered up to now,
  this mixing was suppressed by the small size of the charm quark Yukawa
  coupling. The corresponding partial width is larger than in the other
  scenarios because $\tst_1$ is still dominantly $\tst_R$ as can be seen
  from Fig.~\ref{fig:T2mu}. (See also the discussion in the vicinity of
  this figure.) Although one might expect significant contributions to
  \eqref{eq:albet} from off-diagonal entries in the higgsino and gaugino
  `Yukawa' matrices, we have checked that the contribution from the
  $\left(\bm{\mathcal{U}}_{R}\right)^{\dagger}_{\tilde{t}_{1}2}
  \left(\mathbf{B}^{u}_{\tz_{1}}\right)_{22}$ term is still two orders
  of magnitude larger than any other entry in \eqref{eq:albet}. In
  contrast, Scenario $(5b)$ shows no change from $(2b)$, because the
  boundary condition is exactly the same in the two cases.

\item Finally, the boundary condition for $\bdm^2_D$ is the only
  difference between Scenarios $(5a)$ and $(6a)$. Since $\bdm^2_D$ only
  affects the decay rate via its effect on $\bdm^2_{U,Q}$ (see item
  3. above), the change that it causes is too small to be seen in the
  Table for this case where the flavour-changing partial width is so
  large.  Scenarios $(6b)$ and $(3b)$ coincide because the GUT scale
  boundary conditions coincide since $\bdm^2_U$ and $\bdm^2_D$ are unit
  matrices, and so unaffected by any rotations.
\end{enumerate}

\begin{figure}\begin{centering}
\includegraphics[width=.7\textwidth]{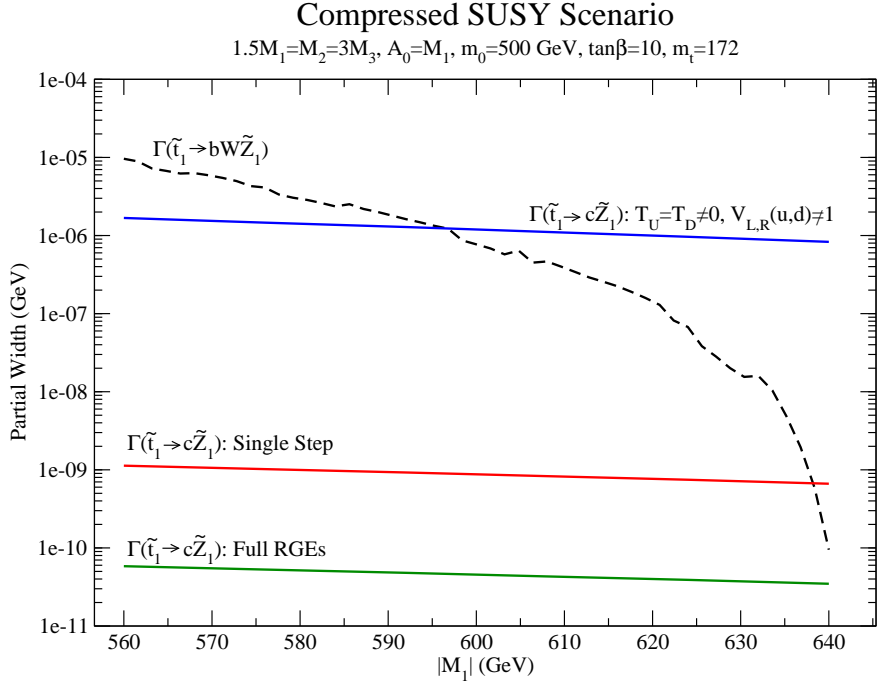}
\caption{Comparison of the partial width of the three-body tree level
decay of the stop with the two-body loop decay in the compressed SUSY
scenario discussed in the text. The dashed (black) line shows the
partial width for the three-body
decay $\tst_1\to \mathit{bW\tilde{Z}_{1}}$ while the other lines
represent calculations of rates for $\tst_1\to c\tz_1$ decay. For the
red line in the range $10^{-9}$ we use the single-step RGE integration,
the green line around $10^{-10}$ is the full calculation and the upper
line in the range $10^{-6}$ shows the enhancement as a result of taking
$\mathbf{T}_{U,D}=\mathit{diag}\{498^{2},500^{2},502^{2}\}$~GeV$^{2}$ in
the general basis using \eqref{eq:genrot}.}
\label{fig:compscan}
\end{centering}
\end{figure}

In our pedagogical examples in Table~\ref{tab:widthTcomp}, we considered
the case with very large GUT scale splitting in the squark mass matrices
which are likely excluded, especially when the squark mass matrices
and the corresponding Yukawa couplings are unaligned. However,
flavour-violation effects can be large even for very small GUT scale
splitting in the squark mass matrices. We illustrate this for the
compressed SUSY scenario, proposed by Martin \cite{martcomp} where
efficient neutralino annihilation to top pairs via the exchange of a
light squark leads to the observed cold dark matter relic density. We
use mSUGRA-like GUT scale inputs, where $m_0=500$~GeV, $A_0=M_{1}$,
$\tan\beta=10$ and $\mu>0$, but the gaugino masses are split so that
$1.5M_{1}=M_{2}=3M_{3}$, with $\bdm_{U,D}^2=diag\{498^2, 500^2,
502^2\}$~GeV$^2$ at $Q=M_{\rm GUT}$.  We show the partial widths for the
three body and the flavour-violating two body decays of $\tst_1$
as a function of $|M_1|$ in Fig.~\ref{fig:compscan}.
Over the whole of this region the two-body flavour-conserving decays of
$\tst_1$ as well as its decays to sneutrinos are kinematically
forbidden, but the three-body decay $\tst_1 \to
\mathit{bW\tilde{Z}_{1}}$ is allowed until its kinematic boundary at
$|M_1|\sim 640$~GeV. The (black) dashed line in the figure shows the
partial width for this three-body decay. The partial width for the
two-body decay, for the case of universal GUT scale masses, calculated
using the single-step approximation is in the range $10^{-9}$~GeV and
competes with the three-body decay near the extreme edge of phase
space. However, the corresponding result of our complete calculation is
essentially always smaller than the width for the three-body decay,
although it may compete with the four-body decay rate of $\tst_1$ at
large $|M_{1}|$ \cite{djouadi}.

The highest horizontal line in Fig.~\ref{fig:compscan} shows the result
of the complete calculation of
$\Gamma(\tst_1 \to c\tz_1)$ for a non-universal model with
$\mathbf{T}_{U,D}=\mathit{diag}\{498^{2},500^{2},502^{2}\}$ (in the
general basis of \eqref{eq:genrot}) and $c_{U,D}=0$. We see that
despite the very small splitting between diagonal entries of the squark
mass matrices at $Q=M_{\rm GUT}$, the rate for the flavour-violating two
body decay is enhanced by three orders of magnitude, and can be
competitive with the rate for the decay $\tst_1 \to \mathit{bW\tilde{Z}_{1}}$.
We  also mention that if we take this same splitting,
but use $\mathbf{V}_{R}(u,d)=\mathbf{V}_{L}(u)=\dblone$, we recover the
result with no splitting to within around $4\%$.

Our discussion of flavour-violating squark decays in this section has
revolved around the introduction of flavour-violation via a completely
arbitrary choice of the ${\bf T}_{U,D,Q}$ matrices in
(\ref{eq:GUTbound}). It has been suggested that flavour-violation can be
introduced into the theory in a controlled way through the minimal
flavour violation (MFV) ans\"atz \cite{mfv}. The general idea stems from
the observation that, except for Yukawa couplings, the Lagrangian of the
Standard Model is invariant under {\it independent} rotations in flavour
space of the electroweak singlet up-type, singlet down-type, and the
doublet quarks (along with independent rotations among the singlet, and
also doublet, lepton flavours). If we regard the Yukawa couplings as
spurion fields, and assume that {\it these are the only source of
flavour-violation} even in extensions of the SM, we obtain the MFV
ansatz.

In the case of the MSSM, since $\bdm_{U,D}^2$ and $\left({\bf
  f}_{u,d}^T{\bf f}_{u,d}^*\right)^n$ transform the same way under the
  flavour rotations just discussed, as do $\bdm^2_Q$ and $\left({\bf
  f}_{u,d}^*{\bf f}_{u,d}^T\right)^n$, it is easy to see that
  $\bdm^2_{U,D}$ must take the same form (\ref{eq:GUTboundsferm}) with
  ${\bf T}_{U,D}=\bm{0}$. In
  contrast, for the doublet squark SSB mass matrix we would have,
\begin{equation}
\bdm^2_Q=m_{Q0}^2\left[\dblone+t_u {\bf f}_{u}^*{\bf f}_{u}^T +
             t_d {\bf f}_{d}^*{\bf f}_{d}^T +\cdots\right]\;,
\label{eq:mfvmq}
\end{equation}
where the ellipses denote quartic and higher terms of the form,
$\left({\bf f}_{u,d}^*{\bf f}_{u,d}^T\right)^2$, $\left({\bf
f}_{u}^*{\bf f}_{u}^T\right)\left({\bf f}_{d}^*{\bf f}_{d}^T\right)$ and
$\left({\bf f}_{d}^*{\bf f}_{d}^T\right)\left({\bf f}_{u}^*{\bf
f}_{u}^T\right)$, {\it etc.}
A similar analysis shows that,
\begin{equation}
{\bf a}_u={\bf f}_u\left[A_{u0}\dblone+\alpha_{u1}{\bf f}^\dagger_u{\bf f}_u
 +\beta_{u1}{\bf f}^\dagger_d{\bf f}_d +\cdots\right]\;,
\end{equation}
where the ellipses denote higher powers of ${\bf  f}^\dagger_{u,d}{\bf
  f}_{u,d}$. A similar formula applies for ${\bf a}_d$. MFV also enters
  the slepton sector in an analogous manner.  

It is interesting to examine how $\Gamma(\tst_1\to c\tz_1)$ is
affected if instead of choosing {\it ad hoc} values of ${\bf T}_{U,D,Q}$
as we did above, we choose these according the the MFV ans\"atz. Toward
this end, we once again start with our canonical mSUGRA point, and study
how this width varies as we switch on non-vanishing values for the MFV
parameters. Since our purpose is only to illustrate this variation, we
restrict ourselves to varying only the parameters that enter the squark
mass matrices, and leave ${\bf a}_{\bullet}$-matrices at their mSUGRA
values.
\begin{figure}\begin{centering}
\includegraphics[width=.7\textwidth]{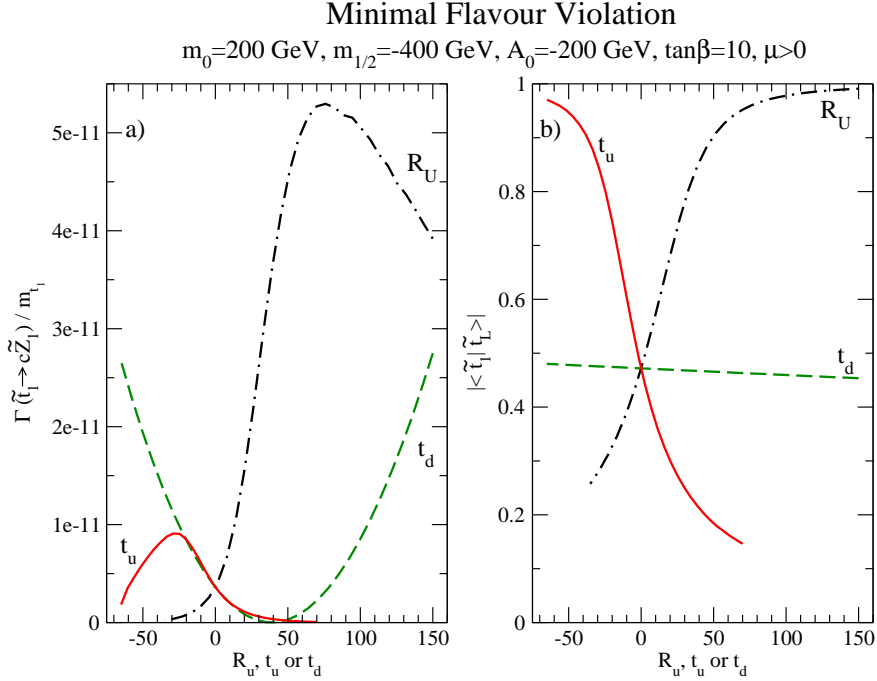}
\caption{{\it a})~Variation of $\Gamma(\tst_1\to c\tz_1)/m_{\tst_1}$
with MFV parameters $R_U$ (dot-dashed), $t_u$ (solid) and $t_d$ (dashed)
that determine the form of the squark mass matrices as described in the
text. We introduce flavour-violation into the mSUGRA model with our
canonical choice of parameters shown on the figure. For each curve, only
the one MFV parameter labelling that curve has a non-zero value. {\it
b}) The variation of $|\langle \tst_1|\tst_L\rangle|$, the
$\tst_L$ content in the lightest top squark,  which would equal
to $|\cos\theta_t|$ in the absence of any inter-generational mixing. }
\label{fig:mfv}
\end{centering}
\end{figure}
The variation of $\Gamma(\tst_1\to c\tz_1)/m_{\tst_1}$ with $R_u$, $t_u$
and $t_d$ is shown in Fig.~\ref{fig:mfv}{\it a}. The reader may be
surprised by the large variation of the ratio (which removes the trivial
growth of the width with $m_{\tst_1}$) from the variation in $R_U$ since
we have emphasized that the SSB parameters {\it do not introduce any
flavour violation in this case.} To understand this, we have shown the
top-squark ``mixing angle'' $|\langle \tst_1|\tst_L\rangle|$ in frame
{\it b}). Since inter-generation mixing is very small, $\tst_1$ is
dominantly $\tst_L$ and $\tst_R$, with tiny admixtures of other
squarks. We see that the partial width roughly tracks this mixing
angle which is a reflection of the fact that for these parameters,
the $\tst_L-\tc_L$ mixing term
is considerably larger than the $\tc_L-\tst_R$ mixing term (see the
discussion near (\ref{eq:epsilon})).
The fall-off in the width
toward the right end of the dot-dashed curve is because the splitting
between $m^{2}_{\tst_1}$ and the (2,2) element of $\bm{\mathcal{M}}^2_{LL}$
increases, leading to a suppression of $\epsilon$.

Flavour violation truly enters via the SSB parameters for non-zero
values of $t_u$ and $t_d$. For the $t_u\not=0$ case, the
flavour-violating effects in the up-squark/quark sector will be small
because the up quark Yukawa coupling matrix ${\bf f}_u$ is not-very-off
diagonal (since it's off-diagonal entries arise only from the evolution
to the GUT scale) and so is approximately aligned with the ${\bf m}^2_Q$
matrix at $Q=M_{\rm GUT}$.\footnote{We would, however, expect larger
flavour-violation in the down type sector, since the down-type Yukawa
matrix is not similarly aligned, and because the top Yukawa coupling is
larger.} The variation of the branching fraction in the solid curve with
$t_u$ largely tracks the intra-generation $t$-squark mixing, just as in
the $R_U$ case discussed in the last paragraph. The solid curve in the
left frame turns over because the GUT scale value of
$\left(\bdm^2_U\right)_{33}$, and hence $m_{\tst_1}^2$, reduces so as to
suppress $\epsilon$.

We have argued that the variation of the partial width for the $\tst \to
c\tz_1$ decay for the dot-dashed and solid curves in the figure mostly
tracks intra-generation $t$-squark mixing and does not derive from large
flavour violation of the SSB parameters. This is sharply different from
the result for the $t_d\not=0$ case, shown by the dashed curve in the
figure, where ${\bf m}_Q^2$ receives relatively large off-diagonal
contributions from the term involving the down Yukawa coupling matrix
(which is {\it not diagonal} even at $Q=m_t$) in
(\ref{eq:mfvmq}). Unlike the other two cases where the large top Yukawa
coupling significantly affects the (3,3) entry of either
$\bm{\mathcal{M}}_{RR}$ or $\bm{\mathcal{M}}_{LL}$, in this case the
$\tst_1$ mass and the intra-generational mixing are both essentially
constant for the entire range of $t_d$ in the figure, and the variation
of the width is truly the effect of the additional flavour-violation.
This is why the width continues to grow for very large values of
$|t_d|$. The reason for the drop in the width for intermediate positive
values of $t_d$ is an accidental cancellation between the two
contributions to $\epsilon$ in (\ref{eq:epsilon}).

Before closing this section, we remark that the width for
  flavour-violating decay of $\tst_1$ is very sensitive to the
  parameters $R_U$, $t_u$ and $t_d$. Even for values of
  these parameters $\sim 1-10$, 
  the change in the width from its mSUGRA value is
  ${\cal O}$(10-100)\%, and much larger if these parameters take on large
  values (though this may be constrained by low energy data). This is a
  reflection of the sensitivity of flavour physics predictions to small
  changes in the model which would have little impact on the usually
  studied collider signals for supersymmetry. The flip side of this is
  that we must view any restriction of parameter regions from low energy
  constraints from flavour physics in proper perspective, since these will
  almost surely be sensitive to the underlying flavour structure that
  has a negligible effect on direct searches for SUSY, either at colliders
  or via dark matter detection experiments.

\section{Summary}\label{summary}

Renormalization group methods allow us to extract predictions from theories
with simple physical principles operating at energy scales many orders
of magnitude larger than the highest energies accessible in
experiments. Due to effects arising from renormalization, these same
simple principles lead to a complex pattern of predictions at experimentally
accessible energies. RGEs have played a central role in the analysis of
many supersymmetric models, generally assumed to reduce to the MSSM
(quite likely  augmented by right-handed neutrino superfields, and
perhaps also additional Higgs singlets) at energy scales in between the 
weak scale and the GUT or Planck scales, where almost certainly
additional new physics would be anticipated. 

This is the second and last of a series of two papers where we have
reexamined the RGEs for the MSSM including threshold corrections to the
one-loop RGEs (which are comparable to or larger than the usually included
two-loop effects) along with flavour effects that arise below the scale
of supersymmetry breaking. In Paper~I \cite{rge1}, where we studied the
scale-dependence of the dimensionless couplings of the MSSM, we showed
that we have to extend the system of RGEs to include, in addition to the
usually studied RGEs for gauge couplings and Yukawa coupling matrices, RGEs
for the couplings of gauginos and higgsinos to matter fermions and their
superpartners. Not only do these couplings evolve independently of their
supersymmetric analogues below the scale of the highest SUSY threshold
where SUSY-breaking effects come into play, the gaugino-fermion-sfermion
coupling is also no longer flavour-independent (since it is no longer
protected from feeling flavour-breaking effects below the scale of SUSY
breaking) and so develops into a matrix in flavour-space. In this paper,
we complete this program by extending the analysis of Paper~I to the
RGEs for the dimensionful parameters of the MSSM.

Toward this end, we have first adapted the RGEs for the dimensionful
parameters of a general ({\it i.e.} non-supersymmetric) gauge field
theory \cite{luo} that includes interactions of two-component spinor and
real scalar fields with one another and with gauge fields, and written
these in a form suitable for the derivation of the RGEs for the
parameters of the theory with four-component Dirac and Majorana
spinors, together with real or complex scalar fields that we find more
convenient for our analysis. We use non-supersymmetric methods because
we want to include threshold corrections which, of course, break
supersymmetry. The details of our method (including some minor corrections
to the RGEs in the literature) along with our results for the RGEs for
the dimensionful parameters for a non-supersymmetric (albeit not
completely general) field theory are found in Sec~\ref{sec:formal}. Our
procedure for decoupling heavy particles, which is essentially the same
as in Paper~I, is described in Sec.~\ref{sec:dec} with particular
attention to some complications that arise when decoupling Higgs bosons,
and especially squarks. In Sec.~\ref{sec:mssm}, we use these general
equations to derive the RGEs for the dimensionful parameters of the
MSSM. The complete set of RGEs is listed in Appendix~\ref{sec:rgeapp}.

In Sec.~\ref{sec:flav} we discuss examples of numerical solutions to the
MSSM RGEs for $\mu$ and the dimensionful SSB parameters, focussing on
flavour-violation and threshold correction effects, including a
discussion of new
technical complications that arise in the treatment of radiative EWSB. 
We have also presented the most general parametrization of high scale
SSB parameters valid assuming that the physics of SUSY breaking is
flavour-blind. 
Unlike in Paper~I, where SUSY model-dependence of the dimensionless
couplings arises only via the location of the sparticle thresholds, 
the dimensionful parameters show considerable dependence on the
underlying model. For instance, even if the gaugino mass parameters
all originate in a common parameter $m_{1/2}$, as is the
case in all SUSY GUT models where the SUSY-breaking VEV does not also break
the GUT symmetry, the resulting gaugino unification condition,
$$\frac{M_2}{M_1}=\frac{\alpha_2}{\alpha_1}\;,$$ receives threshold
corrections $\sim$ 10\% if the scale of SUSY scalars is around
$10^7$~GeV, with gaugino and higgsino masses at the TeV scale.  In
Fig.~\ref{fig:SUGRAtri} -- Fig.~\ref{fig:rsmu}, we show illustrative
examples of the scale dependence of trilinear scalar coupling and SSB squark mass
matrices, both for mSUGRA as well as for non-universal models where high
scale SSB
parameters {\it do not include} a new source of flavour-violation, while in
Fig.~\ref{fig:T2mu} we show the elements of the singlet squark matrices
${\bf m}_U^2$ and ${\bf m}_D^2$ for a scenario where a large --- indeed
phenomenologically unrealistic --- flavour-violation is introduced via
GUT scale squark mass matrices.

In Sec.~\ref{sec:stop} we apply the results that we have obtained in
this paper to the examination of flavour-violating decays of squarks.
Since flavour-conserving couplings of squarks to neutralinos are
presumably much larger than the corresponding flavour-violating
couplings, the branching fraction for flavour-violating squark decays is
likely to be small, unless the two body decays of squarks to neutralinos
(or charginos) and quarks of the same generation are all kinematically
forbidden. With this in mind, we re-visit the decay $\tst_1\to c\tz_1$
that has received the most attention in the
literature. In this connection, we find that within the mSUGRA model
(where this decay has been most extensively studied) the commonly-used
``single-step'' approximation to obtain the flavour-violating
$\tst_1\tz_1 c$ coupling typically over-estimates the decay rate by a
factor of 10-25, and could lead to a qualitatively wrong picture for
event-topologies from top-squark pair production for $m_{\tst_1}\sim
100-300$~GeV. We have also examined the rate for this decay in a number
of non-universal scenarios for SSB parameters, with and without
flavour-violation. We then saw that the decay rate is sensitive to the
individual matrices ${\bf V}_{L,R}(u,d)$ that enter via the
diagonalization of the Yukawa coupling matrices, and not just to the KM
matrix. Indeed, this dependence of physics on the separate matrices is
the generic situation, while the dependence of physics on just the KM
combination of these that we have become used to from studies within the
SM or the mSUGRA frameworks, is true only in very special situations:
see Table~\ref{tab:widthTcomp}.\footnote{Although this is well-known to
many authors, we stress this here because there has been occasional
confusion about this issue. For a different example, see p. 215 of
Ref.~\cite{wss}.} We have also examined this decay rate in models with
non-universal SSB parameters but no new source of flavour-violation, or
where flavour violation is introduced in a controlled way via the
so-called ``minimal flavour violation'' ans\"atz. Interestingly, as
shown in Fig.~\ref{fig:mfv}, the decay rate changes by a qualitatively
similar magnitude as we vary the model parameters, in both classes of
models. The reason for this is that the variation of the width due to
changes in the (flavour-conserving) $\tst_L-\tst_R$ mixing is comparable
to the true flavour-violating contributions to this decay rate in MFV
scenarios. A determination of the left-right mixing in the $t$-squark
sector (which will be difficult at the LHC \cite{thetat} but possible at
an $e^+e^-$ collider with sufficient centre-of-mass energy
\cite{sopczak}) should allow us to readily distinguish between the
scenarios.

To sum up, in this series of two papers, we have presented the RGEs for
the, in general, complex parameters of the MSSM including one-loop
threshold effects (necessary for two-loop accuracy) as well as
flavour-mixing effects. The complete set of RGEs is listed in the
Appendices of these two papers, and will facilitate the examination of
the flavour phenomenology in SUSY models with arbitrary ans\"atze for
flavour violation via the Yukawa coupling matrices, as well as via the
SSB sector.

\section*{Acknowledgments} 
We are grateful to H.~Baer, D.~Casta\~no, V.~Cirigliano, A.~Dedes, S.~Martin,
K.~Melnikov, A.~Mustafayev and M.~Vaughn for clarifying communications
and discussions.  We thank A.~Mustafayev for his comments on the
manuscript. This research was supported in part by a grant from
the United States Department of Energy.

\appendix

\section{Numerical Instabilities Associated with Matrix Diagonalization}\label{sec:orthfix}

\subsection{The problem}
We have emphasized that in order to properly implement particle
decoupling into the RGEs, we have to be in the mass basis of the
particles being decoupled.  Our procedure for decoupling squarks,
therefore, requires us to evaluate the unitary transformation from the
given current basis to a new current basis that coincides with the
squark mass basis (approximated, as discussed in the main text, to be
the basis in which the SSB squark mass squared matrices are
diagonal). Below the scale $Q$ where at least one squark has decoupled,
we not only have the rotations $\mathbf{V}_{L,R}(u,d)$ (unitary matrices
by construction) which connect the current basis with the basis in which
the Yukawas are diagonal at $m_{t}$, but also the squark rotations
$\mathbf{R}_{\bullet}$ which connect the current basis to the ``squark
mass basis'' that are obtained by numerically diagonalizing the SSB
matrices ${\bf m}_\bullet^2$. $\mathbf{R}_{\bullet}$ is of course the
matrix of the orthogonal eigenvectors of ${\bf m}^2_\bullet$.  If there
is a degeneracy of eigenvalues, the orthogonal eigenvectors are not
uniquely defined. This leads to a practical problem when we {\it numerically}
solve for the eigenvectors in the case that two eigenvalues of any SSB
squark mass matrix with large off-diagonal components are degenerate to
within $\sim 1$\%. In this case, the corresponding eigenvectors, because
of (system-dependent) numerical errors are not exactly orthogonal, and
the corresponding matrix $\mathbf{R}_{\bullet}$ is not precisely
unitary.\footnote{Using the \texttt{g77 FORTRAN} compiler with
\textsf{Macintosh Intel Macbook}, together with the subroutine
\texttt{CG} in the \texttt{EISPACK} collection of subroutines, we found
that $\mathbf{R}_{\bullet}^\dagger \mathbf{R}_{\bullet}$ deviated from
$\dblone$ to about one part in $10^{10}$ compared to a part in $10^{18}$
for $\mathbf{V}_{L,R}(u,d)^\dagger\mathbf{V}_{L,R}(u,d)$ for
$\mathbf{V}_{L,R}(u,d)$ of the form (\ref{eq:rotmat}). We obtain
a similar size deviation from identity using the subroutine
\texttt{ZGEEV} in the \texttt{LAPACK} collection of subroutines}

The deviation from unitarity is very small, a part in $10^{10}$ in our
case, but is nonetheless orders of magnitude larger than what we can
tolerate when calculating the smallest off-diagonal elements of ${\bf
m}^2_\bullet$. To understand why our calculation is sensitive to this
seemingly tiny level of noise, let us imagine what would happen if we
attempted to evolve the off-diagonal elements of ${\bf m}_U^2$ from
$Q=M_{\rm GUT}$ in a basis where the Yukawa coupling matrices all have
large off-diagonal elements at the GUT scale. (This is not what we
actually do, but we could imagine doing so since we know that we are
well above all SUSY and Higgs field thresholds where the choice of basis
should be irrelevant.) In the mSUGRA
framework, the squark mass matrices are all given by ${\bf
m}_\bullet^2=m_0^2\dblone$ at $Q=M_{GUT}$ in any basis. Then, from 
(\ref{appeq:rgemup}) we see that these would
develop off-diagonal components $\sim \textit{few}\times f^2
\times m_0^2  \simeq f^2 \times 4\times
10^4$ for the case shown in Fig.~\ref{fig:SUGRAmup},
where $f^2$ denotes the size of the off-diagonal
element of ${\bf f}_u^T{\bf f}_u^*$. In this rough estimate we have
assumed that the loop factor $1/(16\pi^2)$ is compensated for by the
large logarithm.  In the general current basis where ${\bf f}_u$ has
comparable off-diagonal and diagonal elements, $f^2\sim 1$, and the
magnitude of the off-diagonal elements of ${\bf m}^2_U$ 
are ${\cal O}(10^4)$~GeV$^2$. Rotating to our
standard current basis should yield the result in
Fig.~\ref{fig:SUGRAmup}. In particular, we should obtain $\left|{\bf
m}^2_U\right|_{12} \sim 10^{-9}$~GeV$^2$ because there would be large
cancellations arising from the unitarity of $\mathbf{R}_U$ that would
suppress this matrix element. If instead the unitarity of $\mathbf{R}$
holds only to a part in $10^{10}$ because of numerical errors in
obtaining the eigenvectors, we will find that because the cancellations
are not perfect all off-diagonal elements of ${\bf m}_U^2$ will have a
magnitude that is at least $\textit{few} \times 100^{2}\times 10^{-10}
\sim \textit{few} \times 10^{-6}$~GeV$^2$, much larger than the magnitude of the
(1,2) element in Fig.~\ref{fig:SUGRAmup}. We note here that the noise
that leads to the non-unitarity of ${\bf V}_{L,R}(u,d)$ matrices at the
$10^{-18}$ level is completely irrelevant. 

\subsection{The solution}
The non-unitarity of $\mathbf{R}_\bullet$ is only an issue when the
off-diagonal entries of the squark mass matrix, in the basis where the
Yukawas are diagonal at $m_{t}$, are small compared to the diagonal
entries. Since we, therefore, only need to consider matrices that are already
approximately diagonal, we can associate the eigenvectors,
$(\mathbf{e}_{1},\mathbf{e}_{2},\mathbf{e}_{3})$, with the approximate
eigenvalues $((\bdm^{2}_{\bullet})_{11},(\bdm^{2}_{\bullet})_{22},
(\bdm^{2}_{\bullet})_{33})$, respectively.
%
As an illustration, let us take a case where the
$(\bdm^{2}_{\bullet})_{23}$ entry is the off-diagonal entry with the
largest magnitude, and $(\bdm^{2}_{\bullet})_{12}$ the one with the
smallest. We know the ordering quite unambiguously because above all
squark thresholds we do not need to rotate by the matrices
$\mathbf{R}_\bullet$ that potentially are the origin of the noise. We
need to ensure that the $(\bdm^{2}_{\bullet})_{12}$ entry does not
suffer from any numerical noise due to the diagonalisation. This leads
us to fix $\mathbf{e}_{1}\cdot \mathbf{e}_{2}=0$ and move any
non-orthogonality of the eigenvectors into $\mathbf{e}_{2}\cdot
\mathbf{e}_{3}$ so that the noise moves to 
$(\bdm^{2}_{\bullet})_{23}$, the off-diagonal
element with the largest magnitude. To accomplish this, we slightly
modify (by parts in $10^{10}$, the limit of accuracy of the
diagonalization routines) {\it only} the eigenvector $\mathbf{e}_{2}$
from its value as given by the diagonalization routines, thereby leaving
$\mathbf{e}_{1}\cdot\mathbf{e}_{3}$ unaffected. 

\begin{figure}\begin{centering}
\includegraphics[width=.7\textwidth]{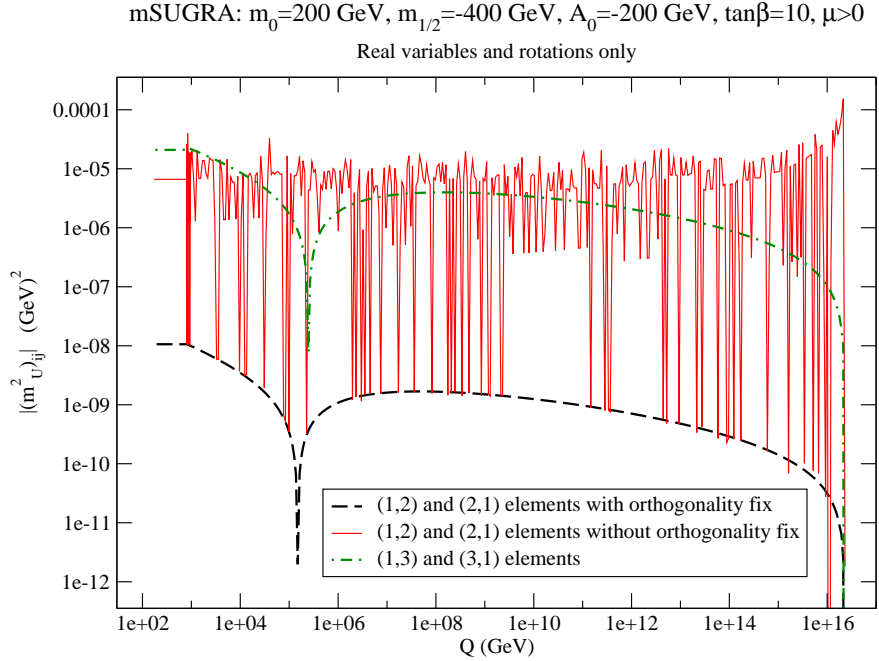}
\caption{The scale-dependence of the (1,2) element of the Hermitian
$\bdm^{2}_{U}$ matrix for our sample mSUGRA point (in the basis
specified by (\ref{eq:genrot})) calculated using two
different procedures discussed in the text. The dashed (black) line shows
the magnitude of the smallest element, \textit{i.e.} $|\left({\bf
  m}^2_U\right)_{12}| =  |\left({\bf
  m}^2_U\right)_{21}|$, when
we have used our procedure to ensure that the corresponding eigenvectors
are orthogonal.
The solid (red) 
line shows the same element when we do
not pay attention to the orthogonality of the eigenvectors of ${\bf
m}_U^2$. The lighter dot-dashed (green) line shows the magnitude of the
$(1,3)$ element, and provides a scale for the size of the numerical
noise discussed in the text. The noise in this curve is too small to be
visible. All elements are zero at the GUT scale.}
\label{fig:mu12bad}
\end{centering}
\end{figure}

To completely clarify what we have just described, although (as we have
already stated) we do not need to rotate
to the squark mass basis until we reach the highest squark threshold, 
we plot, in Fig.~\ref{fig:mu12bad}, the result for 
$|\bdm^{2}_{U}|_{12}$ obtained by the two different 
methods mentioned above, in the
basis where the up-type Yukawas are diagonal at $m_{t}$, over the whole
range $M_{Z}<Q<M_{\mathrm{GUT}}$.
The dashed (black) line shows the result where we have no rotation by
$\mathbf{R}_U$ from $Q=M_{\rm GUT}$ until the highest squark threshold,
beyond which we implement our method for ensuring that the error from
the non-orthogonality of the eigenvectors of ${\bf m}^2_U$ only shows up
in the (2,3) element. Indeed we see that this curve is smooth over its
entire range of $Q$. 
The solid (red) line shows the result of carrying out the squark rotation
without our fix of the eigenvectors over the entire range of $Q$. Note
that there is significant noise all the way down to the low $Q$ region
where only some of the squarks have decoupled. This noise is largest at
$Q=M_{\rm GUT}$ where the eigenvalues of ${\bf m}_U^2$ are degenerate,
and settles down to $10^{-5}$~GeV$^2$, not far from our estimate
above. The important thing is that the frozen value of this element is
significantly different in the two cases, as a result of this noise,
just before squark decoupling. It is for this range of $Q$ (where the
mass matrices that enter flavour-changing processes involving squarks will be
evaluated) that we must reduce the numerical error as far as possible.
The magnitude of the $(1,3)$ element of ${\bf m}_U^2$ is
shown for comparison by the dot-dashed (green) curve. It has no visible
noise because the corresponding eigenvalues are sufficiently split, 
and the corresponding eigenvectors are orthogonal to a very high
accuracy. 

The reader will be struck by the fact that the random downward
fluctuations in the solid curve are roughly bounded by the dashed
(black) curve which shows the correct magnitude of the matrix
element. The reason for this is that the {\it fluctuations} whose
typical magnitude is $\sim 10^{-5}$~GeV$^2$ need to {\it randomly}
fluctuate down by four orders of magnitude to even reach the 
dashed (black) line, and even more to go below, the chance for which is very
small. Indeed it is because we have shown results for the case where the
SSB squark mass matrices are real that we see these fluctuations go down
to even the level of the dashed (black) line. For the more general case
the chance for both the real and the imaginary part of any matrix
element to {\it simultaneously} fluctuate downward by this large
magnitude is very small, so that the calculated magnitude (not shown
here) is always larger than $10^{-6}$~GeV$^2$.

\begin{figure}\begin{centering}
\includegraphics[width=.7\textwidth]{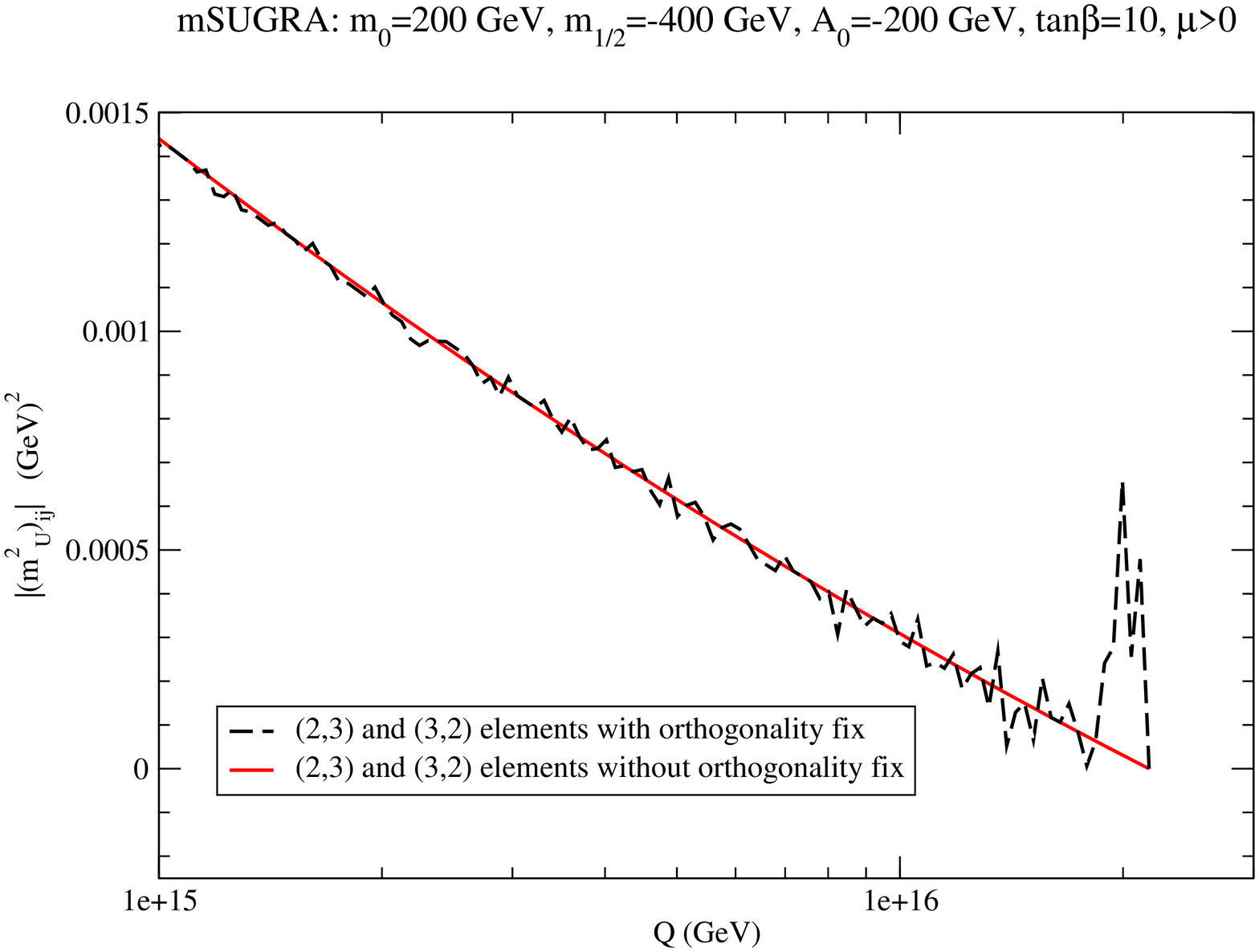}
\caption{Scale dependence of the magnitude of the $(2,3)$ entry of
$\bdm^{2}_{U}$ for the same mSUGRA point as Fig.~\ref{fig:mu12bad}. We
focus on the running at the extreme high scale, and compare the noise in
the magnitude of just this element, both before and after fixing the
orthogonality of the eigenvectors as described in the text. As in
Fig.~\ref{fig:mu12bad}, the solid (red) line shows the result before the
orthogonality fix while the dashed (black) line shows the result after
this fix
when this error has been moved to the $(2,3)$  element which
now randomly fluctuates close to $M_{\rm GUT}$ where the eigenvalues of
${\bf m}_U^2$ are roughly degenerate.}
\label{fig:mu23bad}
\end{centering}
\end{figure}

After our fix of the eigenvectors, any error from the non-unitarity of
${\bf R}_U$ is shifted to the largest off-diagonal element, and the only
residue of the resulting noise that remains is in the magnitude of this
element for scales close to $M_{\rm GUT}$ \mbox{---} where the
eigenvalues are closest \mbox{---} as seen in Fig.~\ref{fig:mu23bad}. At
lower scales, the eigenvalues split, and the noise level (whose magnitude
remains the same as in Fig.~\ref{fig:mu12bad}) becomes insignificant.
Again, the solid (red) line shows the evolution of the magnitude of the
$(2,3)$ element before fixing the eigenvectors, while the dashed black
line shows the same thing after the orthogonality fix. We see that the
numerical noise has indeed moved to the $(2,3)$ element which now shows
fluctuations, but only close to $M_{\rm GUT}$ where the eigenvalues of
${\bf m}_U^2$ are roughly degenerate.

\section{Renormalization Group Equations of Dimensionful Parameters}\label{sec:rgeapp}

This appendix contains the RGEs with full thresholds for the
dimensionful couplings of the MSSM with $R$-parity conservation. Note
that we write these RGEs in the current basis in which the SSB sfermion
mass matrices, but not the quark Yukawa matrices, are diagonal. In any
other basis they must be modified to account for the rotation from this
basis to the sfermion mass basis so that the sfermions can be properly
decoupled. In this case the squark $\theta_{\tq_k}$'s become matrices
$\bm{\Theta}_{q_k}$ as discussed in Sec~\ref{subsec:sqdec}, where further
details may be found. The RGEs for the superpotential parameter $\mu$ and the
gaugino SSB mass parameters are,
\begin{equation}\label{app:rgemu}\begin{split}
{\left(4\pi\right)}^2\frac{d\mu}{dt}=&\frac{1}{2}\mu\sh\left[3\suk\bftur^\dagger_{kl}\bftur_{lk}+3\sdk\bftdr^\dagger_{kl}\bftdr_{lk}+\sek\bfter^\dagger_{kl}\bfter_{lk}\right.\\
&\left.\qquad\quad+3\sqk\bftuq_{kl}\bftuq^\dagger_{lk}+3\sqk\bftdq_{kl}\bftdq^\dagger_{lk}+\slk\bftel_{kl}\bftel^\dagger_{lk}\right]\\
&+\frac{1}{4}\mu\sh\left[\left(3\swi\mgthusq+\sbi\mgtphusq\right)\left(\sn^2\h+\cs^2\Hh\right)\right.\\
&\left.\qquad\qquad+\left(3\swi\mgthdsq+\sbi\mgtphdsq\right)\left(\cs^2\h+\sn^2\Hh\right)\right]\\
&+\sn\cs\left(-\h+\Hh\right)\left[3\swi\gthu\left(M_2+iM'_{2}\right)\gthd+\sbi\gtphu\left(M_1+iM'_{1}\right)\gtphd\right]\\
&-\mu\sh\left(\frac{3}{2}g'^2+\frac{9}{2}g^2_2\right)\;, \\
\end{split}\end{equation}
\begin{equation}\label{app:rgem1}\begin{split}
{\left(4\pi\right)}^2\frac{dM_1}{dt}=&M_1\sbi\left[\frac{1}{3}\sqk\bgtpq_{kl}\bgtpq^\dagger_{lk}+\slk\bgtpl_{kl}\bgtpl^\dagger_{lk}+\frac{8}{3}\suk\bgtpur^\dagger_{kl}\bgtpur_{lk}\right.\\
&\left.\qquad\quad+\frac{2}{3}\sdk\bgtpdr^\dagger_{kl}\bgtpdr_{lk}+2\sek\bgtper^\dagger_{kl}\bgtper_{lk}\right.\\
&\left.\qquad\quad+\sh\mgtphusq\left(\sn^2\h+\cs^2\Hh\right)+\sh\mgtphdsq\left(\cs^2\h+\sn^2\Hh\right)\right]\\[5pt]
&+2\sn\cs\left(-\h+\Hh\right)\sh\left[\gtphd\mu^*\gtphu+(\gtphd)^{*}\mu(\gtphu)^{*}\right]\;,
\end{split}\end{equation}
\begin{equation}\begin{split}
{\left(4\pi\right)}^2\frac{dM'_1}{dt}=&M'_1\sbi\left[\frac{1}{3}\sqk\bgtpq_{kl}\bgtpq^\dagger_{lk}+\slk\bgtpl_{kl}\bgtpl^\dagger_{lk}+\frac{8}{3}\suk\bgtpur^\dagger_{kl}\bgtpur_{lk}\right.\\
&\left.\qquad\quad+\frac{2}{3}\sdk\bgtpdr^\dagger_{kl}\bgtpdr_{lk}+2\sek\bgtper^\dagger_{kl}\bgtper_{lk}\right.\\
&\left.\qquad\quad+\sh\mgtphusq\left(\sn^2\h+\cs^2\Hh\right)+\sh\mgtphdsq\left(\cs^2\h+\sn^2\Hh\right)\right]\\[5pt]
&+2i\sn\cs\left(-\h+\Hh\right)\sh\left[\gtphd\mu^*
\gtphu-(\gtphd)^{*}\mu(\gtphu)^{*}\right]\;,
\end{split}\end{equation}
\begin{equation}\label{app:rgem2}\begin{split}
{\left(4\pi\right)}^2\frac{dM_2}{dt}=&M_2\swi\left[3\sqk\bgtq_{kl}\bgtq^\dagger_{lk}+\slk\bgtl_{kl}\bgtl^\dagger_{lk}+\sh\mgthusq\left(\sn^2\h+\cs^2\Hh\right)\right.\\
&\left.\qquad\quad+\sh\mgthdsq\left(\cs^2\h+\sn^2\Hh\right)\right]\\[5pt]
&+2\sn\cs\left(-\h+\Hh\right)\sh\left[\gthd\mu^*\gthu+(\gthd)^{*}\mu(\gthu)^{*}\right]-12\swi M_2g^2_2\;,
\end{split}\end{equation}
\begin{equation}\label{app:rgem2p}\begin{split}
{\left(4\pi\right)}^2\frac{dM'_2}{dt}=&M'_2\swi\left[3\sqk\bgtq_{kl}\bgtq^\dagger_{lk}+\slk\bgtl_{kl}\bgtl^\dagger_{lk}+\sh\mgthusq\left(\sn^2\h+\cs^2\Hh\right)\right.\\
&\left.\qquad\quad+\sh\mgthdsq\left(\cs^2\h+\sn^2\Hh\right)\right]\\[5pt]
&+2i\sn\cs\left(-\h+\Hh\right)\sh\left[\gthd\mu^*\gthu-(\gthd)^{*}\mu(\gthu)^{*}\right]-12\swi M'_2g^2_2\;,
\end{split}\end{equation}
\begin{equation}\begin{split}
{\left(4\pi\right)}^2\frac{dM_3}{dt}=&M_3\sgl\left[2\sqk\bgtsq_{kl}\bgtsq^\dagger_{lk}+\suk\bgtsur^\dagger_{kl}\bgtsur_{lk}+\sdk\bgtsdr^\dagger_{kl}\bgtsdr_{lk}-18g^2_2\right]\;,
\end{split}\end{equation}
\begin{equation}\label{app:rgem3p}\begin{split}
{\left(4\pi\right)}^2\frac{dM'_3}{dt}=&M'_3\sgl\left[2\sqk\bgtsq_{kl}\bgtsq^\dagger_{lk}+\suk\bgtsur^\dagger_{kl}\bgtsur_{lk}+\sdk\bgtsdr^\dagger_{kl}\bgtsdr_{lk}-18g^2_2\right]\;.
\end{split}\end{equation}

\newpage
The following RGEs are only valid above $Q=m_{H}$, where
$\theta_{h}=\theta_{H}=1$. We separate the two regimes of different
Higgs boson content to simplify the resulting formulae, and to make
explicit the parameters which remain in the theory below the heavy Higgs
decoupling scale, $Q=m_{H}$.
\begin{equation}\begin{split}\label{eq:RGEau}
{\left(4\pi\right)}^2\frac{d\au_{ij}}{dt}=&\suk\au_{ik}\left[-\frac{2\glp^{2}}{3}\delta_{kj}+2\left[\fuul^{\dagger}\fuul\right]_{kj}\right]\\
&+\sul\sqk\left[-2\left(\frac{\glp^{2}}{9}+\frac{4\gthl^{2}}{3}\right)\delta_{ik}\delta_{lj}+6\fuhu_{ij}\fuhu^{\dagger}_{lk}\right]\au_{kl}\\
&+\sqk\left[\left(\frac{\glp^{2}}{6}-\frac{3\gtwl^{2}}{2}\right)\delta_{ik}+4\left[\fuur\fuur^{\dagger}\right]_{ik}\right]\au_{kj}\\
&+2\sdk\ad_{ik}\left[\fdul^{\dagger}\fuul\right]_{kj}\\
&+\frac{2}{3}\sbi\left(M_{1}-iM'_{1}\right)\left(\sh(\gtphu)^{*}\bgtpq^{*}_{ik}\bftur_{kj}-\frac{4}{3}\bgtpq^{*}_{ik}(\bdf_{u})_{kl}\bgtpur^{*}_{kj}\right.\\
&\left.\qquad\qquad\qquad\qquad\quad-4\sh\bftuq_{ik}\bgtpur^{*}_{kj}(\gtphu)^{*}\right)\\
&-6\swi\sh\left(M_{2}-iM'_{2}\right)(\gthu)^{*}\bgtq^{*}_{ik}\bftur_{kj}\\
&-\frac{32}{3}\sgl\left(M_{3}-iM'_{3}\right)\bgtsq^{*}_{ik}(\bdf_{u})_{kl}\bgtsur^{*}_{kj}\\
&+\suk\au_{ik}\left[\frac{8}{9}\sbi\bgtpur^{T}_{kl}\bgtpur^{*}_{lj}+\frac{8}{3}\sgl\bgtsur^{T}_{kl}\bgtsur^{*}_{lj}+2\sh\bftur^{\dagger}_{kl}\bftur_{lj}\right]\\
&+\left[3(\bdf^{\dagger}_{u})_{kl}(\bdf_{u})_{lk}+\frac{1}{2}\sh\sbi\mgtphusq+\frac{3}{2}\sh\swi\mgthusq\right]\au_{ij}\\
&+\sql\left[\sh\bftuq_{ik}\bftuq^{\dagger}_{kl}+\sh\bftdq_{ik}\bftdq^{\dagger}_{kl}+\frac{1}{18}\sbi\bgtpq^{*}_{ik}\bgtpq^{T}_{kl}\right.\\
&\left.\qquad\quad+\frac{3}{2}\swi\bgtq^{*}_{ik}\bgtq^{T}_{kl}+\frac{8}{3}\sgl\bgtsq^{*}_{ik}\bgtsq^{T}_{kl}\right]\au_{lj}\\
&-3\left\{\left(\frac{1}{36}\sqi+\frac{4}{9}\suj+\frac{1}{4}\right)g'^{2}+\frac{3}{4}\left(\sqi+1\right)g^{2}_{2}+\frac{4}{3}\left(\sqi+\suj\right)g^{2}_{3}\right\}\au_{ij}\;,
\end{split}\end{equation}
\begin{equation}\begin{split}\label{eq:RGEmtsfuhu}
{\left(4\pi\right)}^2\frac{d\mtsfuhu_{ij}}{dt}=&\frac{2\glp^{2}}{3}\suk\delta_{kj}\mtsfuhu_{ik}\\
&+\sul\sqk\left[-2\left(\frac{\glp^{2}}{9}+\frac{4\gthl^{2}}{3}\right)\delta_{ik}\delta_{lj}+6\fuhu_{ij}\fuhu^{\dagger}_{lk}\right]\mtsfuhu_{kl}\\
&-\sqk\left[\left(\frac{\glp^{2}}{6}-\frac{3\gtwl^{2}}{2}\right)\delta_{ik}+2\left[\fddr\fddr^{\dagger}\right]_{ik}\right]\mtsfuhu_{kj}\\
&-2\sdk\mtsfdhd_{ik}\left[\fddl^{\dagger}\fudl\right]_{kj}\\
&-\frac{2}{3}\sbi\sh\mu^{*}\gtphd\left(4\bftuq_{ik}\bgtpur^{*}_{kj}-\bgtpq^{*}_{ik}\bftur_{kj}\right)\\
&-6\sh\swi\mu^{*}\gthd\bgtq^{*}_{ik}\bftur_{kj}+4\sh\mu^{*}\bftdq_{ik}(\bdf^{\dagger}_{d})_{kl}\bftur_{lj}\\
&+\suk\mtsfuhu_{ik}\left[\frac{8}{9}\sbi\bgtpur^{T}_{kl}\bgtpur^{*}_{lj}+\frac{8}{3}\sgl\bgtsur^{T}_{kl}\bgtsur^{*}_{lj}\right.\\
&\left.\qquad\qquad\qquad\quad+2\sh\bftur^{\dagger}_{kl}\bftur_{lj}\right]\\
&+\left[3(\bdf^{\dagger}_{d})_{kl}(\bdf_{d})_{lk}+(\bdf^{\dagger}_{e})_{kl}(\bdf_{e})_{lk}+\frac{1}{2}\sbi\sh\mgtphdsq+\frac{3}{2}\swi\sh\mgthdsq\right]\mtsfuhu_{ij}\\
&+\sql\left[\sh\bftuq_{ik}\bftuq^{\dagger}_{kl}+\sh\bftdq_{ik}\bftdq^{\dagger}_{kl}+\frac{1}{18}\sbi\bgtpq^{*}_{ik}\bgtpq^{T}_{kl}\right.\\
&\left.\qquad\quad+\frac{3}{2}\swi\bgtq^{*}_{ik}\bgtq^{T}_{kl}+\frac{8}{3}\sgl\bgtsq^{*}_{ik}\bgtsq^{T}_{kl}\right]\mtsfuhu_{lj}\\
&-3\left\{\left(\frac{1}{36}\sqi+\frac{4}{9}\suj+\frac{1}{4}\right)g'^{2}+\frac{3}{4}\left(\sqi+1\right)g^{2}_{2}\right.\\
&\qquad\qquad\qquad\qquad\qquad\qquad\qquad\qquad\left.+\frac{4}{3}\left(\sqi+\suj\right)g^{2}_{3}\right\}\mtsfuhu_{ij}\;,
\end{split}\end{equation}
\begin{equation}\begin{split}
{\left(4\pi\right)}^2\frac{d\ad_{ij}}{dt}=&\sqk\left[-\left(\frac{\glp^{2}}{6}+\frac{3\gtwl^{2}}{2}\right)\delta_{ik}+4\left[\fddr\fddr^{\dagger}\right]_{ik}\right]\ad_{kj}\\
&+\sdl\sqk\left[2\left(\frac{\glp^{2}}{18}-\frac{4\gthl^{2}}{3}\right)\delta_{ik}\delta_{lj}+6\fdhd_{ij}\fdhd^{\dagger}_{lk}\right]\ad_{kl}\\
&+2\sel\slk\fdhd_{ij}\fehd^{\dagger}_{lk}\ae_{kl}\\
&+\sdk\ad_{ik}\left[-\frac{\glp^{2}}{3}\delta_{kj}+2\left[\fddl^{\dagger}\fddl\right]_{kj}\right]+2\suk\au_{ik}\left[\fudl^{\dagger}\fddl\right]_{kj}\\
&+\frac{2}{3}\sbi\left(M_{1}-iM'_{1}\right)\left(-\sh(\gtphd)^{*}\bgtpq^{*}_{ik}\bftdr_{kj}+\frac{2}{3}\bgtpq^{*}_{ik}(\bdf_{d})_{kl}\bgtpdr^{*}_{lj}\right.\\
&\left.\qquad\qquad\qquad\qquad\quad-2\sh\bftdq_{ik}\bgtpdr^{*}_{kj}(\gtphd)^{*}\right)\\
&-6\swi\sh\left(M_{2}-iM'_{2}\right)(\gthd)^{*}\bgtq^{*}_{ik}\bftdr_{kj}\\
&-\frac{32}{3}\sgl\left(M_{3}-iM'_{3}\right)\bgtsq^{*}_{ik}(\bdf_{d})_{kl}\bgtsdr^{*}_{lj}\\
&+\sql\left[\sh\bftuq_{ik}\bftuq^{\dagger}_{kl}+\sh\bftdq_{ik}\bftdq^{\dagger}_{kl}+\frac{1}{18}\sbi\bgtpq^{*}_{ik}\bgtpq^{T}_{kl}\right.\\
&\left.\qquad\quad+\frac{3}{2}\swi\bgtq^{*}_{ik}\bgtq^{T}_{kl}+\frac{8}{3}\sgl\bgtsq^{*}_{ik}\bgtsq^{T}_{kl}\right]\ad_{lj}\\
&+\left[3(\bdf_{d})_{kl}(\bdf^{\dagger}_{d})_{lk}+(\bdf_{e})_{kl}(\bdf^{\dagger}_{e})_{lk}+\frac{1}{2}\sbi\sh\mgtphdsq+\frac{3}{2}\swi\sh\mgthdsq\right]\ad_{ij}\\
&+\sdk\ad_{ik}\left[\frac{2}{9}\sbi\bgtpdr^{T}_{kl}\bgtpdr^{*}_{lj}+\frac{8}{3}\sgl\bgtsdr^{T}_{kl}\bgtsdr^{*}_{lj}+2\sh\bftdr^{\dagger}_{kl}\bftdr_{lj}\right]\\
&-3\left\{\left(\frac{1}{36}\sqi+\frac{1}{9}\sdj+\frac{1}{4}\right)g'^{2}+\frac{3}{4}\left(\sqi+1\right)g^{2}_{2}+\frac{4}{3}\left(\sqi+\sdj\right)g^{2}_{3}\right\}\ad_{ij}\;,
\end{split}\end{equation}
\begin{equation}\begin{split}
{\left(4\pi\right)}^2\frac{d\mtsfdhd_{ij}}{dt}=&\frac{\glp^{2}}{3}\sdk\delta_{kj}\mtsfdhd_{ik}\\
&+\sdl\sqk\left[2\left(\frac{\glp^{2}}{18}-\frac{4\gthl^{2}}{3}\right)\delta_{ik}\delta_{lj}+6\fdhd_{ij}\fdhd^{\dagger}_{lk}\right]\mtsfdhd_{kl}\\
&+2\sel\slk\fdhd_{ij}\fehd^{\dagger}_{lk}\mtsfehd_{kl}\\
&+\sqk\left[\left(\frac{\glp^{2}}{6}+\frac{3\gtwl^{2}}{2}\right)\delta_{ik}-2\left[\fuur\fuur^{\dagger}\right]_{ik}\right]\mtsfdhd_{kj}\\
&-2\suk\mtsfuhu_{ik}\left[\fuul^{\dagger}\fdul\right]_{kj}\\
&-\frac{2}{3}\sbi\sh\mu^{*}\gtphu\left(2\bftdq_{ik}\bgtpdr^{*}_{kj}+\bgtpq^{*}_{ik}\bftdr_{kj}\right)\\
&-6\sh\swi\mu^{*}\gthu\bgtq^{*}_{ik}\bftdr_{kj}+4\sh\mu^{*}\bftuq_{ik}(\bdf^{\dagger}_{u})_{kl}\bftdr_{lj}\\
&+\sdk\mtsfdhd_{ik}\left[\frac{2}{9}\sbi\bgtpdr^{T}_{kl}\bgtpdr^{*}_{lj}+\frac{8}{3}\sgl\bgtsdr^{T}_{kl}\bgtsdr^{*}_{lj}\right.\\
&\left.\qquad\qquad\qquad\quad+2\sh\bftdr^{\dagger}_{kl}\bftdr_{lj}\right]\\
&+\left[3(\bdf_{u})_{kl}(\bdf^{\dagger}_{u})_{lk}+\frac{1}{2}\sbi\sh\mgtphusq+\frac{3}{2}\swi\sh\mgthusq\right]\mtsfdhd_{ij}\\
&+\sql\left[\sh\bftuq_{ik}\bftuq^{\dagger}_{kl}+\sh\bftdq_{ik}\bftdq^{\dagger}_{kl}+\frac{1}{18}\sbi\bgtpq^{*}_{ik}\bgtpq^{T}_{kl}\right.\\
&\left.\qquad\quad+\frac{3}{2}\swi\bgtq^{*}_{ik}\bgtq^{T}_{kl}+\frac{8}{3}\sgl\bgtsq^{*}_{ik}\bgtsq^{T}_{kl}\right]\mtsfdhd_{lj}\\
&-3\left\{\left(\frac{1}{36}\sqi+\frac{1}{9}\sdj+\frac{1}{4}\right)g'^{2}+\frac{3}{4}\left(\sqi+1\right)g^{2}_{2}\right.\\
&\qquad\qquad\qquad\qquad\qquad\qquad\qquad\qquad\left.+\frac{4}{3}\left(\sqi+\sdj\right)g^{2}_{3}\right\}\mtsfdhd_{ij}\;, 
\end{split}\end{equation}
\begin{equation}\begin{split}
{\left(4\pi\right)}^2\frac{d\ae_{ij}}{dt}=&\slk\left[\left(\frac{\glp^{2}}{2}-\frac{3\gtwl^{2}}{2}\right)\delta_{ik}+4\left[\feer\feer^{\dagger}\right]_{ik}\right]\ae_{kj}\\
&+\sel\slk\left[-\glp^{2}\delta_{ik}\delta_{lj}+2\fehd_{ij}\fehd^{\dagger}_{lk}\right]\ae_{kl}+6\sdl\sqk\fehd_{ij}\fdhd^{\dagger}_{lk}\ad_{kl}\\
&+\sek\ae_{ik}\left[-\glp^{2}\delta_{kj}+2\left[\feel^{\dagger}\feel\right]_{kj}\right]\\
&+2\sbi\left(M_{1}-iM'_{1}\right)\left(\sh(\gtphd)^{*}\bgtpl^{*}_{ik}\bfter_{kj}-2\bgtpl^{*}_{ik}(\bdf_{e})_{kl}\bgtper^{*}_{lj}\right.\\
&\left.\qquad\qquad\qquad\qquad\quad-2\sh\bftel_{ik}\bgtper^{*}_{kj}(\gtphd)^{*}\right)\\
&-6\swi\sh\left(M_{2}-iM'_{2}\right)(\gthd)^{*}\bgtl^{*}_{ik}\bfter_{kj}\\
&+\sll\left[\sh\bftel_{ik}\bftel^{\dagger}_{kl}+\frac{1}{2}\sbi\bgtpl^{*}_{ik}\bgtpl^{T}_{kl}+\frac{3}{2}\swi\bgtl^{*}_{ik}\bgtl^{T}_{kl}\right]\ae_{lj}\\
&+\left[3(\bdf_{d})_{kl}(\bdf^{\dagger}_{d})_{lk}+(\bdf_{e})_{kl}(\bdf^{\dagger}_{e})_{lk}+\frac{1}{2}\sbi\sh\mgtphdsq+\frac{3}{2}\swi\sh\mgthdsq\right]\ae_{ij}\\
&+\sek\ae_{ik}\left[2\sbi\bgtper^{T}_{kl}\bgtper^{*}_{lj}+2\sh\bfter^{\dagger}_{kl}\bfter_{lj}\right]\\
&-3\left\{\left(\frac{1}{4}\sLi+\sej+\frac{1}{4}\right)g'^{2}+\frac{3}{4}\left(\sLi+1\right)g^{2}_{2}\right\}\ae_{ij}\;,
\end{split}\end{equation}
\begin{equation}\begin{split}
{\left(4\pi\right)}^2\frac{d\mtsfehd_{ij}}{dt}=&\glp^{2}\sek\delta_{kj}\mtsfehd_{ik}\\
&+\sel\slk\left[-\glp^{2}\delta_{ik}\delta_{lj}+2\fehd_{ij}\fehd^{\dagger}_{lk}\right]\mtsfehd_{kl}\\
&+6\sdl\sqk\fehd_{ij}\fdhd^{\dagger}_{lk}\mtsfdhd_{kl}\\
&-\slk\left[\left(\frac{\glp^{2}}{2}-\frac{3\gtwl^{2}}{2}\right)\delta_{ik}\right]\mtsfehd_{kj}\\
&-2\sbi\sh\mu^{*}\gtphu\left(2\bftel_{ik}\bgtper^{*}_{kj}-\bgtpl^{*}_{ik}\bfter_{kj}\right)\\
&-6\sh\swi\mu^{*}\gthu\bgtl^{*}_{ik}\bfter_{kj}\\
&+\sek\mtsfehd_{ik}\left[2\sbi\bgtper^{T}_{kl}\bgtper^{*}_{lj}+2\sh\bfter^{\dagger}_{kl}\bfter_{lj}\right]\\
&+\left[3(\bdf_{u})_{kl}(\bdf^{\dagger}_{u})_{lk}+\frac{1}{2}\sbi\sh\mgtphusq+\frac{3}{2}\swi\sh\mgthusq\right]\mtsfehd_{ij}\\
&+\sll\left[\sh\bftel_{ik}\bftel^{\dagger}_{kl}+\frac{1}{2}\sbi\bgtpl^{*}_{ik}\bgtpl^{T}_{kl}+\frac{3}{2}\swi\bgtl^{*}_{ik}\bgtl^{T}_{kl}\right]\mtsfehd_{lj}\\
&-3\left\{\left(\frac{1}{4}\sLi+\sej+\frac{1}{4}\right)g'^{2}+\frac{3}{4}\left(\sLi+1\right)g^{2}_{2}\right\}\mtsfehd_{ij}\;, 
\end{split}\end{equation}
\begin{equation}\begin{split}
{\left(4\pi\right)}^{2}\frac{d\left(m^{2}_{H_{u}}+\left|\tilde{\mu}\right|^{2}\right)}{dt}=&\frac{3}{2}\left[\glp^{2}+\gtwl^{2}\right]\left(m^{2}_{H_{u}}+\left|\tilde{\mu}\right|^{2}\right)-\glp^{2}\left(m^{2}_{H_{d}}+\left|\tilde{\mu}\right|^{2}\right)\\
&+\suk\sul\left[-2\glp^{2}\delta_{lk}+6\left[\fuult\fuuls\right]_{lk}\right]\musq_{kl}\\
&+\sqk\sql\left[\glp^{2}\delta_{lk}+6\left[\fuurs\fuurt\right]_{lk}\right]\mqsq_{kl}\\
&+\sdk\sdl\glp^{2}\delta_{lk}\mdsq_{kl}-\slk\sll\glp^{2}\delta_{lk}\mlsq_{kl}\\
&+\sek\sel\glp^{2}\delta_{lk}\mesq_{kl}+6\suk\sql\aus_{lk}\aut_{kl}\\
&+6\sql\sdk\nbmtfdhds_{lk}\nbmtsfdhdt_{kl}+2\sll\sek\nbmtfehds_{lk}\nbmtsfehdt_{kl}\\
&-2\sh\left|\mu\right|^{2}\left\{\sbi\mgtphusq+3\swi\mgthusq\right\}\\
&-2\sh\left\{\sbi\left(M^{2}_{1}+M'^{2}_{1}\right)\mgtphusq+3\swi\left(M^{2}_{2}+M'^{2}_{2}\right)\mgthusq\right\}\\
&-\left(\frac{3g'^{2}}{2}+\frac{9g^{2}_{2}}{2}\right)\left(m^{2}_{H_{u}}+\left|\tilde{\mu}\right|^{2}\right)\\
&+\left\{\left[6\bdf^{*}_{u}\bdf^{T}_{u}\right]_{kk}+\sbi\sh\mgtphusq+3\swi\sh\mgthusq\right\}\left(m^{2}_{H_{u}}+\left|\tilde{\mu}\right|^{2}\right)\;,\\
\end{split}\end{equation}
\begin{equation}\begin{split}
{\left(4\pi\right)}^{2}\frac{d\left(m^{2}_{H_{d}}+\left|\tilde{\mu}\right|^{2}\right)}{dt}=&-\glp^{2}\left(m^{2}_{H_{u}}+\left|\tilde{\mu}\right|^{2}\right)+\frac{3}{2}\left[\glp^{2}+\gtwl^{2}\right]\left(m^{2}_{H_{d}}+\left|\tilde{\mu}\right|^{2}\right)\\
&+2\suk\sul\glp^{2}\delta_{lk}\musq_{kl}\\
&+\sqk\sql\left[-\glp^{2}\delta_{lk}+6\left[\fddrs\fddrt\right]_{lk}\right]\mqsq_{kl}\\
&+\sdk\sdl\left[-\glp^{2}\delta_{lk}+6\left[\fddlt\fddls\right]_{lk}\right]\mdsq_{kl}\\
&+\slk\sll\left[\glp^{2}\delta_{lk}+2\left[\feers\feert\right]_{lk}\right]\mlsq_{kl}\\
&+\sek\sel\left[-\glp^{2}\delta_{lk}+2\left[\feelt\feels\right]_{lk}\right]\mesq_{kl}\\
&+6\suk\sql\nbmtfuhus_{lk}\nbmtsfuhut_{kl}+6\sql\sdk\ads_{lk}\adt_{kl}\\
&+2\sll\sek\aes_{lk}\aet_{kl}\\
&-2\sh\left|\mu\right|^{2}\left\{\sbi\mgtphdsq+3\swi\mgthdsq\right\}\\
&-2\sh\left\{\sbi \left(M^{2}_{1}+M'^{2}_{1}\right)\mgtphdsq+3\swi\left(M^{2}_{2}+M'^{2}_{2}\right)\mgthdsq\right\}\\
&-\left(\frac{3g'^{2}}{2}+\frac{9g^{2}_{2}}{2}\right)\left(m^{2}_{H_{d}}+\left|\tilde{\mu}\right|^{2}\right)\\
&+\left\{\left[6\bdf^{*}_{d}\bdf^{T}_{d}+2\bdf^{*}_{e}\bdf^{T}_{e}\right]_{kk}+\sbi\sh\mgtphdsq+3\swi\sh\mgthdsq\right\}\left(m^{2}_{H_{d}}+\left|\tilde{\mu}\right|^{2}\right)\;,\\
\end{split}\end{equation}
\begin{equation}\begin{split}
{\left(4\pi\right)}^2\frac{d\mqsq_{ij}}{dt}=&\left\{\frac{1}{3}\glp^{2}\delta_{ij}+2\left[\fuurs\fuurt\right]_{ij}\right\}\left(m^{2}_{H_{u}}+\left|\tilde{\mu}\right|^{2}\right)\\
&+\left\{-\frac{1}{3}\glp^{2}\delta_{ij}+2\left[\fddrs\fddrt\right]_{ij}\right\}\left(m^{2}_{H_{d}}+\left|\tilde{\mu}\right|^{2}\right)\\
&-\frac{2}{3}\suk\glp^{2}\delta_{ij}\musq_{kk}+\frac{1}{3}\sqk\glp^{2}\delta_{ij}\mqsq_{kk}\\
&+\sqk\sql\left(\frac{\glp^{2}}{18}+\frac{3\gtwl^{2}}{2}+\frac{8\gthl^{2}}{3}\right)\delta_{ik}\delta_{lj}\mqsq_{kl}\\
&+\frac{1}{3}\sdk\glp^{2}\delta_{ij}\mdsq_{kk}-\frac{1}{3}\slk\glp^{2}\delta_{ij}\mlsq_{kk}+\frac{1}{3}\sek\glp^{2}\delta_{ij}\mesq_{kk}\\
&+2\suk\sul\fuhus_{ik}\musq_{kl}\fuhut_{lj}+2\sdk\sdl\fdhds_{ik}\mdsq_{kl}\fdhdt_{lj}\\
&+2\suk\aus_{ik}\aut_{kj}+2\suk\nbmtfuhus_{ik}\nbmtsfuhut_{kj}\\
&+2\sdk\ads_{ik}\adt_{kj}+2\sdk\nbmtfdhds_{ik}\nbmtsfdhdt_{kj}\\
&-\frac{2}{9}\sbi\left(M^{2}_{1}+M'^{2}_{1}\right)\bgtpq_{ik}\bgtpq^{\dagger}_{kj}-6\swi\left(M^{2}_{2}+M'^{2}_{2}\right)\bgtq_{ik}\bgtq^{\dagger}_{kj}\\
&-\frac{32}{3}\sgl\left(M^{2}_{3}+M'^{2}_{3}\right)\bgtsq_{ik}\bgtsq^{\dagger}_{kj}\\
&-4\sh\left|\mu\right|^{2}\left[\bftuq^{*}_{ik}\bftuq^{T}_{kj}+\bftdq^{*}_{ik}\bftdq^{T}_{kj}\right]\\
&-3\left(\sqi+\sqj\right)\left(\frac{1}{36}g'^{2}+\frac{3}{4}g^{2}_{2}+\frac{4}{3}g^{2}_{3}\right)\mqsq_{ij}\\
&+\sql\left[\frac{1}{18}\sbi\bgtpq_{ik}\bgtpq^{\dagger}_{kl}+\frac{3}{2}\swi\bgtq_{ik}\bgtq^{\dagger}_{kl}+\frac{8}{3}\sgl\bgtsq_{ik}\bgtsq^{\dagger}_{kl}\right.\\
&\left.\qquad\quad+\sh\bftuq^{*}_{ik}\bftuq^{T}_{kl}+\sh\bftdq^{*}_{ik}\bftdq^{T}_{kl}\right]\mqsq_{lj}\\
&+\sqk\mqsq_{ik}\left[\frac{1}{18}\sbi\bgtpq_{kl}\bgtpq^{\dagger}_{lj}+\frac{3}{2}\swi\bgtq_{kl}\bgtq^{\dagger}_{lj}+\frac{8}{3}\sgl\bgtsq_{kl}\bgtsq^{\dagger}_{lj}\right.\\
&\left.\qquad\qquad\qquad\ +\sh\bftuq^{*}_{kl}\bftuq^{T}_{lj}+\sh\bftdq^{*}_{kl}\bftdq^{T}_{lj}\right]\;,\\
\end{split}\end{equation}
\begin{equation}\begin{split}\label{appeq:rgemup}
{\left(4\pi\right)}^2\frac{d\musq_{ij}}{dt}=&\left\{-\frac{4}{3}\glp^{2}\delta_{ij}+4\left[\fuult\fuuls\right]_{ij}\right\}\left(m^{2}_{H_{u}}+\left|\tilde{\mu}\right|^{2}\right)+\frac{4}{3}\glp^{2}\delta_{ij}\left(m^{2}_{H_{d}}+\left|\tilde{\mu}\right|^{2}\right)\\
&+\frac{8}{3}\suk\glp^{2}\delta_{ij}\musq_{kk}+\frac{8}{3}\suk\sul\left[\frac{1}{3}\glp^{2}+\gthl^{2}\right]\delta_{ik}\delta_{lj}\musq_{kl}\\
&-\frac{4}{3}\sqk\glp^{2}\delta_{ij}\mqsq_{kk}-\frac{4}{3}\sdk\glp^{2}\delta_{ij}\mdsq_{kk}+\frac{4}{3}\slk\glp^{2}\delta_{ij}\mlsq_{kk}\\
&-\frac{4}{3}\sek\glp^{2}\delta_{ij}\mesq_{kk}+4\sqk\sql\fuhut_{ik}\fuhus_{lj}\mqsq_{kl}\\
&+4\sqk\aut_{ik}\aus_{kj}+4\sqk\nbmtsfuhut_{ik}\nbmtfuhus_{kj}\\
&-\frac{32}{9}\sbi\left(M^{2}_{1}+M'^{2}_{1}\right)\bgtpur^{\dagger}_{ik}\bgtpur_{kj}-\frac{32}{3}\sgl\left(M^{2}_{3}+M'^{2}_{3}\right)\bgtsur^{\dagger}_{ik}\bgtsur_{kj}\\
&-8\sh\left|\mu\right|^{2}\bftur^{T}_{ik}\bftur^{*}_{kj}-3\left(\sui+\suj\right)\left(\frac{4}{9}g'^{2}+\frac{4}{3}g^{2}_{3}\right)\musq_{ij}\\
&+\sul\left[\frac{8}{9}\sbi\bgtpur^{\dagger}_{ik}\bgtpur_{kl}+\frac{8}{3}\sgl\bgtsur^{\dagger}_{ik}\bgtsur_{kl}+2\sh\bftur^{T}_{ik}\bftur^{*}_{kl}\right]\musq_{lj}\\
&+\suk\musq_{ik}\left[\frac{8}{9}\sbi\bgtpur^{\dagger}_{kl}\bgtpur_{lj}+\frac{8}{3}\sgl\bgtsur^{\dagger}_{kl}\bgtsur_{lj}+2\sh\bftur^{T}_{kl}\bftur^{*}_{lj}\right]\;,\\
\end{split}\end{equation}
\begin{equation}\begin{split}
{\left(4\pi\right)}^2\frac{d\mdsq_{ij}}{dt}=&\frac{2}{3}\glp^{2}\delta_{ij}\left(m^{2}_{H_{u}}+\left|\tilde{\mu}\right|^{2}\right)+\left\{-\frac{2}{3}\glp^{2}\delta_{ij}+4\left[\fddlt\fddls\right]_{ij}\right\}\left(m^{2}_{H_{d}}+\left|\tilde{\mu}\right|^{2}\right)\\
&-\frac{4}{3}\suk\glp^{2}\delta_{ij}\musq_{kk}+\frac{2}{3}\sqk\glp^{2}\delta_{ij}\mqsq_{kk}+\frac{2}{3}\sdk\glp^{2}\delta_{ij}\mdsq_{kk}\\
&+\frac{2}{3}\sdk\sdl\left[\frac{1}{3}\glp^{2}+4\gthl^{2}\right]\delta_{ik}\delta_{lj}\mdsq_{kl}-\frac{2}{3}\slk\glp^{2}\delta_{ij}\mlsq_{kk}\\
&+\frac{2}{3}\sek\glp^{2}\delta_{ij}\mesq_{kk}+4\sqk\sql\fdhdt_{ik}\fdhds_{lj}\mqsq_{kl}\\
&+4\sqk\adt_{ik}\ads_{kj}+4\sqk\nbmtsfdhdt_{ik}\nbmtfdhds_{kj}\\
&-\frac{8}{9}\sbi\left(M^{2}_{1}+M'^{2}_{1}\right)\bgtpdr^{\dagger}_{ik}\bgtpdr_{kj}-\frac{32}{3}\sgl\left(M^{2}_{3}+M'^{2}_{3}\right)\bgtsdr^{\dagger}_{ik}\bgtsdr_{kj}\\
&-8\sh\left|\mu\right|^{2}\bftdr^{T}_{ik}\bftdr^{*}_{kj}-3\left(\sdi+\sdj\right)\left(\frac{1}{9}g'^{2}+\frac{4}{3}g^{2}_{3}\right)\mdsq_{ij}\\
&+\sdl\left[\frac{2}{9}\sbi\bgtpdr^{\dagger}_{ik}\bgtpdr_{kl}+\frac{8}{3}\sgl\bgtsdr^{\dagger}_{ik}\bgtsdr_{kl}+2\sh\bftdr^{T}_{ik}\bftdr^{*}_{kl}\right]\mdsq_{lj}\\
&+\sdk\mdsq_{ik}\left[\frac{2}{9}\sbi\bgtpdr^{\dagger}_{kl}\bgtpdr_{lj}+\frac{8}{3}\sgl\bgtsdr^{\dagger}_{kl}\bgtsdr_{lj}+2\sh\bftdr^{T}_{kl}\bftdr^{*}_{lj}\right]\;,\\
\end{split}\end{equation}
\begin{equation}\begin{split}
{\left(4\pi\right)}^2\frac{d\mlsq_{ij}}{dt}=&-\glp^{2}\delta_{ij}\left(m^{2}_{H_{u}}+\left|\tilde{\mu}\right|^{2}\right)+\left\{\glp^{2}\delta_{ij}+2\left[\feers\feert\right]_{ij}\right\}\left(m^{2}_{H_{d}}+\left|\tilde{\mu}\right|^{2}\right)\\
&+2\suk\glp^{2}\delta_{ij}\musq_{kk}-\sqk\glp^{2}\delta_{ij}\mqsq_{kk}-\sdk\glp^{2}\delta_{ij}\mdsq_{kk}\\
&+\slk\glp^{2}\delta_{ij}\mlsq_{kk}+\slk\sll\left(\frac{\glp^{2}}{2}+\frac{3\gtwl^{2}}{2}\right)\delta_{ik}\delta_{lj}\mlsq_{kl}\\
&-\sek\glp^{2}\delta_{ij}\mesq_{kk}+2\sek\sel\fehds_{ik}\mesq_{kl}\fehdt_{lj}\\
&+2\sek\aes_{ik}\aet_{kj}+2\sek\nbmtfehds_{ik}\nbmtsfehdt_{kj}\\
&-2\sbi\left(M^{2}_{1}+M'^{2}_{1}\right)\bgtpl_{ik}\bgtpl^{\dagger}_{kj}-6\swi\left(M^{2}_{2}+M'^{2}_{2}\right)\bgtl_{ik}\bgtl^{\dagger}_{kj}\\
&-4\sh\left|\mu\right|^{2}\bftel^{*}_{ik}\bftel^{T}_{kj}-3\left(\sLi+\sLj\right)\left(\frac{1}{4}g'^{2}+\frac{3}{4}g^{2}_{2}\right)\mlsq_{ij}\\
&+\sll\left[\frac{1}{2}\bgtpl_{ik}\bgtpl^{\dagger}_{kl}\sbi+\frac{3}{2}\bgtl_{ik}\bgtl^{\dagger}_{kl}\swi+\bftel^{*}_{ik}\bftel^{T}_{kl}\sh\right]\mlsq_{lj}\\
&+\slk\mlsq_{ik}\left[\frac{1}{2}\bgtpl_{kl}\bgtpl^{\dagger}_{lj}\sbi+\frac{3}{2}\bgtl_{kl}\bgtl^{\dagger}_{lj}\swi+\bftel^{*}_{kl}\bftel^{T}_{lj}\sh\right]\;,\\
\end{split}\end{equation}
\begin{equation}\begin{split}
{\left(4\pi\right)}^2\frac{d\mesq_{ij}}{dt}=&2\glp^{2}\delta_{ij}\left(m^{2}_{H_{u}}+\left|\tilde{\mu}\right|^{2}\right)+\left\{-2\glp^{2}\delta_{ij}+4\left[\feelt\feels\right]_{ij}\right\}\left(m^{2}_{H_{d}}+\left|\tilde{\mu}\right|^{2}\right)\\
&-4\suk\glp^{2}\delta_{ij}\musq_{kk}+2\sqk\glp^{2}\delta_{ij}\mqsq_{kk}+2\sdk\glp^{2}\delta_{ij}\mdsq_{kk}\\
&-2\slk\glp^{2}\delta_{ij}\mlsq_{kk}+2\sek\glp^{2}\delta_{ij}\mesq_{kk}\\
&+2\sek\sel\glp^{2}\delta_{lj}\delta_{ik}\mesq_{kl}+4\slk\sll\fehdt_{ik}\mlsq_{kl}\fehds_{lj}\\
&+4\slk\aet_{ik}\aes_{kj}+4\slk\nbmtsfehdt_{ik}\nbmtfehds_{kj}\\
&-8\sbi\left(M^{2}_{1}+M'^{2}_{1}\right)\bgtper^{\dagger}_{ik}\bgtper_{kj}-8\sh\left|\mu\right|^{2}\bfter^{T}_{ik}\bfter^{*}_{kj}\\
&-3\left(\sei+\sej\right)g'^{2}\mesq_{ij}\\
&+\sel\left[2\bgtper^{\dagger}_{ik}\bgtper_{kl}\sbi+2\bfter^{T}_{ik}\bfter^{*}_{kl}\sh\right]\mesq_{lj}\\
&+\sek\mesq_{ik}\left[2\bgtper^{\dagger}_{kl}\bgtper_{lj}\sbi+2\bfter^{T}_{kl}\bfter^{*}_{lj}\sh\right]\;.\\
\end{split}\end{equation}
\newpage
Below the scale $Q=m_{H}$, as discussed in Sec.~\ref{subsec:tri}, the trilinear couplings to the doublet $\mathsf{h}$, and the mass parameter $m^{2}_{\mathsf{h}}$ remain in the theory, with RGEs given by,
\vspace{1cm}

\noindent${\left(4\pi\right)}^2\frac{d\left[\sn\au_{ij}-\cs\mtsfuhu_{ij}\right]}{dt}$
\begin{equation}\label{eq:triu}\begin{split}
=&\h\suk\left[\sn\au_{ik}-\cs\mtsfuhu_{ik}\right]\left[\frac{2\glp^{2}}{3}\left(\cs^{2}-\sn^{2}\right)\delta_{kj}+2\sn^{2}\left[\fuul^{\dagger}\fuul\right]_{kj}\right]\\
&+\sul\sqk\left[-2\left(\frac{\glp^{2}}{9}+\frac{4\gthl^{2}}{3}\right)\delta_{ik}\delta_{lj}+6\fuhu_{ij}\fuhu^{\dagger}_{lk}\right]\left[\sn\au_{kl}-\cs\mtsfuhu_{kl}\right]\\
&+2\h\sqk\left[\left(\frac{\glp^{2}}{12}-\frac{3\gtwl^{2}}{4}\right)\left(\sn^{2}-\cs^{2}\right)\delta_{ik}+2\sn^{2}\left[\fuur\fuur^{\dagger}\right]_{ik}-\cs^{2}\left[\fddr\fddr^{\dagger}\right]_{ik}\right]\\
&\qquad\qquad\qquad\qquad\qquad\qquad\qquad\qquad\qquad\qquad\qquad\qquad\qquad\times\left[\sn\au_{kj}-\cs\mtsfuhu_{kj}\right]\\
&+\frac{2}{3}\sbi\sn\left(M_{1}-iM'_{1}\right)\\
&\qquad\quad\times\left(\sh(\gtphu)^{*}\bgtpq^{*}_{ik}\bftur_{kj}-\frac{4}{3}\bgtpq^{*}_{ik}(\bdf_{u})_{kl}\bgtpur^{*}_{kj}-4\sh\bftuq_{ik}\bgtpur^{*}_{kj}(\gtphu)^{*}\right)\\
&-\frac{32}{3}\sgl\sn\left(M_{3}-iM'_{3}\right)\bgtsq^{*}_{ik}(\bdf_{u})_{kl}\bgtsur^{*}_{lj}-6\swi\sh\sn\left(M_{2}-iM'_{2}\right)(\gthu)^{*}\bgtq^{*}_{ik}\bftur_{kj}\\
&+\frac{2}{3}\sbi\sh\cs\mu^{*}\gtphd\left(4\bftuq_{ik}\bgtpur^{*}_{kj}-\bgtpq^{*}_{ik}\bftur_{kj}\right)+6\sh\swi\cs\mu^{*}\gthd\bgtq^{*}_{ik}\bftur_{kj}\\
&-4\sh\cs\mu^{*}\bftdq_{ik}(\bdf^{\dagger}_{d})_{kl}\bftur_{lj}\\
&+\suk\left[\sn\au_{ik}-\cs\mtsfuhu_{ik}\right]\left[\frac{8}{9}\sbi\bgtpur^{T}_{kl}\bgtpur^{*}_{lj}+\frac{8}{3}\sgl\bgtsur^{T}_{kl}\bgtsur^{*}_{lj}+2\sh\bftur^{\dagger}_{kl}\bftur_{lj}\right]\\
&+\h\left[3\sn^{2}(\bdf^{\dagger}_{u})_{kl}(\bdf_{u})_{lk}+\cs^{2}\left\{3(\bdf^{\dagger}_{d})_{kl}(\bdf_{d})_{lk}+(\bdf^{\dagger}_{e})_{kl}(\bdf_{e})_{lk}\right\}\right]\left[\sn\au_{ij}-\cs\mtsfuhu_{ij}\right]\\
&+\frac{1}{2}\h\sh\left[\cs^{2}\left\{\sbi\mgtphdsq+3\swi\mgthdsq\right\}+\sn^{2}\left\{\sbi\mgtphusq+3\swi\mgthusq\right\}\right]\\
&\qquad\qquad\qquad\qquad\qquad\qquad\qquad\qquad\qquad\qquad\qquad\times\left[\sn\au_{ij}-\cs\mtsfuhu_{ij}\right]\\
&+\sql\left[\sh\bftuq_{ik}\bftuq^{\dagger}_{kl}+\sh\bftdq_{ik}\bftdq^{\dagger}_{kl}+\frac{1}{18}\sbi\bgtpq^{*}_{ik}\bgtpq^{T}_{kl}+\frac{3}{2}\swi\bgtq^{*}_{ik}\bgtq^{T}_{kl}\right.\\
&\left.\qquad\quad+\frac{8}{3}\sgl\bgtsq^{*}_{ik}\bgtsq^{T}_{kl}\right]\left[\sn\au_{lj}-\cs\mtsfuhu_{lj}\right]\\
&-3\left\{\left(\frac{1}{36}\sqi+\frac{4}{9}\suj+\frac{1}{4}\h\right)g'^{2}+\frac{3}{4}\left(\sqi+\h\right)g^{2}_{2}+\frac{4}{3}\left(\sqi+\suj\right)g^{2}_{3}\right\}\\
&\qquad\qquad\qquad\qquad\qquad\qquad\qquad\qquad\qquad\qquad\qquad\qquad\times\left[\sn\au-\cs\mtsfuhu\right]_{ij}\;,
\end{split}\end{equation}
\noindent${\left(4\pi\right)}^2\frac{d\left[\cs\ad_{ij}-\sn\mtsfdhd_{ij}\right]}{dt}$
\begin{equation}\begin{split}
=&2\h\sqk\left[-\left(\frac{\glp^{2}}{12}+\frac{3\gtwl^{2}}{4}\right)\left(\cs^{2}-\sn^{2}\right)\delta_{ik}-\sn^{2}\left[\fuur\fuur^{\dagger}\right]_{ik}+2\cs^{2}\left[\fddr\fddr^{\dagger}\right]_{ik}\right]\\
&\qquad\qquad\qquad\qquad\qquad\qquad\qquad\qquad\qquad\qquad\qquad\qquad\qquad\times\left[\cs\ad_{kj}-\sn\mtsfdhd_{kj}\right]\\
&+\sdl\sqk\left[2\left(\frac{\glp^{2}}{18}-\frac{4\gthl^{2}}{3}\right)\delta_{ik}\delta_{lj}+6\fdhd^{\dagger}_{lk}\fdhd_{ij}\right]\left[\cs\ad_{kl}-\sn\mtsfdhd_{kl}\right]\\
&+2\sel\slk\fdhd_{ij}\fehd^{\dagger}_{lk}\left[\cs\ae_{kl}-\sn\mtsfehd_{kl}\right]\\
&+\h\sdk\left[\cs\ad_{ik}-\sn\mtsfdhd_{ik}\right]\left[-\frac{\glp^{2}}{3}\left(\cs^{2}-\sn^{2}\right)\delta_{kj}+2\cs^{2}\left[\fddl^{\dagger}\fddl\right]_{kj}\right]\\
&-4\sh\sn\mu^{*}\bftuq_{ik}(\bdf^{\dagger}_{u})_{kl}\bftdr_{lj}+\frac{2}{3}\sbi\sh\sn\mu^{*}\gtphu\left(2\bftdq_{ik}\bgtpdr^{*}_{kj}+\bgtpq^{*}_{ik}\bftdr_{kj}\right)\\
&+6\sh\swi\sn\mu^{*}\gthu\bgtq^{*}_{ik}\bftdr_{kj}\\
&+\frac{2}{3}\sbi\cs\left(M_{1}-iM'_{1}\right)\\
&\qquad\qquad\times\left(-\sh(\gtphd)^{*}\bgtpq^{*}_{ik}\bftdr_{kj}+\frac{2}{3}\bgtpq^{*}_{ik}(\bdf_{d})_{kl}\bgtpdr^{*}_{lj}-2\sh\bftdq_{ik}\bgtpdr^{*}_{kj}(\gtphd)^{*}\right)\\
&-\frac{32}{3}\sgl\cs\left(M_{3}-iM'_{3}\right)\bgtsq^{*}_{ik}(\bdf_{d})_{kl}\bgtsdr^{*}_{lj}-6\swi\sh\cs\left(M_{2}-iM'_{2}\right)(\gthd)^{*}\bgtq^{*}_{ik}\bftdr_{kj}\\
&+\sql\left[\sh\bftuq_{ik}\bftuq^{\dagger}_{kl}+\sh\bftdq_{ik}\bftdq^{\dagger}_{kl}+\frac{1}{18}\sbi\bgtpq^{*}_{ik}\bgtpq^{T}_{kl}+\frac{3}{2}\swi\bgtq^{*}_{ik}\bgtq^{T}_{kl}\right.\\
&\left.\qquad\quad +\frac{8}{3}\sgl\bgtsq^{*}_{ik}\bgtsq^{T}_{kl}\right]\left[\cs\ad_{lj}-\sn\mtsfdhd_{lj}\right]\\
&+\h\left[3\sn^{2}(\bdf_{u})_{kl}(\bdf^{\dagger}_{u})_{lk}+\cs^{2}\left\{3(\bdf_{d})_{kl}(\bdf^{\dagger}_{d})_{lk}+(\bdf_{e})_{kl}(\bdf^{\dagger}_{e})_{lk}\right\}\right]\left[\cs\ad_{ij}-\sn\mtsfdhd_{ij}\right]\\
&+\frac{1}{2}\h\sh\left[\cs^{2}\left\{\sbi\mgtphdsq+3\swi\mgthdsq\right\}+\sn^{2}\left\{\sbi\mgtphusq+3\swi\mgthusq\right\}\right]\\
&\qquad\qquad\qquad\qquad\qquad\qquad\qquad\qquad\qquad\qquad\qquad\times\left[\cs\ad_{ij}-\sn\mtsfdhd_{ij}\right]\\
&+\sdk\left[\cs\ad_{ik}-\sn\mtsfdhd_{ik}\right]\left[\frac{2}{9}\sbi\bgtpdr^{T}_{kl}\bgtpdr^{*}_{lj}+\frac{8}{3}\sgl\bgtsdr^{T}_{kl}\bgtsdr^{*}_{lj}+2\sh\bftdr^{\dagger}_{kl}\bftdr_{lj}\right]\\
&-3\left\{\left(\frac{1}{36}\sqi+\frac{1}{9}\sdj+\frac{1}{4}\h\right)g'^{2}+\frac{3}{4}\left(\sqi+\h\right)g^{2}_{2}+\frac{4}{3}\left(\sqi+\sdj\right)g^{2}_{3}\right\}\\
&\qquad\qquad\qquad\qquad\qquad\qquad\qquad\qquad\qquad\qquad\qquad\qquad\times\left[\cs\ad-\sn\mtsfdhd\right]_{ij}\;,
\end{split}\end{equation}
\noindent${\left(4\pi\right)}^2\frac{d\left[\cs\ae_{ij}-\sn\mtsfehd_{ij}\right]}{dt}$
\begin{equation}\begin{split}
=&2\h\slk\left[\left(\frac{\glp^{2}}{4}-\frac{3\gtwl^{2}}{4}\right)\left(\cs^{2}-\sn^{2}\right)\delta_{ik}+2\cs^{2}\left[\feer\feer^{\dagger}\right]_{ik}\right]\left[\cs\ae_{kj}-\sn\mtsfehd_{kj}\right]\\
&+\sel\slk\left[-\glp^{2}\delta_{ik}\delta_{lj}+2\fehd^{\dagger}_{lk}\fehd_{ij}\right]\left[\cs\ae_{kl}-\sn\mtsfehd_{kl}\right]\\
&+6\sdl\sqk\fehd_{ij}\fdhd^{\dagger}_{lk}\left[\cs\ad_{kl}-\sn\mtsfdhd_{kl}\right]\\
&+\h\sek\left[\cs\ae_{ik}-\sn\mtsfehd_{ik}\right]\left[-\glp^{2}\left(\cs^{2}-\sn^{2}\right)\delta_{kj}+2\cs^{2}\left[\feel^{\dagger}\feel\right]_{kj}\right]\\
&+2\sbi\sh\sn\mu^{*}\gtphu\left(2\bftel_{ik}\bgtper^{*}_{kj}-\bgtpl^{*}_{ik}\bfter_{kj}\right)+6\sh\swi\sn\mu^{*}\gthu\bgtl^{*}_{ik}\bfter_{kj}\\
&+2\sbi\cs\left(M_{1}-iM'_{1}\right)\\
&\qquad\qquad\times\left(\sh(\gtphd)^{*}\bgtpl^{*}_{ik}\bfter_{kj}-2\bgtpl^{*}_{ik}(\bdf_{e})_{kl}\bgtper^{*}_{lj}-2\sh\bftel_{ik}\bgtper^{*}_{kj}(\gtphd)^{*}\right)\\
&-6\swi\sh\cs\left(M_{2}-iM'_{2}\right)(\gthd)^{*}\bgtl^{*}_{ik}\bfter_{kj}\\
&+\sll\left[\sh\bftel_{ik}\bftel^{\dagger}_{kl}+\frac{1}{2}\sbi\bgtpl^{*}_{ik}\bgtpl^{T}_{kl}+\frac{3}{2}\swi\bgtl^{*}_{ik}\bgtl^{T}_{kl}\right]\left[\cs\ae_{lj}-\sn\mtsfehd_{lj}\right]\\
&+\h\left[3\sn^{2}(\bdf_{u})_{kl}(\bdf^{\dagger}_{u})_{lk}+\cs^{2}\left\{3(\bdf_{d})_{kl}(\bdf^{\dagger}_{d})_{lk}+(\bdf_{e})_{kl}(\bdf^{\dagger}_{e})_{lk}\right\}\right]\left[\cs\ae_{ij}-\sn\mtsfehd_{ij}\right]\\
&+\frac{1}{2}\h\sh\left[\cs^{2}\left\{\sbi\mgtphdsq+3\swi\mgthdsq\right\}+\sn^{2}\left\{\sbi\mgtphusq+3\swi\mgthusq\right\}\right]\\
&\qquad\qquad\qquad\qquad\qquad\qquad\qquad\qquad\qquad\qquad\qquad\times\left[\cs\ae_{ij}-\sn\mtsfehd_{ij}\right]\\
&+\sek\left[\cs\ae_{ik}-\sn\mtsfehd_{ik}\right]\left[2\sbi\bgtper^{T}_{kl}\bgtper^{*}_{lj}+2\sh\bfter^{\dagger}_{kl}\bfter_{lj}\right]\\
&-3\left\{\left(\frac{1}{4}\sLi+\sej+\frac{1}{4}\h\right)g'^{2}+\frac{3}{4}\left(\sLi+\h\right)g^{2}_{2}\right\}\left[\cs\ae-\sn\mtsfehd\right]_{ij}\;,
\end{split}\end{equation}
\noindent${\left(4\pi\right)}^{2}\frac{d\left[\sn^{2}\left(m^{2}_{H_{u}}+\left|\tilde{\mu}\right|^{2}\right)+\cs^{2}\left(m^{2}_{H_{d}}+\left|\tilde{\mu}\right|^{2}\right)-\sn\cs\left(b+b^{*}\right)\right]}{dt}$
\begin{equation}\begin{split}
=&\frac{3}{2}\h\left[\glp^{2}+\gtwl^{2}\right]\left(\cs^{2}-\sn^{2}\right)^{2}\left[\sn^{2}\left(m^{2}_{H_{u}}+\left|\tilde{\mu}\right|^{2}\right)+\cs^{2}\left(m^{2}_{H_{d}}+\left|\tilde{\mu}\right|^{2}\right)-\sn\cs\left(b+b^{*}\right)\right]\\
&+\suk\sul\left[-2\glp^{2}\left(\sn^{2}-\cs^{2}\right)\delta_{lk}+6\sn^{2}\left[\fuult\fuuls\right]_{lk}\right]\musq_{kl}\\
&+\sqk\sql\left[\glp^{2}\left(\sn^{2}-\cs^{2}\right)\delta_{lk}+6\sn^{2}\left[\fuurs\fuurt\right]_{lk}+6\cs^{2}\left[\fddrs\fddrt\right]_{lk}\right]\mqsq_{kl}\\
&+\sdk\sdl\left[\glp^{2}\left(\sn^{2}-\cs^{2}\right)\delta_{lk}+6\cs^{2}\left[\fddlt\fddls\right]_{lk}\right]\mdsq_{kl}\\
&+\slk\sll\left[-\glp^{2}\left(\sn^{2}-\cs^{2}\right)\delta_{lk}+2\cs^{2}\left[\feers\feert\right]_{lk}\right]\mlsq_{kl}\\
&+\sek\sel\left[\glp^{2}\left(\sn^{2}-\cs^{2}\right)\delta_{lk}+2\cs^{2}\left[\feelt\feels\right]_{lk}\right]\mesq_{kl}\\
&+6\suk\sql\left[\sn\au_{lk}-\cs\mtsfuhu_{lk}\right]\left[\sn\au^{\dagger}_{kl}-\cs\mtsfuhu^{\dagger}_{kl}\right]\\
&+6\sql\sdk\left[\cs\ad_{lk}-\sn\mtsfdhd_{lk}\right]\left[\cs\ad^{\dagger}_{kl}-\sn\mtsfdhd^{\dagger}_{kl}\right]\\
&+2\sll\sek\left[\cs\ae_{lk}-\sn\mtsfehd_{lk}\right]\left[\cs\ae^{\dagger}_{kl}-\sn\mtsfehd^{\dagger}_{kl}\right]\\
&-2\sh\left|\mu\right|^{2}\left\{\sbi\left[\sn^{2}\mgtphusq+\cs^{2}\mgtphdsq\right]+3\swi\left[\sn^{2}\mgthusq+\cs^{2}\mgthdsq\right]\right\}\\
&-2\sh\left\{\sbi\left(M^{2}_{1}+M'^{2}_{1}\right)\left[\sn^{2}\mgtphusq+\cs^{2}\mgtphdsq\right]+3\swi\left(M^{2}_{2}+M'^{2}_{2}\right)\left[\sn^{2}\mgthusq+\cs^{2}\mgthdsq\right]\right\}\\
&-\frac{1}{2}\left\{-4\sh\sbi\sn\cs\mu^{*}\gtphu\gtphd\left(M_{1}+iM'_{1}\right)-12\sh\swi\sn\cs\mu^{*}\gthu\gthd\left(M_{2}+iM'_{2}\right)\right\}\\
&-\frac{1}{2}\left\{-4\sh\sbi\sn\cs\mu(\gtphu)^{*}(\gtphd)^{*}\left(M_{1}-iM'_{1}\right)-12\sh\swi\sn\cs\mu(\gthu)^{*}(\gthd)^{*}\left(M_{2}-iM'_{2}\right)\right\}\\
&-\h\left(\frac{3g'^{2}}{2}+\frac{9g^{2}_{2}}{2}\right)\left[\sn^{2}\left(m^{2}_{H_{u}}+\left|\tilde{\mu}\right|^{2}\right)+\cs^{2}\left(m^{2}_{H_{d}}+\left|\tilde{\mu}\right|^{2}\right)-\sn\cs\left(b+b^{*}\right)\right]\\
&+\h\left[\sn^{2}\left\{\left[6\bdf^{*}_{u}\bdf^{T}_{u}\right]_{kk}+\sbi\sh\mgtphusq+3\swi\sh\mgthusq\right\}\right.\\
&\left.\qquad\qquad\qquad+\cs^{2}\left\{\left[6\bdf^{*}_{d}\bdf^{T}_{d}+2\bdf^{*}_{e}\bdf^{T}_{e}\right]_{kk}+\sbi\sh\mgtphdsq+3\swi\sh\mgthdsq\right\}\right]\\
&\qquad\qquad\qquad\qquad\qquad\qquad\qquad\times\left[\sn^{2}\left(m^{2}_{H_{u}}+\left|\tilde{\mu}\right|^{2}\right)+\cs^{2}\left(m^{2}_{H_{d}}+\left|\tilde{\mu}\right|^{2}\right)-\sn\cs\left(b+b^{*}\right)\right]\;.\\
\end{split}\end{equation}
\newpage
With $\theta_{H}=0$, RGEs for the remaining SSB scalar mass parameters take the form,
\begin{equation}\begin{split}\label{app:mQ2rgelow}
{\left(4\pi\right)}^2\frac{d\mqsq_{ij}}{dt}=&\h\left\{-\frac{1}{3}\left(\cs^{2}-\sn^{2}\right)\glp^{2}\delta_{ij}+2\sn^{2}\left[\fuurs\fuurt\right]_{ij}+2\cs^{2}\left[\fddrs\fddrt\right]_{ij}\right\}\\
&\qquad\qquad\qquad\qquad\quad\times\left[\sn^{2}\left(m^{2}_{H_{u}}+\left|\tilde{\mu}\right|^{2}\right)+\cs^{2}\left(m^{2}_{H_{d}}+\left|\tilde{\mu}\right|^{2}\right)-\sn\cs\left(b+b^{*}\right)\right]\\
&-\frac{2}{3}\suk\glp^{2}\delta_{ij}\musq_{kk}+\frac{1}{3}\sqk\glp^{2}\delta_{ij}\mqsq_{kk}\\
&+\sqk\sql\left(\frac{\glp^{2}}{18}+\frac{3\gtwl^{2}}{2}+\frac{8\gthl^{2}}{3}\right)\delta_{ik}\delta_{lj}\mqsq_{kl}\\
&+\frac{1}{3}\sdk\glp^{2}\delta_{ij}\mdsq_{kk}-\frac{1}{3}\slk\glp^{2}\delta_{ij}\mlsq_{kk}+\frac{1}{3}\sek\glp^{2}\delta_{ij}\mesq_{kk}\\
&+2\suk\sul\fuhus_{ik}\musq_{kl}\fuhut_{lj}+2\sdk\sdl\fdhds_{ik}\mdsq_{kl}\fdhdt_{lj}\\
&+2\suk\h\left[\sn\aus_{ik}-\cs\mtfuhus_{ik}\right]\left[\sn\aut_{kj}-\cs\mtsfuhut_{kj}\right]\\
&+2\sdk\h\left[\cs\ads_{ik}-\sn\mtfdhds_{ik}\right]\left[\cs\adt_{kj}-\sn\mtsfdhdt_{kj}\right]\\
&-\frac{2}{9}\sbi\left(M^{2}_{1}+M'^{2}_{1}\right)\bgtpq_{ik}\bgtpq^{\dagger}_{kj}-6\swi\left(M^{2}_{2}+M'^{2}_{2}\right)\bgtq_{ik}\bgtq^{\dagger}_{kj}\\
&-\frac{32}{3}\sgl\left(M^{2}_{3}+M'^{2}_{3}\right)\bgtsq_{ik}\bgtsq^{\dagger}_{kj}\\
&-4\sh\left|\mu\right|^{2}\left[\bftuq^{*}_{ik}\bftuq^{T}_{kj}+\bftdq^{*}_{ik}\bftdq^{T}_{kj}\right]\\
&-3\left(\sqi+\sqj\right)\left(\frac{1}{36}g'^{2}+\frac{3}{4}g^{2}_{2}+\frac{4}{3}g^{2}_{3}\right)\mqsq_{ij}\\
&+\sql\left[\frac{1}{18}\sbi\bgtpq_{ik}\bgtpq^{\dagger}_{kl}+\frac{3}{2}\swi\bgtq_{ik}\bgtq^{\dagger}_{kl}+\frac{8}{3}\sgl\bgtsq_{ik}\bgtsq^{\dagger}_{kl}\right.\\
&\qquad\quad\left.+\sh\bftuq^{*}_{ik}\bftuq^{T}_{kl}+\sh\bftdq^{*}_{ik}\bftdq^{T}_{kl}\right]\mqsq_{lj}\\
&+\sqk\mqsq_{ik}\left[\frac{1}{18}\sbi\bgtpq_{kl}\bgtpq^{\dagger}_{lj}+\frac{3}{2}\swi\bgtq_{kl}\bgtq^{\dagger}_{lj}+\frac{8}{3}\sgl\bgtsq_{kl}\bgtsq^{\dagger}_{lj}\right.\\
&\qquad\qquad\qquad\ \ \left.+\sh\bftuq^{*}_{kl}\bftuq^{T}_{lj}+\sh\bftdq^{*}_{kl}\bftdq^{T}_{lj}\right]\;,\\
\end{split}\end{equation}
\begin{equation}\begin{split}
{\left(4\pi\right)}^2\frac{d\musq_{ij}}{dt}=&\h\left\{\frac{4}{3}\left(\cs^{2}-\sn^{2}\right)\glp^{2}\delta_{ij}+4\sn^{2}\left[\fuult\fuuls\right]_{ij}\right\}\\
&\qquad\qquad\qquad\qquad\quad\times\left[\sn^{2}\left(m^{2}_{H_{u}}+\left|\tilde{\mu}\right|^{2}\right)+\cs^{2}\left(m^{2}_{H_{d}}+\left|\tilde{\mu}\right|^{2}\right)-\sn\cs\left(b+b^{*}\right)\right]\\
&+\frac{8}{3}\suk\glp^{2}\delta_{ij}\musq_{kk}+\frac{8}{3}\suk\sul\left[\frac{1}{3}\glp^{2}+\gthl^{2}\right]\delta_{ik}\delta_{lj}\musq_{kl}\\
&-\frac{4}{3}\sqk\glp^{2}\delta_{ij}\mqsq_{kk}-\frac{4}{3}\sdk\glp^{2}\delta_{ij}\mdsq_{kk}+\frac{4}{3}\slk\glp^{2}\delta_{ij}\mlsq_{kk}\\
&-\frac{4}{3}\sek\glp^{2}\delta_{ij}\mesq_{kk}+4\sqk\sql\fuhut_{ik}\fuhus_{lj}\mqsq_{kl}\\
&+4\sqk\h\left[\sn\aut_{ik}-\cs\mtsfuhut_{ik}\right]\left[\sn\aus_{kj}-\cs\mtfuhus_{kj}\right]\\
&-\frac{32}{9}\sbi\left(M^{2}_{1}+M'^{2}_{1}\right)\bgtpur^{\dagger}_{ik}\bgtpur_{kj}-\frac{32}{3}\sgl\left(M^{2}_{3}+M'^{2}_{3}\right)\bgtsur^{\dagger}_{ik}\bgtsur_{kj}\\
&-8\sh\left|\mu\right|^{2}\bftur^{T}_{ik}\bftur^{*}_{kj}-3\left(\sui+\suj\right)\left(\frac{4}{9}g'^{2}+\frac{4}{3}g^{2}_{3}\right)\musq_{ij}\\
&+\sul\left[\frac{8}{9}\sbi\bgtpur^{\dagger}_{ik}\bgtpur_{kl}+\frac{8}{3}\sgl\bgtsur^{\dagger}_{ik}\bgtsur_{kl}+2\sh\bftur^{T}_{ik}\bftur^{*}_{kl}\right]\musq_{lj}\\
&+\suk\musq_{ik}\left[\frac{8}{9}\sbi\bgtpur^{\dagger}_{kl}\bgtpur_{lj}+\frac{8}{3}\sgl\bgtsur^{\dagger}_{kl}\bgtsur_{lj}+2\sh\bftur^{T}_{kl}\bftur^{*}_{lj}\right]\;,\\
\end{split}\end{equation}
\begin{equation}\begin{split}
{\left(4\pi\right)}^2\frac{d\mdsq_{ij}}{dt}=&\h\left\{-\frac{2}{3}\left(\cs^{2}-\sn^{2}\right)\glp^{2}\delta_{ij}+4\cs^{2}\left[\fddlt\fddls\right]_{ij}\right\}\\
&\qquad\qquad\qquad\qquad\quad\times\left[\sn^{2}\left(m^{2}_{H_{u}}+\left|\tilde{\mu}\right|^{2}\right)+\cs^{2}\left(m^{2}_{H_{d}}+\left|\tilde{\mu}\right|^{2}\right)-\sn\cs\left(b+b^{*}\right)\right]\\
&-\frac{4}{3}\suk\glp^{2}\delta_{ij}\musq_{kk}+\frac{2}{3}\sqk\glp^{2}\delta_{ij}\mqsq_{kk}+\frac{2}{3}\sdk\glp^{2}\delta_{ij}\mdsq_{kk}\\
&+\frac{2}{3}\sdk\sdl\left[\frac{1}{3}\glp^{2}+4\gthl^{2}\right]\delta_{ik}\delta_{lj}\mdsq_{kl}-\frac{2}{3}\slk\glp^{2}\delta_{ij}\mlsq_{kk}\\
&+\frac{2}{3}\sek\glp^{2}\delta_{ij}\mesq_{kk}+4\sqk\sql\fdhdt_{ik}\fdhds_{lj}\mqsq_{kl}\\
&+4\sqk\h\left[\cs\adt_{ik}-\sn\mtsfdhdt_{ik}\right]\left[\cs\ads_{kj}-\sn\mtfdhds_{kj}\right]\\
&-\frac{8}{9}\sbi\left(M^{2}_{1}+M'^{2}_{1}\right)\bgtpdr^{\dagger}_{ik}\bgtpdr_{kj}-\frac{32}{3}\sgl\left(M^{2}_{3}+M'^{2}_{3}\right)\bgtsdr^{\dagger}_{ik}\bgtsdr_{kj}\\
&-8\sh\left|\mu\right|^{2}\bftdr^{T}_{ik}\bftdr^{*}_{kj}-3\left(\sdi+\sdj\right)\left(\frac{1}{9}g'^{2}+\frac{4}{3}g^{2}_{3}\right)\mdsq_{ij}\\
&+\sdl\left[\frac{2}{9}\sbi\bgtpdr^{\dagger}_{ik}\bgtpdr_{kl}+\frac{8}{3}\sgl\bgtsdr^{\dagger}_{ik}\bgtsdr_{kl}+2\sh\bftdr^{T}_{ik}\bftdr^{*}_{kl}\right]\mdsq_{lj}\\
&+\sdk\mdsq_{ik}\left[\frac{2}{9}\sbi\bgtpdr^{\dagger}_{kl}\bgtpdr_{lj}+\frac{8}{3}\sgl\bgtsdr^{\dagger}_{kl}\bgtsdr_{lj}+2\sh\bftdr^{T}_{kl}\bftdr^{*}_{lj}\right]\;,\\
\end{split}\end{equation}
\begin{equation}\begin{split}
{\left(4\pi\right)}^2\frac{d\mlsq_{ij}}{dt}=&\h\left\{\left(\cs^{2}-\sn^{2}\right)\glp^{2}\delta_{ij}+2\cs^{2}\left[\feers\feert\right]_{ij}\right\}\\
&\qquad\qquad\qquad\qquad\quad\times\left[\sn^{2}\left(m^{2}_{H_{u}}+\left|\tilde{\mu}\right|^{2}\right)+\cs^{2}\left(m^{2}_{H_{d}}+\left|\tilde{\mu}\right|^{2}\right)-\sn\cs\left(b+b^{*}\right)\right]\\
&+2\suk\glp^{2}\delta_{ij}\musq_{kk}-\sqk\glp^{2}\delta_{ij}\mqsq_{kk}-\sdk\glp^{2}\delta_{ij}\mdsq_{kk}\\
&+\slk\glp^{2}\delta_{ij}\mlsq_{kk}+\slk\sll\left(\frac{\glp^{2}}{2}+\frac{3\gtwl^{2}}{2}\right)\delta_{ik}\delta_{lj}\mlsq_{kl}\\
&-\sek\glp^{2}\delta_{ij}\mesq_{kk}+2\sek\sel\fehds_{ik}\mesq_{kl}\fehdt_{lj}\\
&+2\sek\h\left[\cs\aes_{ik}-\sn\mtfehds_{ik}\right]\left[\cs\aet_{kj}-\sn\mtsfehdt_{kj}\right]\\
&-2\sbi\left(M^{2}_{1}+M'^{2}_{1}\right)\bgtpl_{ik}\bgtpl^{\dagger}_{kj}-6\swi\left(M^{2}_{2}+M'^{2}_{2}\right)\bgtl_{ik}\bgtl^{\dagger}_{kj}\\
&-4\sh\left|\mu\right|^{2}\bftel^{*}_{ik}\bftel^{T}_{kj}-3\left(\sLi+\sLj\right)\left(\frac{1}{4}g'^{2}+\frac{3}{4}g^{2}_{2}\right)\mlsq_{ij}\\
&+\sll\left[\frac{1}{2}\bgtpl_{ik}\bgtpl^{\dagger}_{kl}\sbi+\frac{3}{2}\bgtl_{ik}\bgtl^{\dagger}_{kl}\swi+\bftel^{*}_{ik}\bftel^{T}_{kl}\sh\right]\mlsq_{lj}\\
&+\slk\mlsq_{ik}\left[\frac{1}{2}\bgtpl_{kl}\bgtpl^{\dagger}_{lj}\sbi+\frac{3}{2}\bgtl_{kl}\bgtl^{\dagger}_{lj}\swi+\bftel^{*}_{kl}\bftel^{T}_{lj}\sh\right]\;,\\
\end{split}\end{equation}
\begin{equation}\begin{split}
{\left(4\pi\right)}^2\frac{d\mesq_{ij}}{dt}=&\h\left\{-2\left(\cs^{2}-\sn^{2}\right)\glp^{2}\delta_{ij}+4\cs^{2}\left[\feelt\feels\right]_{ij}\right\}\\
&\qquad\qquad\qquad\qquad\quad\times\left[\sn^{2}\left(m^{2}_{H_{u}}+\left|\tilde{\mu}\right|^{2}\right)+\cs^{2}\left(m^{2}_{H_{d}}+\left|\tilde{\mu}\right|^{2}\right)-\sn\cs\left(b+b^{*}\right)\right]\\
&-4\suk\glp^{2}\delta_{ij}\musq_{kk}+2\sqk\glp^{2}\delta_{ij}\mqsq_{kk}+2\sdk\glp^{2}\delta_{ij}\mdsq_{kk}\\
&-2\slk\glp^{2}\delta_{ij}\mlsq_{kk}+2\sek\glp^{2}\delta_{ij}\mesq_{kk}\\
&+2\sek\sel\glp^{2}\delta_{lj}\delta_{ik}\mesq_{kl}+4\slk\sll\fehdt_{ik}\mlsq_{kl}\fehds_{lj}\\
&+4\slk\h\left[\cs\aet_{ik}-\sn\mtsfehdt_{ik}\right]\left[\cs\aes_{kj}-\sn\mtfehds_{kj}\right]\\
&-8\sbi\left(M^{2}_{1}+M'^{2}_{1}\right)\bgtper^{\dagger}_{ik}\bgtper_{kj}-8\sh\left|\mu\right|^{2}\bfter^{T}_{ik}\bfter^{*}_{kj}\\
&-3\left(\sei+\sej\right)g'^{2}\mesq_{ij}\\
&+\sel\left[2\bgtper^{\dagger}_{ik}\bgtper_{kl}\sbi+2\bfter^{T}_{ik}\bfter^{*}_{kl}\sh\right]\mesq_{lj}\\
&+\sek\mesq_{ik}\left[2\bgtper^{\dagger}_{kl}\bgtper_{lj}\sbi+2\bfter^{T}_{kl}\bfter^{*}_{lj}\sh\right]\;.\\
\end{split}\end{equation}

\newpage

\end{document}